\documentclass[twocolumn,groupedaddress,superscriptaddress,english,showpacs,aps,prb]{revtex4}

\usepackage[normalem]{ulem}

\usepackage[latin9]{inputenc}
\usepackage{graphicx}
\usepackage{amsmath}
\usepackage{amssymb}
\usepackage{bbm}
\usepackage{xspace}
\usepackage{color}
\usepackage{footnote}
\usepackage{comment}
\usepackage{placeins}
\usepackage{hyperref}
\usepackage{subfigure}

\graphicspath{{./Figures/}}

\definecolor{dgreen}{rgb}{0.0,0.5,0.0}

\definecolor{orange}{RGB}{252,77,6}
\definecolor{brown}{RGB}{200,127,50}
\definecolor{blue}{RGB}{00,000,100}
\definecolor{blue2}{RGB}{00,000,250}
\definecolor{green1}{RGB}{00,100,00}
\definecolor{green2}{RGB}{00,150,00}
\definecolor{green3}{RGB}{00,200,00}
\definecolor{green4}{RGB}{00,250,00}

\newcommand{\LCJ}{LC junction\xspace}

\newcommand{\cc}{correlated\xspace} 
\newcommand{\ec}{extended central\xspace} 

\newcommand{\beq}{\begin{equation}}
\newcommand{\eeq}{\end{equation}}

\newcommand{\vv}[1]{\boldsymbol{#1}}

\newcommand{\Imm}{{\rm Im\:}}

\newcommand{\kxy}{{{\bf k} }}

\begin{document}

\title{
Charge redistribution in correlated heterostuctures 
within nonequilibrium 
real-space dynamical mean-field theory.
}

\author{Irakli Titvinidze}
\affiliation{Institute of Theoretical and Computational Physics, Graz University
of Technology, 8010 Graz, Austria}
\author{Max E. Sorantin}
\affiliation{Institute of Theoretical and Computational Physics, Graz University
of Technology, 8010 Graz, Austria}
\author{Antonius Dorda}
\affiliation{Institute of Theoretical and Computational Physics, Graz University
of Technology, 8010 Graz, Austria}
\author{Wolfgang von der Linden}
\affiliation{Institute of Theoretical and Computational Physics, Graz University
of Technology, 8010 Graz, Austria}
\author{Enrico Arrigoni}
\affiliation{Institute of Theoretical and Computational Physics, Graz University
of Technology, 8010 Graz, Austria}


\begin{abstract} 
We address the steady-state behavior of a system consisting of several correlated monoatomic layers sandwiched between two metallic leads under the influence of a bias voltage. In particular, we investigate the interplay of the local Hubbard and the long-range Coulomb interaction on the charge redistribution at the interface, in the paramagnetic regime of the system. We provide a detailed study of the importance of the various system parameters, like Hubbard $U$, lead-correlated region coupling strength,  and the applied voltage on the charge distribution in the correlated region and in the adjacent parts of the  leads. In addition, we also present results for the steady-state current density and double occupancies.  Our results indicate that, in a certain range of parameters, the charge on the two layers at the interface between the leads and the correlated region display opposite signs producing a dipolelike layer at the interface. 
Our results are obtained within nonequilibrium (steady-state) real-space dynamical mean-field theory (R-DMFT), with a  self-consistent treatment of the long-range part of the Coulomb interaction by means of the Poisson equation. The latter is solved by the Newton-Raphson method and we find that this significantly reduces the computational cost compared to existing treatment. As impurity solver for R-DMFT we use the auxiliary master equation approach (AMEA), which addresses the impurity problem within a finite auxiliary system coupled to Markovian environments. 
\end{abstract}

\ifx\clength\undefined
\maketitle
\else
\nocomm
\fi

\section{Introduction}

Correlated systems out of equilibrium and especially electronic transport through heterostructures made from different materials, have attracted increasing interest due to the recent impressive experimental progress to fabricate correlated heterostructures\cite{is.og.01,oh.mu.02, ga.ah.02, an.ga.99, oh.hw.04, zh.wa.12} with atomic resolution and, in particular, growing atomically abrupt layers with different electronic structures\cite{is.og.01, ga.ah.02, oh.mu.02}. 

From a theoretical perspective, investigating and understanding the physical processes which govern the behavior of such systems is a great challenge in the field of theoretical solid state physics. For instance, it was shown that, due to the proximity effect, any finite number of Mott-insulating layers become metallic when sandwiched between semi-infinite metallic leads.\cite{ze.fr.09, ha.fr.11, kn.li.11, ma.am.15, ri.an.16, okam.07, okam.08, ec.we.14}  For such a geometry the effect of impact ionization in periodically driven Mott-insulating layers was studied\cite{so.do.18, ec.we.13} as well as resonance phenomena in a system consisting of several correlated and non-correlated mono-atomic layers\cite{ti.do.16}. Another challenging aspect of such systems that was investigated is the capacitance of multilayer systems made from correlated materials.\cite{st.fr.17, ba.yd.14, ha.fr.12} Due to the local Hubbard and long-range Coulomb interaction present in these systems, charge redistribution takes place.\cite{ha.fr.12, fr.zl.07, ba.es.16} The equilibrium situation was addressed, e.g., in  Refs. \onlinecite{fr.zl.07} and \onlinecite{ba.es.16}. In particular, Ref.~\onlinecite{fr.zl.07} studied the charge redistribution and the corresponding thermo-electric properties for a metal-strongly-correlated barrier-metal device where the onsite energies of the correlated region are shifted compared to the metals, while Ref.~\onlinecite{ba.es.16} investigated the behavior of the correlated thin film in a transverse electric field. Finally, Ref.~\onlinecite{ha.fr.12} considered correlated layers described by the Falicov-Kimball model, where one spin-species is immobile, with emphasis on the nonequilibrium situation arising due to an applied bias voltage.

\begin{figure}[t]
\includegraphics[width=\columnwidth]{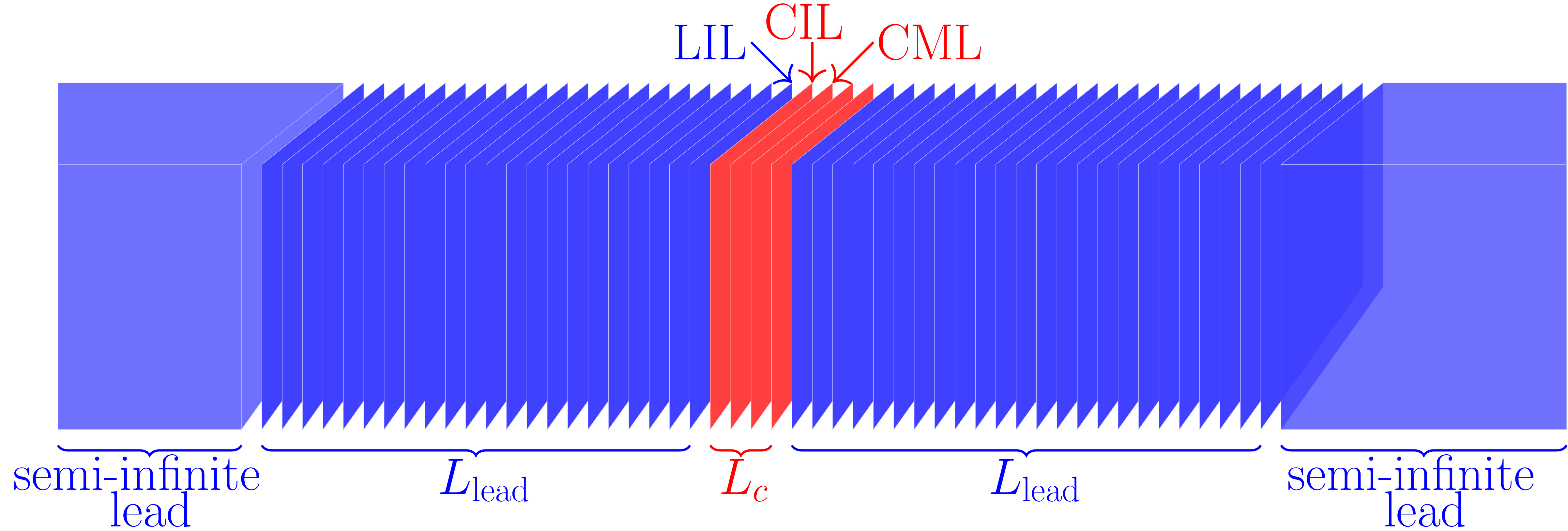}
\caption{(Color online)
A Schematic representation of the system consisting of $L_{c}=4$ correlated interfaces (red) sandwiched between two semi-infinite metallic leads (blue).  In addition to the local Hubbard interaction, present only within the correlated layers, we also take into account long-range Coulomb forces extending into the leads. We take them into account by solving the Poisson equation in an extended region including part of the lead layers ($L_{\rm lead}=23$ for each side). Here LIL, CIL, and CML stands for lead interface layer, correlated interface layer, and correlated middle layer, respectively.}
\label{schematicp}
\end{figure}

Here, we  investigate a system of correlated layers sandwiched between two metallic leads in the paramagnetic phase, see Fig.~\ref{schematicp} for an illustration. Similar to Ref.~\onlinecite{ha.fr.12} we take into account long-range Coulomb interactions, but here we use the Hubbard model where both spin-species are mobile. The goal of the current work is to investigate the influence of local Hubbard and long-range Coulomb interactions on the charge redistribution in a nonequilibrium steady state situation produced by an applied bias voltage.

We obtain that the charge density deviation from the bulk filling on opposite sites of the lead-correlated (LC) junction, have opposite sign in a certain range of parameters, indicating the formation of a dipole-like layer. According to our calculations, such a layer arises for small values of the hybridisation $t_{lc}$ at the LC-junction for all considered interactions and bias voltages. On the other hand, for large $t_{lc}$ it occurs only for weak to intermediate interactions and at low bias voltages.

To describe the behavior of the system we adopt dynamical mean-field theory (DMFT)\cite{ge.ko.96, voll.10, me.vo.89}, which is one of the most powerful methods to investigate high-dimensional strongly correlated electron systems. DMFT was originally developed to describe translationally invariant systems in equilibrium, but was later extended to inhomogeneous systems\cite{po.no.99.sm, po.no.99.ms, po.no.99.ldmft, po.no.99.emfl, free.06, no.ma.10, is.li.09, no.ma.11, okam11, mi.fr.01, free.04, ok.mi.04.ldmft, do.ko.97, do.ko.98, so.yu.08, we.ha.11u, he.co.08, ko.hi.08, ko.hi.09,  no.ka.09,  ko.ba.11, bl.go.11, ki.ki.11, au.as.15, sn.ti.08, sn.ti.11, ti.sc.12, go.ti.10, sc.ti.13, au.ti.15, ze.fr.09, okam.07, okam.08, ec.we.13, ec.we.14, ha.fr.11, ti.do.16}, and also adapted to the nonequilibrium case\cite{ao.ts.14, sc.mo.02u, fr.tu.06, free.08, jo.fr.08, ec.ko.09, okam.07, okam.08, ar.kn.13, ti.do.15, do.ti.16}. In the latter, DMFT is formulated within the nonequilibrium Green's function approach originating from the works of Kubo\cite{kubo.57}, Schwinger\cite{schw.61}, Kadanoff, Baym\cite{ba.ka.61, kad.baym} and Keldysh\cite{keld.65}. 
The only approximation in DMFT is the assumption of a local self-energy. This can be calculated by mapping the original problem onto a single impurity Anderson model (SIAM)\cite{ande.61}, whose parameters are determined 
self-consistently. For  homogeneous systems  the self-energy is the same for each lattice site due to translational symmetry,  and thus one needs to solve only one SIAM problem, per DMFT iteration, while for systems with broken translational invariance, such as the one considered here, one needs to solve many impurity problems to capture the spatial inhomogeneity of the system.
In the current work, the nonequilibrium SIAM problems are solved by using the recently developed auxiliary master equation approach (AMEA)\cite{ar.kn.13, do.nu.14, ti.do.15}, which treats the impurity problem within an auxiliary system consisting of a correlated impurity, a small number $N_B$ of uncorrelated bath sites and two Markovian environments described by a Lindblad master equation. The approach allows for an accurate solution of the steady-state impurity problem already with a small $N_B$.

For the self-consistent solution of the non-linear Poisson equation we used the Newton-Ralphson, which significantly improves the convergence.

The paper is organized as follows: Sec.~\ref{Model_and_Method} decribes the model and method. In particular, in Sec.~\ref{Model} we introduce the Hamiltonian of the system, in Sec.~\ref{RDMFT} we illustrate the application of  real-space dynamical mean-field theory within the nonequilibrium steady-state Green's function  formalism for a system consisting of many layers, in Sec.~\ref{Charge_reconstruction} we give an overview of the solution of the Poisson equation, and finally in Sec.~\ref{Self-consistency_loop} we present the self-consistency loop used to obtain the self-consistent results. Thereafter, in Sec.~\ref{Results} we present our results and our conclusions are presented in  Sec. \ref{Conclusions}.

\section{Model and Method}\label{Model_and_Method}

\subsection{Model}\label{Model}

We consider a system consisting of a \cc region (c) with $L_{c}$ correlated infinite and translationally invariant layers attached to two metallic leads ($\alpha=l,r$), which are semi-infinite in the $z$ direction and translationally invariant in the $xy$ plane (parallel to the correlated layers). The physical situation is depicted in Fig.~\ref{schematicp} and described by the Hamiltonian
\begin{eqnarray}
\label{Hamiltonian}
&&\hspace{-0.75cm}{\cal H}=
-\hspace{-0.5cm}\sum_{z, \langle {\bf r}^{\phantom\dagger},{\bf r}'\rangle, \sigma}\hspace{-0.5cm}t_{z} c_{z,{\bf r},\sigma}^\dagger c_{z,{\bf r}',\sigma}^{\phantom\dagger} 
-\hspace{-0.45cm}\sum_{\langle z, z'\rangle, {\bf r}, \sigma}\hspace{-0.45cm} t_{zz'} c_{z,{\bf r},\sigma}^\dagger c_{z',{\bf r},\sigma}^{\phantom\dagger} \nonumber \\
&&\hspace{-0.3cm}+\hspace{-0.05cm}\sum_{z,{\bf r}}\hspace{-0.05cm} U_z n_{z,{\bf r},\uparrow}n_{z,{\bf r},\downarrow}  
  +\hspace{-0.25cm}\sum_{z,{\bf r},\sigma}\hspace{-0.2cm}\left(v_z^{(0)}+v_z\right)n_{z,{\bf r},\sigma} \, .
\end{eqnarray}
Here $c_{z,{\bf r},\sigma}^\dagger$ creates an electron at site ${\bf r}=(x,y)$  of  layer $z$ with spin ${\sigma}$ and $n_{z, {\bf r}, \sigma}=c_{z, {\bf r}, \sigma}^\dagger c_{z, {\bf r}, \sigma}^{\phantom\dagger}$ denotes the corresponding occupation-number operator. $\langle z, z'\rangle$ stands for neighboring $z$ and $z'$ layers and $\langle {\bf r}^{\phantom\dagger},{\bf r}'\rangle$ stands for neighboring ${\bf r}$ and ${\bf r}'$ sites in the same  layer. 

The first two terms of the Hamiltonian \eqref{Hamiltonian} describe nearest-neighbor intra-layer and inter-layer hoppings, with hopping amplitudes $t_{z}$ and $t_{zz'}$, respectively. The third term introduces the local Hubbard interactions $U_z$, which are nonzero only for the \cc region. The last term describes the onsite energies, whereby $v_z^{(0)}$ is chosen such that we obtain the required bulk filling in the $z$-th layer with the special case of $v_z^{(0)}=-U_z/2$ at half-filling (HF). 
Furthermore, $v_z$  describes the Hartree shift of the onsite energies obtained after the mean-field decoupling of the long-range Coulomb interaction (LRCI),  $V_{ij} (n_i-1)  (n_j-1)$ which is produced by the charge inhomogeneity and has to be determined self-consistently.
In contrast to the local Hubbard interaction, the LRCI affects not only the \cc region, but the leads as well. Therefore, we incorporate parts of the leads, namely $L_{\rm lead}$ layers per side, into the region. Here, $L_{\rm lead}$ has to be chosen large enough such that the (self-consistently determined) electron density $v_z$ converges to the bulk filling of the leads far away from the correlated region. To summarize, the \ec region contains $L=L_{c}+2L_{\rm lead}$ layers. The corresponding indices vary from $-\frac{L-1}{2}$ to $\frac{L-1}{2}$.
$|z| < L_{c}/2$ describes the \cc region,  $L_{c}/2<|z|< L_{c}/2+L_{\rm lead}$ corresponds to the left ($z<0$) and right ($z>0$) leads which we treat explicitly, while $|z|> L_{c}/2+L_{\rm lead}$ corresponds to the semi-infinite lead layers ($z<0$ left lead and $z>0$ right lead). Here we note that the chosen labeling convention leads to half-integer indices for even $L_c$ considered throughout the paper.

We take the Hubbard interaction to be uniform within the \cc region, i.e. $U_{z}=U$ for $|z|< L_{c}/2$ and $U_{z}=0$ on the lead layers ($|z|>L_{c}/2$). We assume isotropic nearest-neighbor hopping parameters  within the \cc region as well as in the leads, respectively. This amounts to the choice $t_{zz'}=t_{z}\equiv t_c$ for the \cc region ($|z| < L_{c}/2$) and $t_{zz'}=t_{z}\equiv t_{\alpha=l,r}$ for the leads ($|z|> L_{c}/2$). Finally, the lead \cc region junction (\LCJ) coupling is the same on both sides $t_{-\frac{L_{c}+1}{2},-\frac{L_{c}-1}{2}} =t_{\frac{L_{c}-1}{2},\frac{L_{c}+1}{2}}\equiv t_{lc}$. We work in units where $e=\hbar=k_b=a=1$, with $a$ denoting the lattice spacing and take $t_c=1$ as unit of energy.

The nonequilibrium situation is reached by applying a bias voltage $V=v_l-v_r$. Here $v_l$ and $v_r$ are the onsite energies far away from the \cc region ($v_{l/r}=v_{z=\pm \infty}$).
Notice that, in general, $V$ is not equal to the difference between the chemical potentials of the leads $\Delta \mu=\mu_l -\mu_r$ due to the contribution from back-scattered electrons.~\cite{potz.89}.

To investigate steady-state properties of our system, we work within the Keldysh Green's function formalism~\cite{kad.baym, schw.61, keld.65, ha.ja, ra.sm.86} and use real-space dynamical mean-field theory (R-DMFT) combined with the Poisson equation to treat the Hubbard interaction and long range coulomb forces, respectively.

\begin{figure}[t!]
\includegraphics[width=0.85\columnwidth]{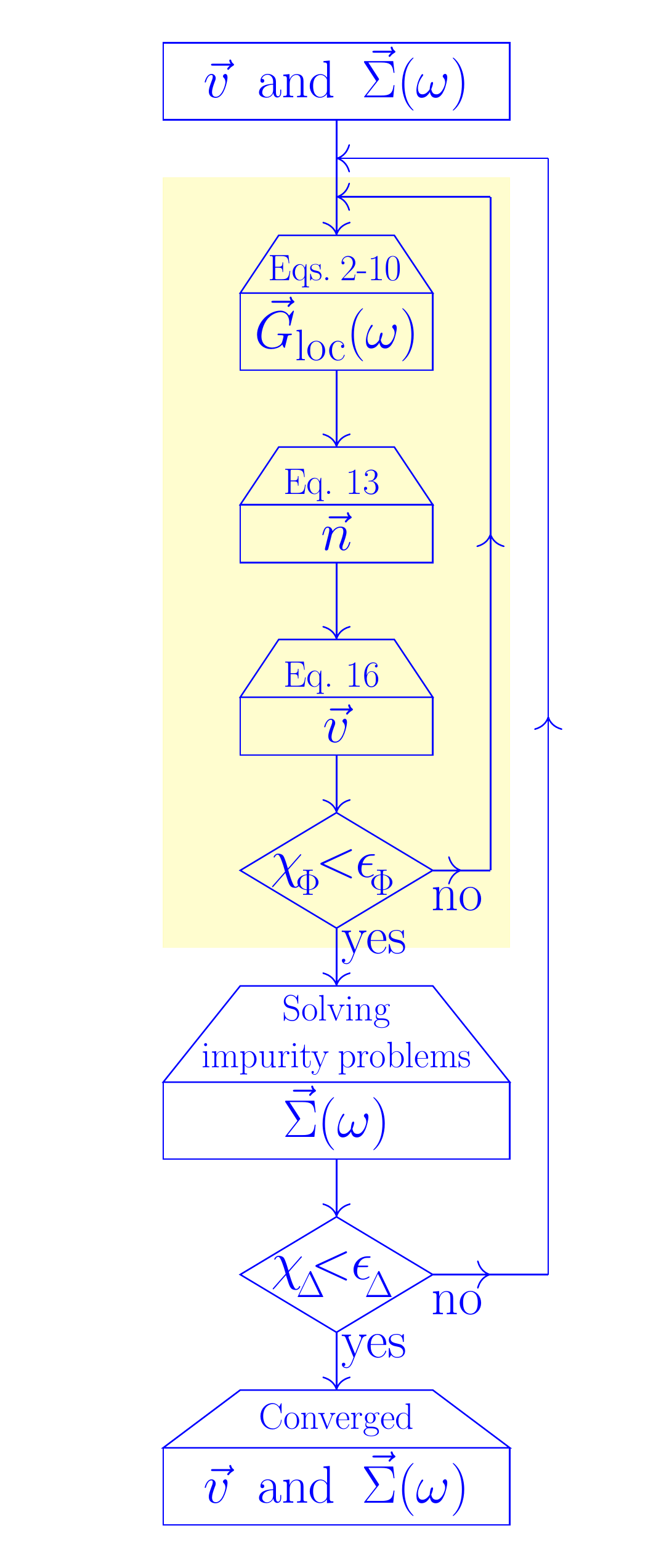}
\caption{(Color online) 
A visualization of the self-consistency loop. The shaded area corresponds to the Poisson loop and we use the vector notation for $z$-dependent quantities, $\vec{v},\vec{n},\vec{\Sigma}(\omega)$ and $\vec{G}_{\rm loc}(\omega)$ introduced in Sec.\ref{Charge_reconstruction}. Moreover, $\chi_{_\Phi}$ and $\chi_{_\Delta}$ are cost functions for the convergence criteria defined in Eqs.\eqref{converge_Posson} and \eqref{converge_DMFT}, respectively. 
}
\label{Fig:convergence}
\end{figure}

\subsection{Real-space Dynamical Mean-Field theory }\label{RDMFT}

Here, we give only a brief overview of the nonequilibrium real-space DMFT approach\cite{sc.mo.02u, fr.tu.06, free.08, jo.fr.08, ec.ko.09, okam.07, okam.08, ti.do.16} together with the employed impurity solver, namely the Auxiliary Master Exquation Approach (AMEA)\cite{ar.kn.13, do.nu.14, ti.do.15, do.so.17}

In the nonequilibrium situation, the model remains translationally invariant along the $xy$ plane (parallel to the layers), which allows to introduce the corresponding momenta $\kxy=(k_x,k_y)$. Moreover, since the steady state Green's functions depend only on the time difference, it is convenient to transform them to the frequency domain $\omega$.

The Green's function for the  \ec region, which consists of  $L=L _{c}+2L_{\rm lead}$ layers, can be expressed via Dyson's equation
\begin{eqnarray}
\label{GR}
&&[{\bf G^{-1}}]^{\gamma}(\omega, \kxy) =[{\bf G}_0^{-1}(\omega, \kxy)]^\gamma - {\boldsymbol \Sigma}^\gamma(\omega)  \,.
\end{eqnarray} 
Here, boldface indicate $L \times L$ matrices, while $\gamma$ stands for retarded (R), advanced (A) and Keldysh (K) components. ${\bf G}^A$ and ${\bf G}^R$ are related via ${\bf G}^A=({\bf G}^R)^\dagger$, while ${\bf G}^K$, in general, is independent of ${\bf G}^{R}$ and needs to be determined separately.

The inverse of the non-interacting Green's function reads 
\begin{align} 
[{\bf G^{-1}_0}]^R_{zz'} (\omega, \kxy) =&t_{zz'}+\delta_{zz'}\hspace{-0.05cm}\left(\omega - v_z-v_{z}^{(0)} - E_z(\kxy)\right) \nonumber\\
&-\delta_{zz'} \Sigma^R_{{\rm hyb},z}(\omega, \kxy)  \, ,
\label{g0R}\\
[{\bf G^{-1}_0}]^K_{zz'} (\omega, \kxy)=&-\delta_{zz'}\Sigma^K_{{\rm hyb},z}(\omega, \kxy)\;.
\label{g0K}
\end{align} 
Where $E_z(\kxy)$ is the dispersion relation for the $z$-th layer of the the \ec region and 
\begin{equation}
\label{Delta}
\Sigma^\gamma_{{\rm hyb},z}(\omega, \kxy)=\delta_{z,-\frac{L-1}{2}}t_l^2 g_l^\gamma (\omega, \kxy)+ \delta_{z,\frac{L-1}{2}}t_r^2 g_r^\gamma (\omega, \kxy) \, ,
\end{equation}
describes the hybridization between the semi-infinite leads and the \ec region. $g_l^\gamma (\omega, \kxy)$ and $g_r^\gamma (\omega, \kxy)$ denote the  Green's functions for the interface layers of the semi-infinite leads disconnected from the \ec region. Their retarded component can be  expressed as\cite{po.no.99.sm, po.no.99.ms, hayd.80}
\begin{align}
g_\alpha^R(\omega, \kxy)=& \frac{\omega -v_\alpha-v_\alpha^{(0)}-E_\alpha(\kxy)}{2t_\alpha^2}  
\nonumber \\
&- i\frac{\sqrt{4t_\alpha^2-(\omega-v_\alpha-v_\alpha^{(0)}-E_\alpha(\kxy))^2}}{2t_\alpha^2} \, .
\label{galphaR}
\end{align}
where $v_{\alpha=l/r}+v_{\alpha=l/r}^{(0)}$ and $E_{\alpha=l/r}(\kxy)$ denote the onsite energies and the dispersion relation for the left/right lead, respectively. The sign of the square-root for negative argument in \eqref{galphaR} must be chosen such that the Green's function has the correct $1/\omega$ behavior for $|\omega|\to \infty$. Since the disconnected leads are separately in equilibrium, we can obtain their Keldysh components from the retarded ones via the fluctuation dissipation theorem\cite{ha.ja}
\begin{equation}
g_\alpha^K(\omega, \kxy)= 2i(1-2f_\alpha(\omega))\;\Imm g_{\alpha}^R(\omega,\kxy) \, .
\label{galphaK}
\end{equation}
Here, $f_\alpha(\omega)$ is the Fermi distribution for the chemical potential $\mu_\alpha$ and temperature $T_\alpha$.

Finally ${\boldsymbol\Sigma}^\gamma_{zz'}(\omega)=\delta_{zz'} \Sigma^\gamma_z(\omega)$ stands for the self-energy matrix, which due to the DMFT approximation is diagonal and $\kxy$-independent. To determine it, we map each correlated layer $z$ to  a (nonequilibrium) single impurity problem (SIAM) with Hubbard interaction $U_z$ and onsite energy $v_z+v_z^{(0)}$, coupled to a self-consistently determined bath. The latter is specified by its hybridization function obtained as (see e.g. Ref.~\onlinecite{voll.10, ti.do.16})
\begin{eqnarray}
\label{DeltaR}
&&\hspace{-0.5cm}\Delta_{z}^R(\omega)=\omega - v_z-v_z^{(0)} -\Sigma^R_z(\omega) -\frac{1}{G_{{\rm loc},z}^R(\omega)} \, , \\
&&\hspace{-0.5cm}\Delta_{z}^K(\omega)=- \Sigma^K_z(\omega)+\frac{G_{{\rm loc},z}^K(\omega)}{|G_{{\rm loc},z}^R(\omega)|^2} \
\label{DeltaK} 
\end{eqnarray}
where the local Green's function is defined as
\begin{equation}
G_{{\rm loc},z}^\gamma(\omega)=\int\limits_{\rm BZ} \frac{d^2{\kxy}}{(2\pi)^2} {\bf G}_{zz}^\gamma(\omega,\kxy) \;.
\label{G_loc}
\end{equation}
To calculate the  diagonal elements of the matrices ${\bf G}^\gamma(\omega,\kxy)$ from Eq. \eqref{GR} we use the recursive Green's function method\cite{th.ki.81, le.ca.13, ec.we.13, ti.do.16} which we generalize to the present situation of Keldysh Green's functions\cite{ti.do.16}.

To describe the lattice structure of the isolated layers we use a Bethe-lattice density of state (DOS). Due to this choice, we can replace $E_{z}(\vv{k})$ by $t_{z}\varepsilon$ and 
$\int \frac{d\vv{k}}{(2\pi)^2}$ by $\int d\varepsilon \rho(\varepsilon)$, where $\varepsilon$ is a dimensionless parameter characterizing the energy and $\rho(\varepsilon)=\frac{1}{\pi}\sqrt{4-\varepsilon^2}$ is the Bethe-lattice DOS.

The corresponding impurity problems are then solved with AMEA which is a state-of-the-art impurity solver particularly suited to address the steady state. AMEA is based upon mapping\cite{ar.kn.13, do.so.17} the SIAM to an open quantum system of finite size, which includes one correlated site, $N_B$ non-interacting bath sites and two Markovian environments, whose dynamics is governed by a Lindblad master equation. The resulting open quantum system can then be solved by numerical many-body techniques such as Krylov-space based\cite{do.nu.14,ti.do.15} methods (which are the ones we use here), matrix product states (MPS)\cite{do.ga.15} or the so called stochastic wave function algorithm\cite{br.ka.97, Quantum_Jumps}.

\subsection{Charge reconstruction}\label{Charge_reconstruction}

To take into account long range Coulomb forces on a mean-field level, we calculate the onsite energies $v_z$ self-consistently by solving the corresponding Poisson equation
\begin{eqnarray}
\label{Poisson}
\frac{\partial}{\partial z}
\left(
\frac{1}{c_z}\frac{\partial v_z}{\partial z}\right) =-\left(n_z - n^{\rm bulk}
\right) \, .
\end{eqnarray}
It is convenient to adopt
von Neumann boundary conditions, which in discretized form amounts to setting the Coulomb potential of the two bulk semi infinite leads equal to the one of the boundary layers of the \ec region : 
\begin{eqnarray}
\label{Poisson_boundary}
v_{l/r}=v_{\mp\frac{L-1}{2}} \, .
\end{eqnarray}
Here $c_z\equiv \frac{1}{\varepsilon_0 \varepsilon_{r,z}}$, $\varepsilon_{r,z}$ is the relative permittivity of layer $z$ and $\varepsilon_0$ is the permittivity of free space. Moreover
\begin{eqnarray}
\label{n_z} 
n_z=1+\frac{1}{2\pi}\int\limits_{-\infty}^\infty d\omega~\Im mG_{{\rm loc},z}^K(\omega)
\end{eqnarray}
is the electron density at layer  $z$ obtained from nonequilibrium R-DMFT and $n^{\rm bulk}$ is the bulk electron density, which we set equal to $1$ (half-filling) throughout this paper.\cite{z_dependent_layers} 

One way to proceed would be to fix the bias voltage $V$ and in the present particle-hole symmetric case $v_{l}=-v_{r}=V/2$. In this case, one should adjust the asymptotic chemical potentials $\mu_l$ and $\mu_r$ of the leads to obtain the correct asymptotic charge neutrality $n_{z\to \pm \infty} \to n^{\rm bulk}=1$. This is numerically demanding. Another alternative is to carry out the calculations for given $\mu_l=-\mu_r=\Delta \mu/2$ and  update the values of the onsite energies in the semi-infinite leads after each iteration according Eq.~\ref{Poisson_boundary}. The bias voltage is then determined by $V=v_l-v_r$ {\it a posteriori}. Here we follow the second strategy as it is numerically more convenient. In fact, we find that the difference between $\Delta \mu$ and $V$ is quite small in most of the calculations presented in this paper (1\% or smaller), except for weak to intermediate $U$ at large $t_{lc}$, as we will discuss below. 

For better readability we introduce a vector notation for the $z$-dependent quantities, namely
\begin{eqnarray*}
&&\hspace{-0.5cm}\vec{v}= \{v_{-\frac{L-1}{2}},\ldots, v_{\frac{L-1}{2}}\}, \\
&&\hspace{-0.5cm}\vec{n}= \{n_{-\frac{L-1}{2}},\ldots, n_{\frac{L-1}{2}}\},\\
&&\hspace{-0.5cm}\vec{G}_{\rm loc}(\omega)= \{G_{{\rm loc}, -\frac{L-1}{2}}^{R},\ldots, G_{{\rm loc},\frac{L-1}{2}}^{R}, G_{{\rm loc}, -\frac{L-1}{2}}^{K},\ldots, G_{{\rm loc},\frac{L-1}{2}}^{K} \}\\
&&\hspace{-0.5cm}\vec{\Sigma}(\omega)= \{\Sigma_{-\frac{L-1}{2}}^{R},\ldots, \Sigma_{\frac{L-1}{2}}^{R}, \Sigma_{-\frac{L-1}{2}}^{K},\ldots, \Sigma_{\frac{L-1}{2}}^{K}\}.
\end{eqnarray*}
Obviously, the elements of $\vec \Sigma$ are zero outside of the correlated region.

The electron densities depend, through $G_{{\rm loc},z}^K(\omega)$ in Eq.~\eqref{n_z}, on the onsite energies as well as on the self-energy. The self-energy in turn is, through the self-consistency in R-DMFT, a functional of the onsite energies and of itself, i.e. $\vec{\Sigma}=\vec{\Sigma}(\vec{v},\vec{\Sigma})$. Thus, we have to solve Eqs. \eqref{Poisson}-\eqref{n_z} together with the R-DMFT equations in a self consistent manner.

For a fixed self-energy $\vec{\Sigma}(\omega)$, we solve Eq.\eqref{Poisson}-\eqref{n_z} by formulating it as a root searching problem which we treat by the Newton-Raphson method. To this end, we define the function
\begin{equation}
\label{Phij}
\Phi_z(\vec{v})=c_z\left[\frac{\partial}{\partial z}\left(\frac{1}{c_z}\frac{\partial v_z}{\partial z}\right) +\left(n_z(\vec{v}, \vec{\Sigma}) - n^{\rm bulk}\right)\right]\,
\end{equation}
of which we seek the zero. Following the Newton-Raphson scheme, we expand
\begin{equation}
\label{Phi_expansion}
\Phi_{j}(\vec{v} + \Delta \vec{v}) = \Phi_{j}(\vec{v}) + \sum_{i} \frac{\partial \Phi_{j}(\vec{v})}{\partial v_{i}}  \Delta v_{i} \, .
\end{equation}
Here $\Delta \vec{v}={\vec{v}}^{(n+1)}-{\vec{v}}^{(n)}$ is the difference between two consecutive iterations in the self-consistent Poisson loop. Assuming $\Phi_{z}(\vec{v} + \Delta \vec{v})\overset{!}{=} 0$, one obtains the following iteration scheme
\begin{equation}
\label{iteration_scheme}
\vec{v}^{(n+1)}=\vec{v}^{(n)}-M^{-1}\vec{\Phi}(\vec{v}) \, ,
\end{equation}
with $\vec{\Phi}=\{\Phi_{-\frac{L-1}{2}},\ldots, \Phi_{\frac{L-1}{2}}\}$ and 
\begin{equation}
\label{Mji}
M_{ji}=\frac{\partial}{\partial v_{i}} \Phi_{j}(\vec{v}) \, .
\end{equation}
For the technical details about the discretization of the Poisson equation and the expression for the matrix elements $M_{ij}$ we refer to Appendix \ref{appendix}.

\begin{center}
\begin{figure*}[t!]
\subfigure[]{
\label{Fig:N_vs_z_U4U8tlc0p2tlc1}
\begin{minipage}[b]{0.75\textwidth}
\centering \includegraphics[width=1\textwidth]{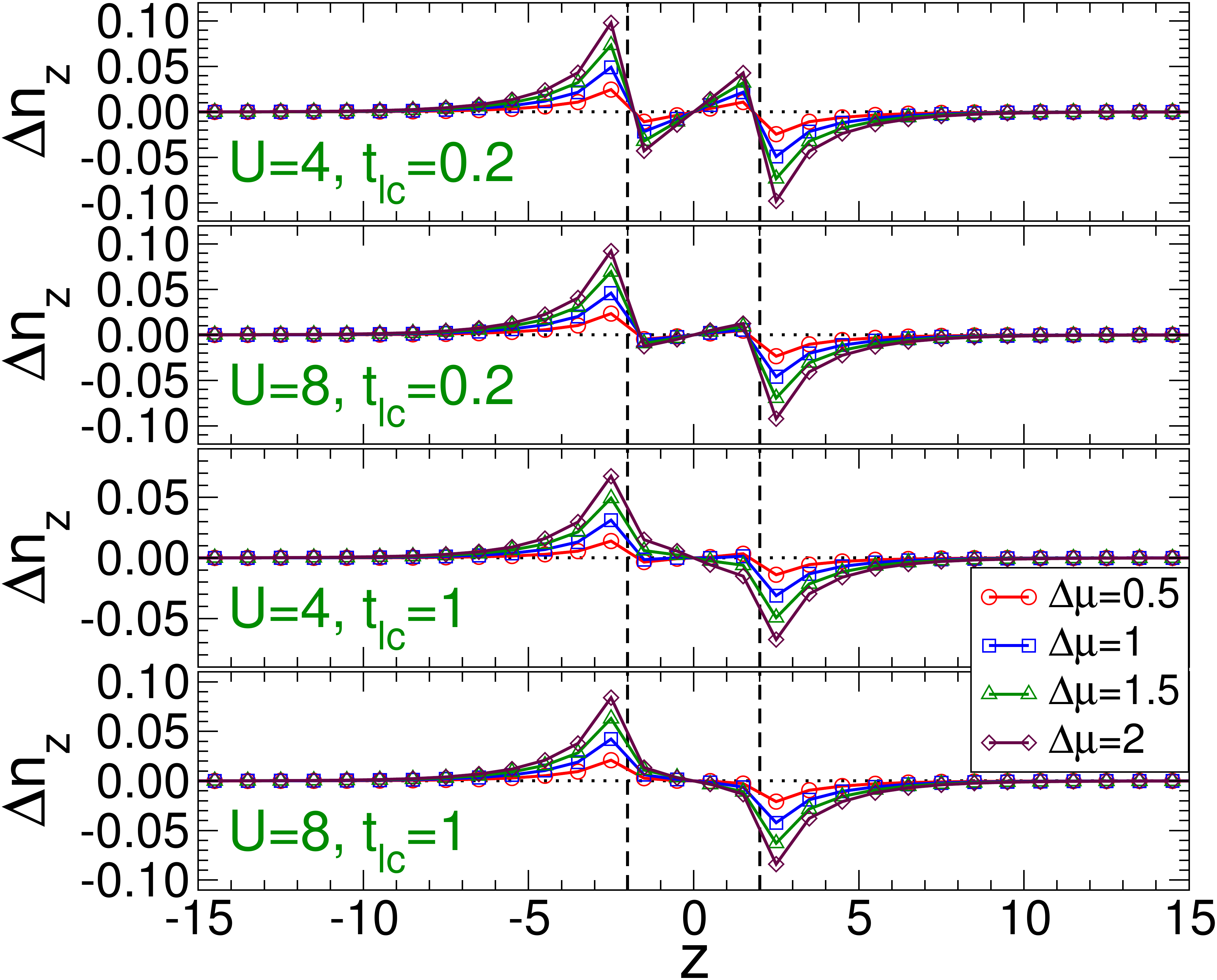}
\end{minipage}}
\\
\subfigure[]{
\label{Fig:N_vs_Phi_U4U8tlc0p2tlc1}
\begin{minipage}[b]{0.45\textwidth}
\centering \includegraphics[width=1\textwidth]{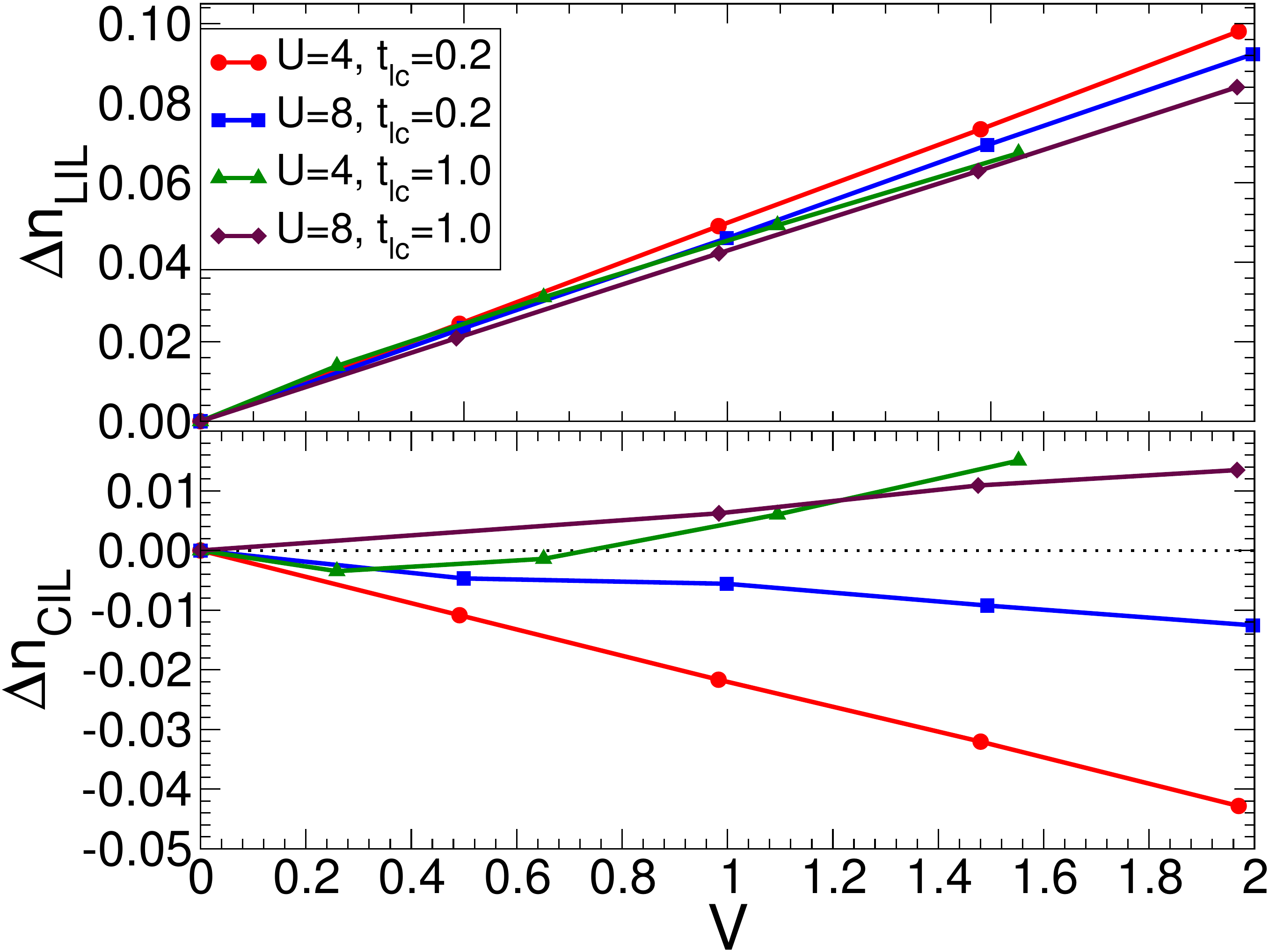}
\end{minipage}}
\hspace{0.025\textwidth}
\subfigure[]{
\label{Fig:tildePhi_vs_Bias_U4U8tlc0p2tlc1}
\begin{minipage}[b]{0.45\textwidth}
\centering \includegraphics[width=1\textwidth]{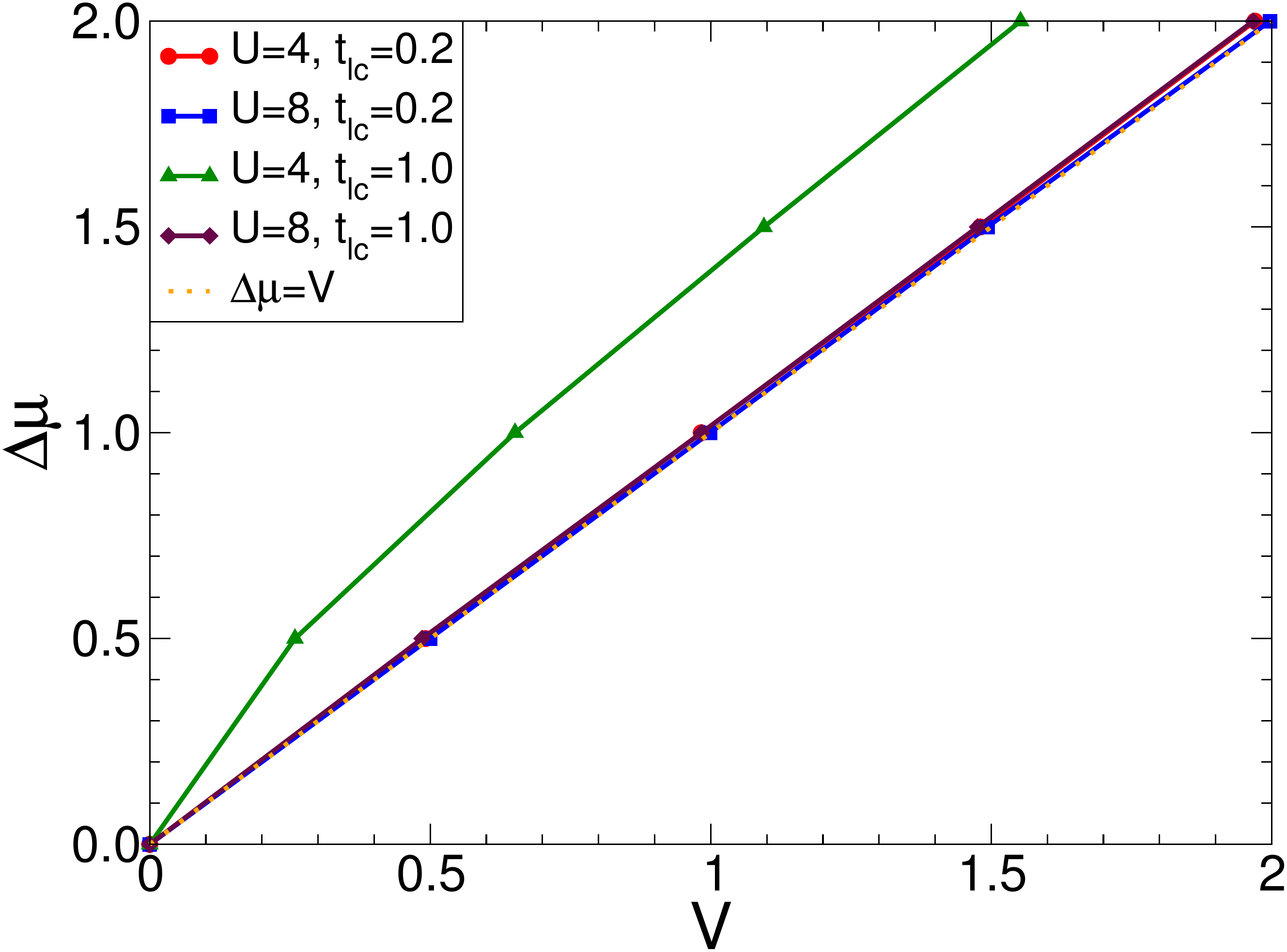}
\end{minipage}}
\caption{(Color online) Charge density deviation from half-filling $\Delta n_z=n_z-1$ (a) as a function of layer index $z$ for Hubbard interactions $U=4,~8$, \LCJ coupling $t_{lc}=t_{rc}=0.2,~1$, and for different values of $\Delta \mu =\mu_l-\mu_r$. We present results for $L_c=4$ correlated layers. Other parameters are $t_c=1$, $t_l=t_r=2$, $v_z^{(0)}=-U_z/2$ and $c=1.5$. 
The black dashed lines separate the correlated region and leads. 
(b) upper panel $\Delta n_z$ for the lead interface layer (LIL) and lower panel $\Delta n_z$  for the \cc region interface layer (CIL) as a function of the bias voltage $V=v_l-v_r$. (c) Dependence of $\Delta \mu$ on the bias voltage $V$. }
\label{Fig:N_U4U8tlc0p2tlc1}
\end{figure*}
\end{center}

\subsection{Self-consistency loop}\label{Self-consistency_loop}

Here, we describe the self-consistency loop used to determine the self-energies $\vec{\Sigma}(\omega)$ together with the onsite energies $\vec{v}$ as self-consistent solution to the R-DMFT equations coupled, through the electronic number densities $\vec{n}(\vec{v},\vec{\Sigma}(\omega))$, with the Poisson equation, Eq.~\ref{Poisson}. An illustration of the algorithm is presented in Fig.~\ref{Fig:convergence}. In short, the iterative solution of the Poisson equation constitutes an inner loop to the R-DMFT self-consistency and is done for fixed self-energies $\vec{\Sigma}(\omega)$ before the determination and solution of the impurity problems, which is more time demanding.

In more detail, we start with an initial guess of the selfenergies $\vec{\Sigma}(\omega)$ and onsite energies $\vec{v}$. Next, the Poisson loop is performed by calculating the electronic densities $\vec{n}$, Eq.\eqref{n_z}, and updating the onsite energies according to Eq.\eqref{iteration_scheme}. These two steps are then iterated until convergence\cite{reminder_nz_depends_on_v} is reached, for which we require
\begin{equation}
\label{converge_Posson}
\chi_{\Phi}\equiv \sqrt{\frac{1}{L}\sum_{i}\Phi_i^2} \leq \epsilon_{\Phi} \, ,
\end{equation}
where $\epsilon_{\Phi}$ is the required accuracy. For each converged Poisson loop we proceed, with the corresponding onsite energies $\vec{v}$, to the R-DMFT iteration which consists of computing the bath hybridization functions, Eq.\eqref{DeltaR}-\eqref{DeltaK}, and solving the corresponding impurity problems thereby obtaining a new set of selfenergies $\vec{\Sigma}(\omega)$. The alternate solution of the Poisson equation and the impurity problems is then iterated until convergence of the R-DMFT loop. We quantify the accuracy of the latter by the weighted difference between the hybridization functions of two consecutive loops\cite{endwith_Sigma}
\begin{equation}
\label{converge_DMFT}
\chi_{\Delta}\equiv \frac{1}{L_c}\sqrt{\sum_{i=L_{\rm lead}+1}^{L_{\rm lead}+L_c}\int\limits_{-\omega_c}^{\omega_c}||\Delta_{i}^{(m)}-\Delta_{i}^{(m-1)}||d\omega} \leq \epsilon_{\Delta}\, ,
\end{equation}
with
\begin{eqnarray*}
||\Delta_{i}^{(m)}-\Delta_{i}^{(m-1)}||=\sum_{\gamma=R,K}\hspace{-0.25cm}\Im m
\{[\Delta_{i}^{\gamma}]^{(m)}-[\Delta_{i}^{\gamma}]^{(m-1)} \}^2 \, .
\end{eqnarray*}

\begin{center}
\begin{figure*}[t!]
\subfigure[]{
\label{Fig:N_vs_z_U4U8Phi2}
\begin{minipage}[b]{0.45\textwidth}
\centering \includegraphics[width=1\textwidth]{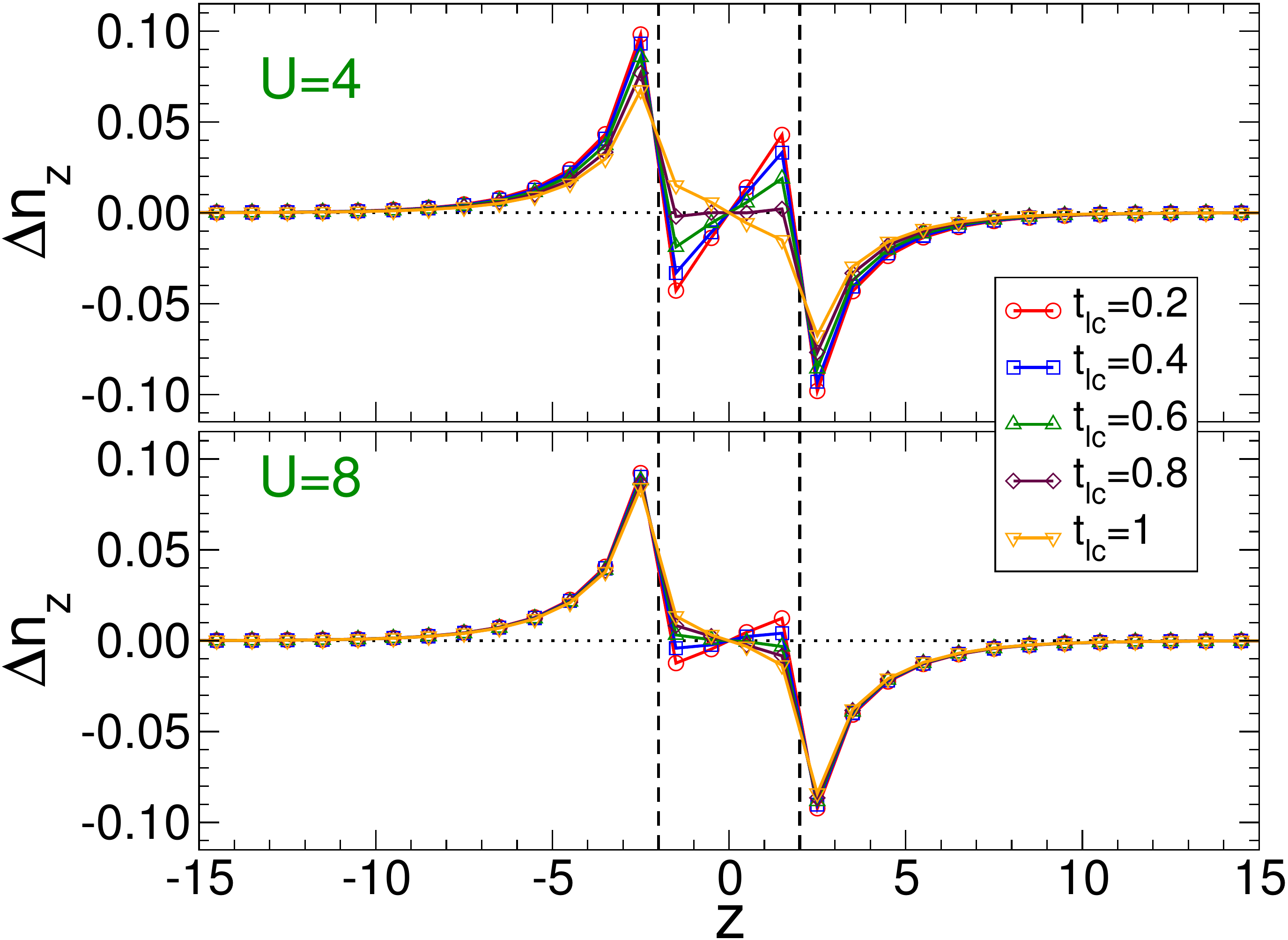}
\end{minipage}}
\hspace{0.025\textwidth}
\subfigure[]{
\label{Fig:N_vs_tlc_Phi2}
\begin{minipage}[b]{0.45\textwidth}
\centering \includegraphics[width=1\textwidth]{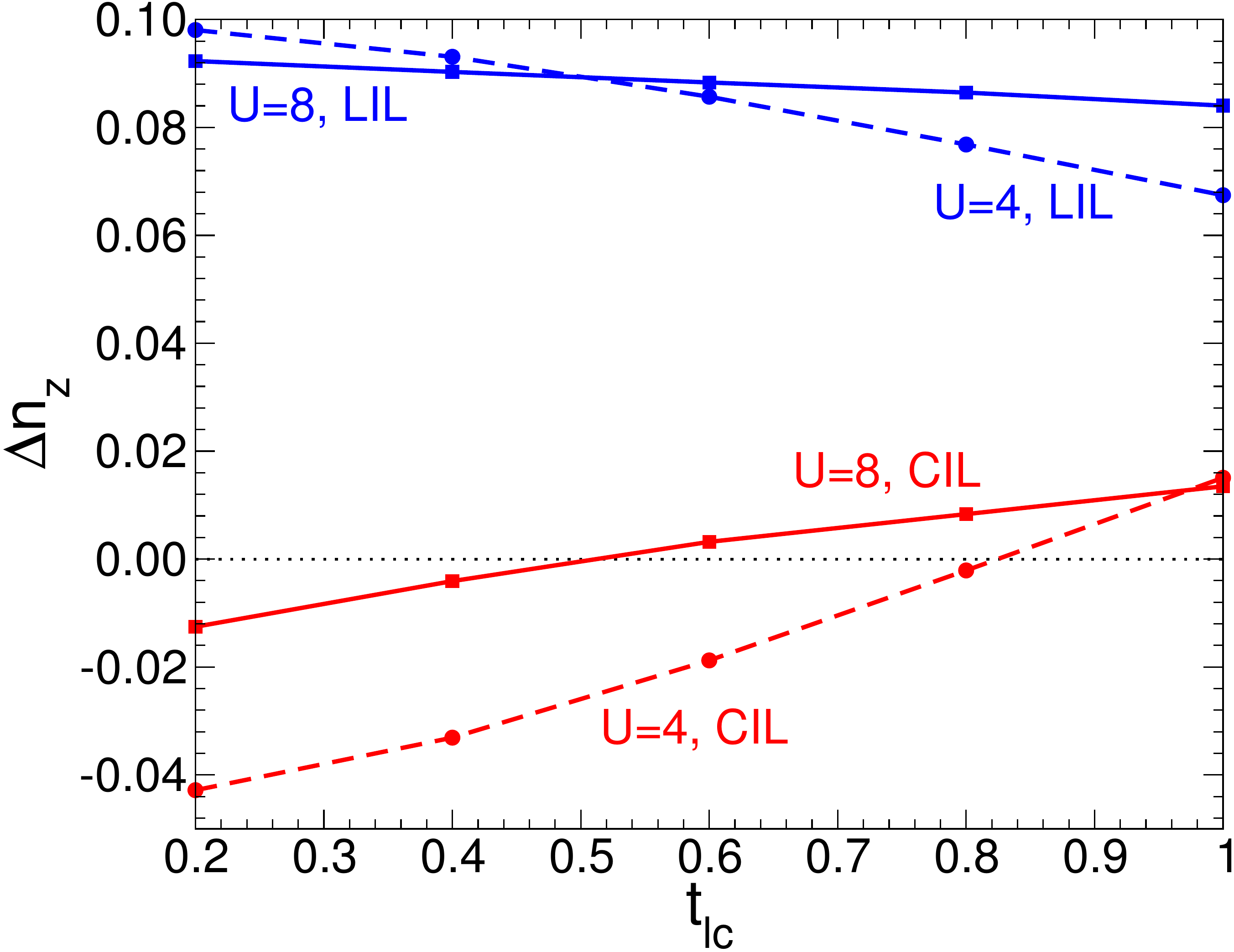}
\end{minipage}
}
\\
\caption{(Color online) (a) $\Delta n_z$ as a function of layer index $z$ for $U=4,~8$, $\Delta \mu =2$ and different values of $t_{lc}$.  (b)~$\Delta n_z$ for the LIL (blue curves) and CIL (red curves) as a function of the \LCJ coupling strength $t_{lc}$. Other parameters are the same as in Fig.~\ref{Fig:N_U4U8tlc0p2tlc1}. }
\label{Fig:N_U4U8Phi2}
\end{figure*}
\end{center}

\begin{center}
\begin{figure*}[t!]
\subfigure[]{
\label{Fig:N_vs_z_Phi2}
\begin{minipage}[b]{0.445\textwidth}
\centering \includegraphics[width=1\textwidth]{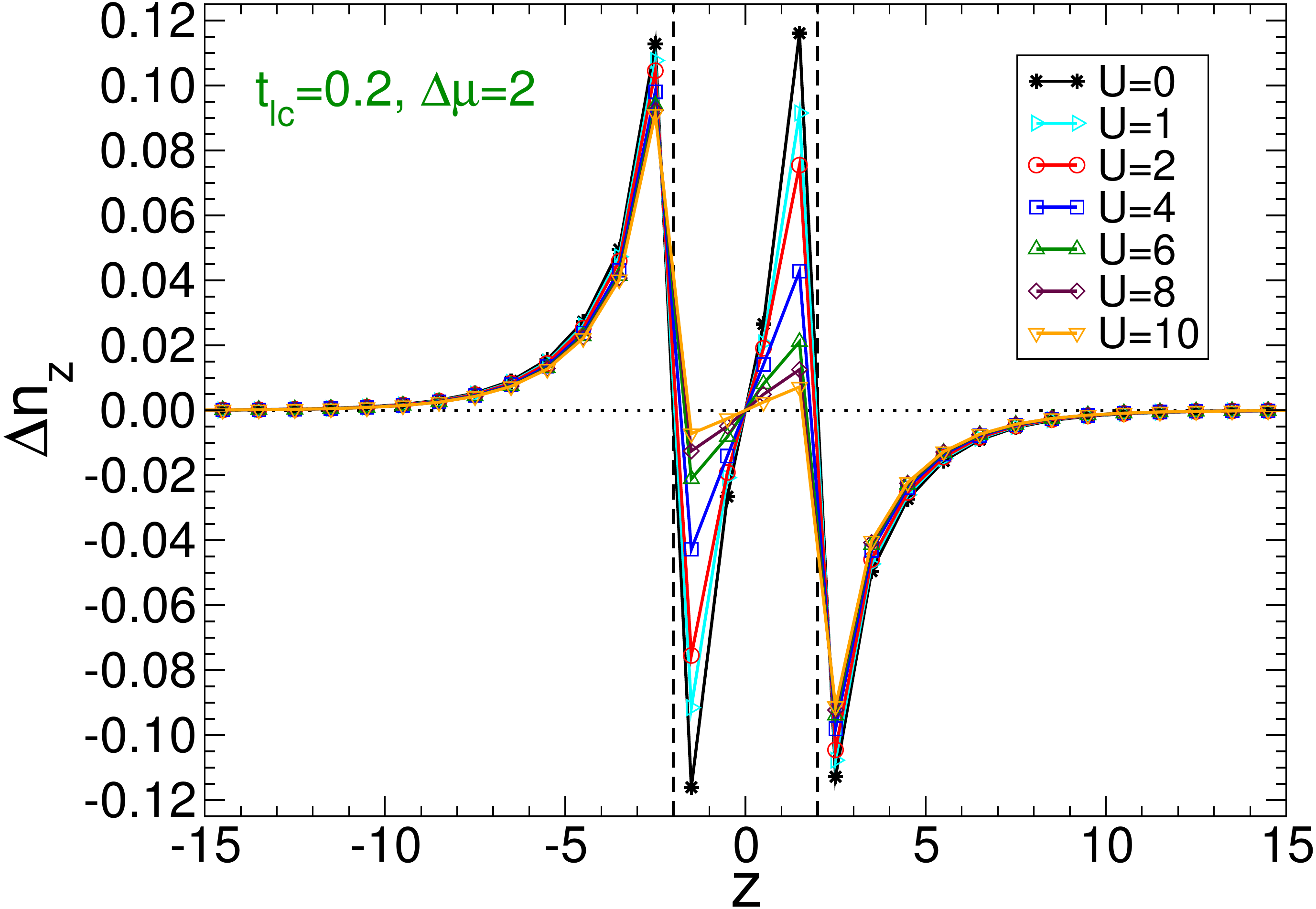}
\end{minipage}}
\hspace{0.025\textwidth}
\subfigure[]{
\label{Fig:N_vs_z_Phi0.5}
\begin{minipage}[b]{0.445\textwidth}
\centering \includegraphics[width=1\textwidth]{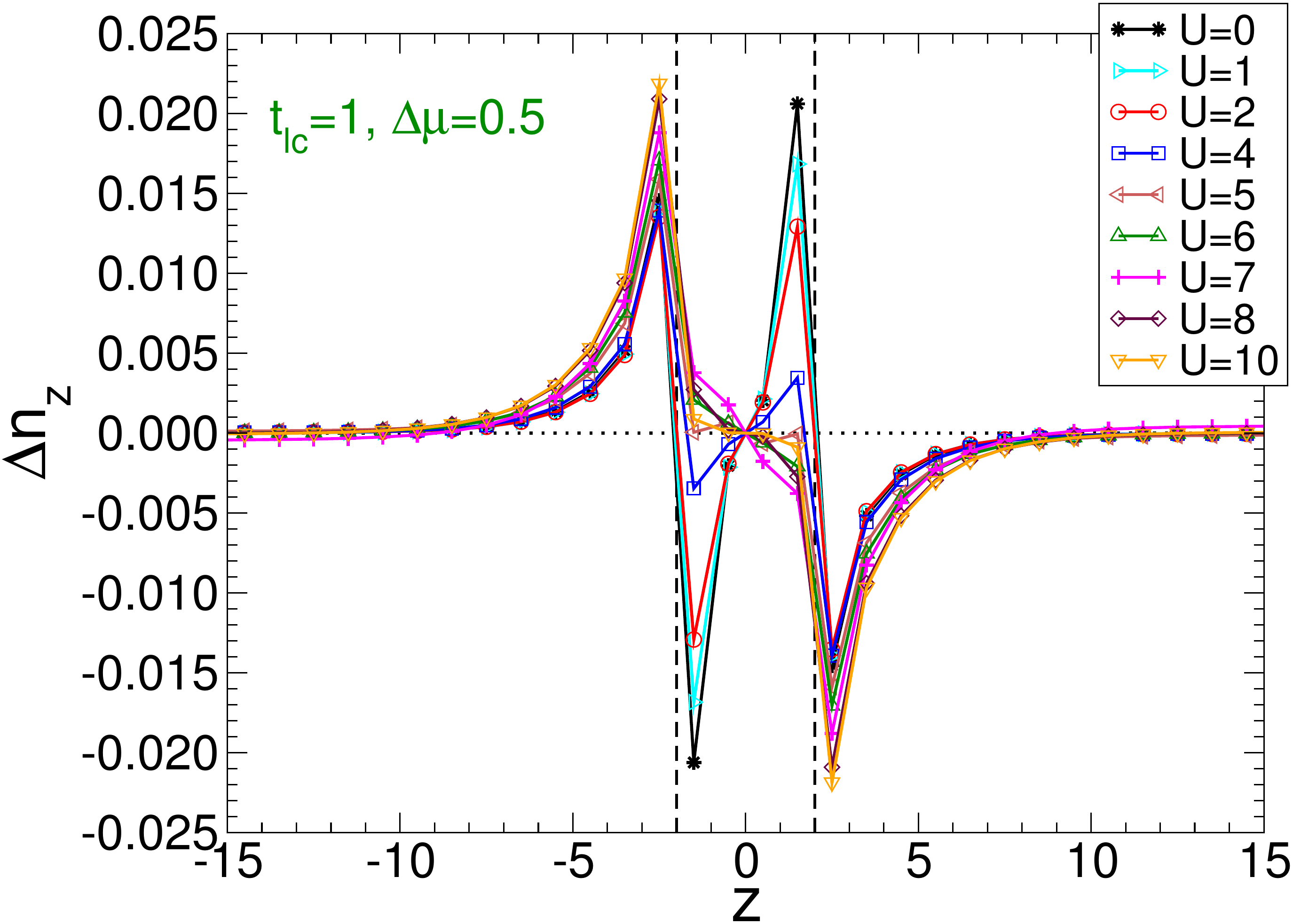}
\end{minipage}}\\
\subfigure[]{
\label{Fig:N_vs_U_Phi2}
\begin{minipage}[b]{0.445\textwidth}
\centering \includegraphics[width=1\textwidth]{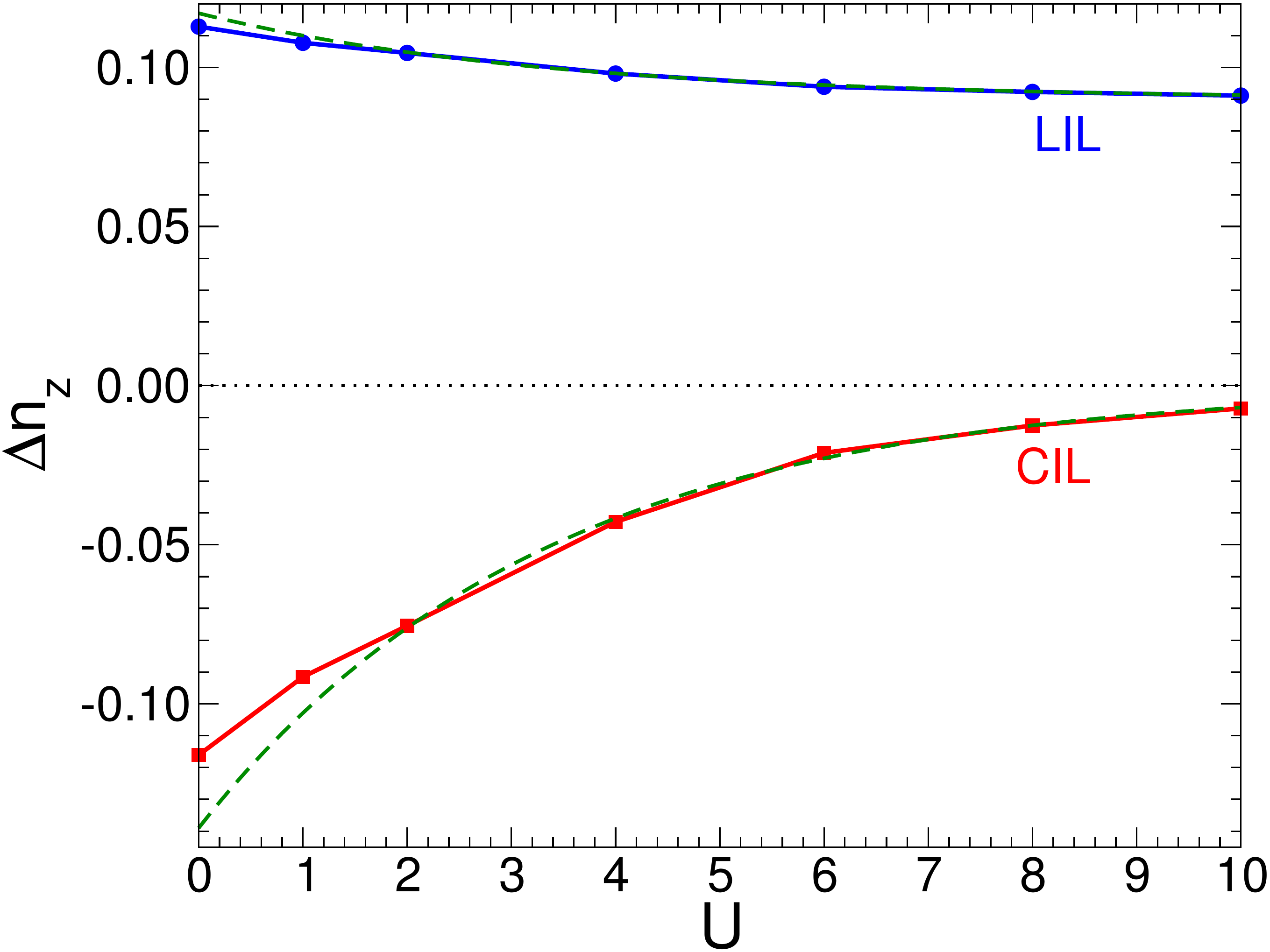}
\end{minipage}}
\hspace{0.025\textwidth}
\subfigure[]{
\label{Fig:N_vs_U_Phi0.5}
\begin{minipage}[b]{0.445\textwidth}
\centering \includegraphics[width=1\textwidth]{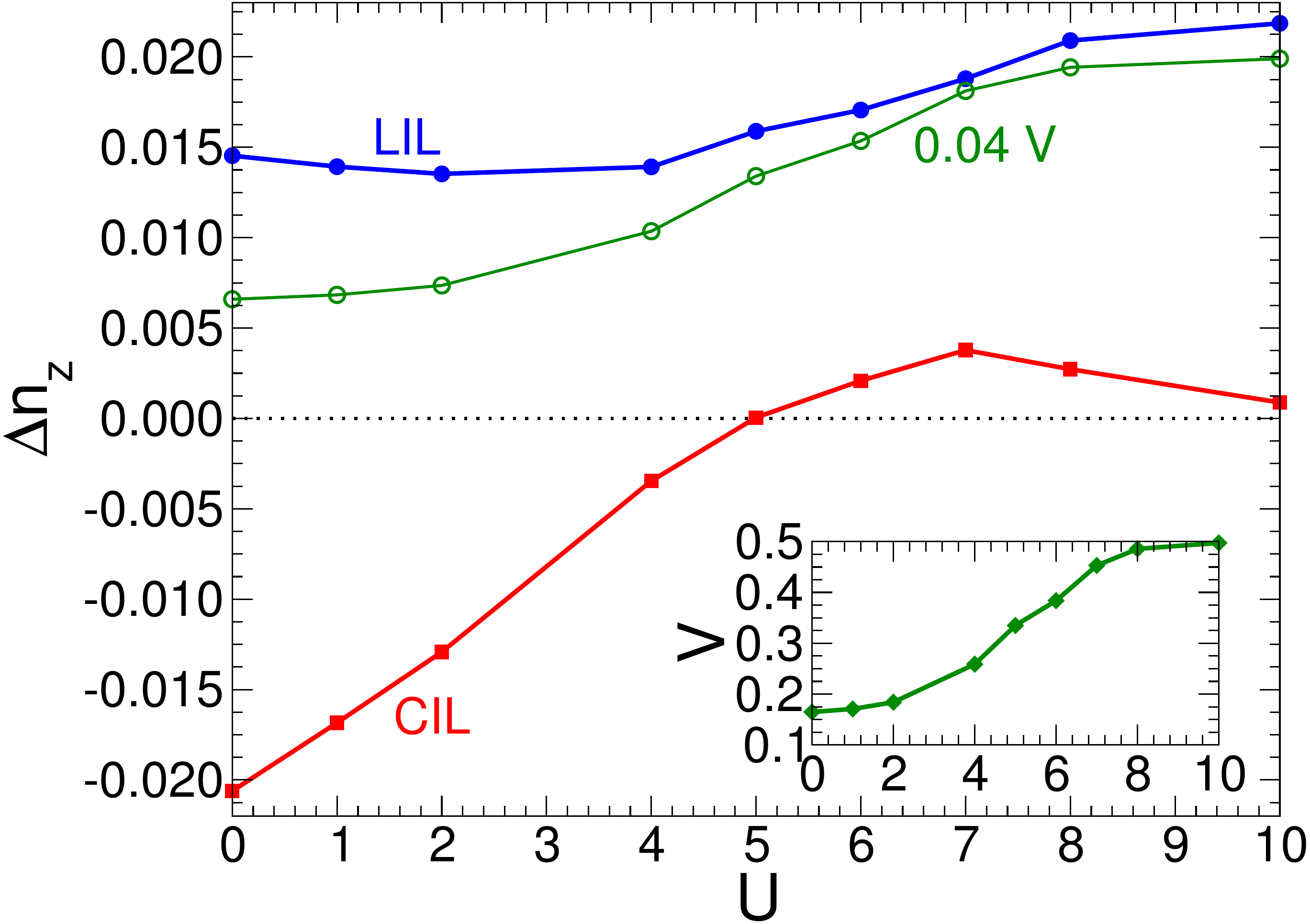}
\end{minipage}}\\
\subfigure[]{
\label{Fig:d_vs_U_Phi2}
\begin{minipage}[b]{0.445\textwidth}
\centering \includegraphics[width=1\textwidth]{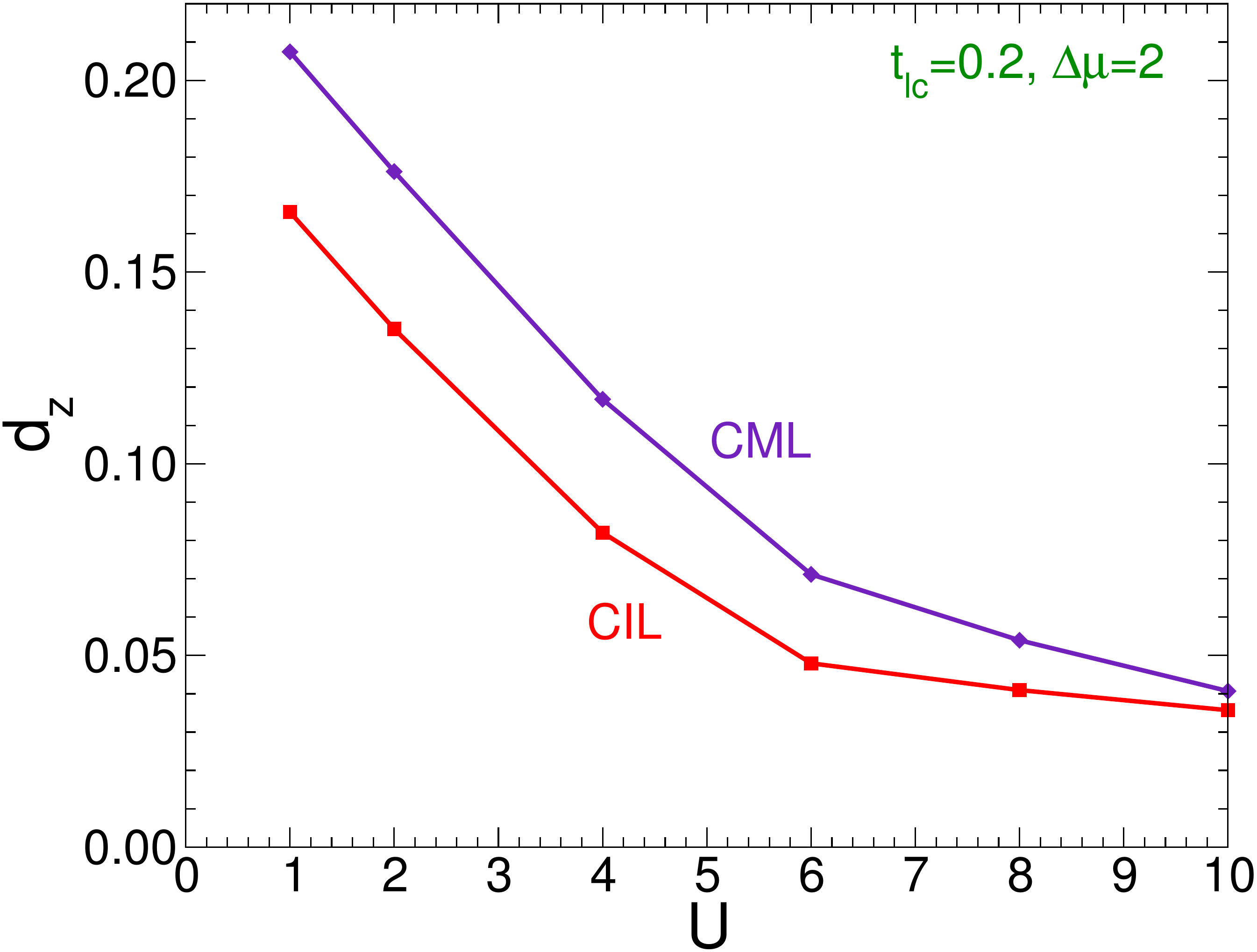}
\end{minipage}}
\hspace{0.025\textwidth}
\subfigure[]{
\label{Fig:d_vs_U_Phi0p5}
\begin{minipage}[b]{0.445\textwidth}
\centering \includegraphics[width=1\textwidth]{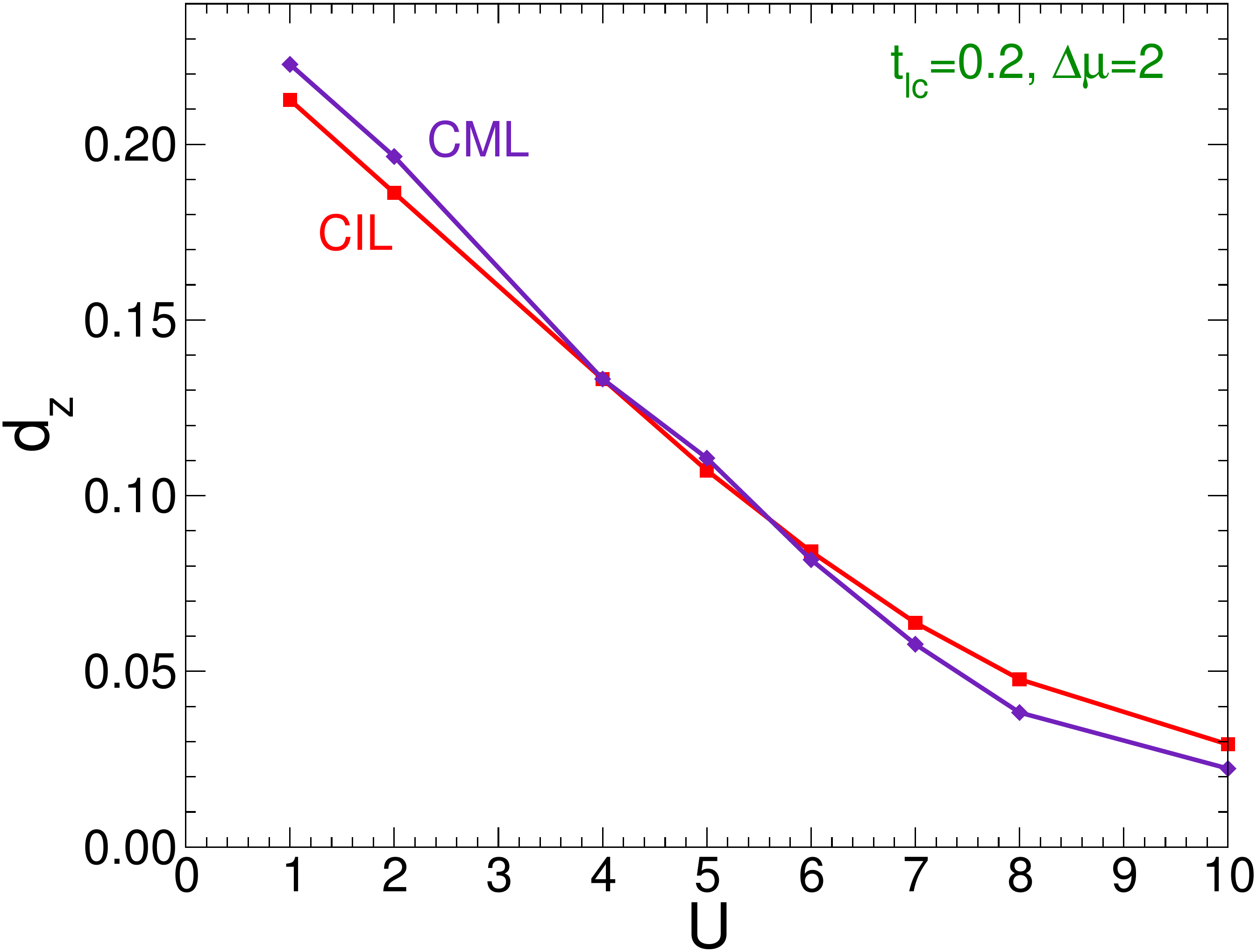}
\end{minipage}}
\caption{(Color online) 
$\Delta n_z$ (a,b) as a function of layer index $z$ for different values of the Hubbard interaction $U$. (c,d) $\Delta n_z$ for the LIL (blue curve) and for the CIL (red curve) as a function of Hubbard interaction $U$. Dashed green lines in (e) show results of the fit for the CIL $\Delta n_{\rm CIL}=A_1 \exp(-A_0 U)$ and for the LIL  $\Delta n_{\rm LIL}=B_2+B_1 \exp(-B_0 U)$, with fit parameters $A_0\sim B_0=0.301$, $A_1=-0.139$, $B_1=0.027$, and $B_2=0.090$. In (d)  we additionally plot the curves 
{\color{green2}$0.04*V$ versus $U$} 
 (green). 
In the inset of (d) we plot the bias voltage as a function of interaction $U$ for fixed value of $\Delta \mu=0.5$ (see details in text).
(e,f) double occupancy $d_z= \langle n_{z,{\bf r},\uparrow} n_{z, {\bf r} \downarrow}\rangle$ for the CIL (red curve) and for the CML(indigo curve) as a function of Hubbard interaction $U$.
The results in (a,c,e) are obtained with $\Delta \mu =2$ and $t_{lc}=t_{rc}=0.2$, while the ones in (b,d,f) with $\Delta \mu =0.5$ and $t_{lc}=t_{rc}=1$. Other parameters are the same as in Fig.~\ref{Fig:N_U4U8tlc0p2tlc1}. 
}
\label{Fig:N_Phi2Phi0p5}
\end{figure*}
\end{center}

\begin{center}
\begin{figure*}[t!]
\subfigure[]{
\label{Fig:Aomega_tlc0p2Phi2}
\begin{minipage}[b]{0.45\textwidth}
\centering \includegraphics[width=1\textwidth]{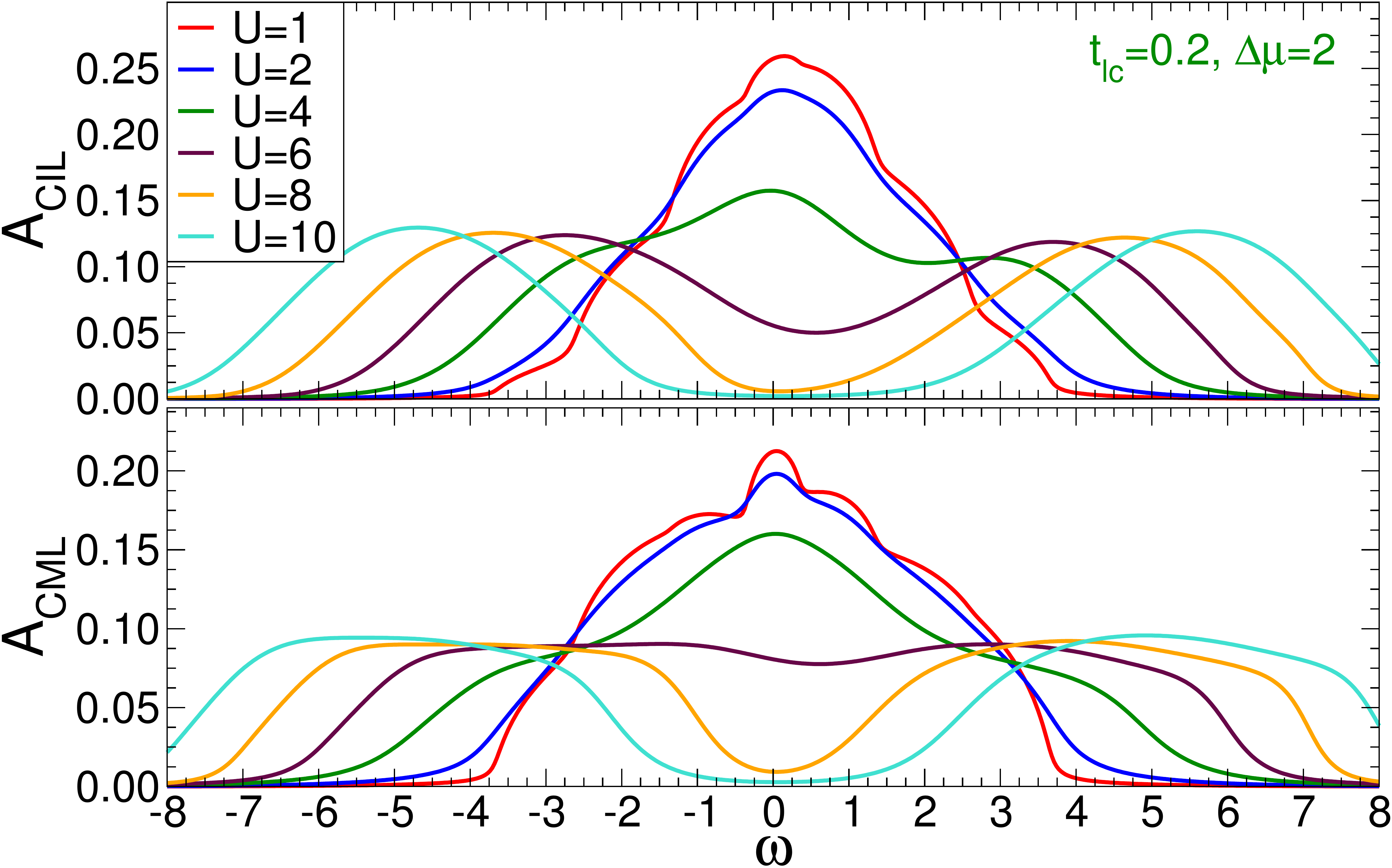}
\end{minipage}
}
\hspace{0.025\textwidth}
\subfigure[]{
\label{Fig:Aomega_tlc1Phi0p5}
\begin{minipage}[b]{0.45\textwidth}
\centering \includegraphics[width=1\textwidth]{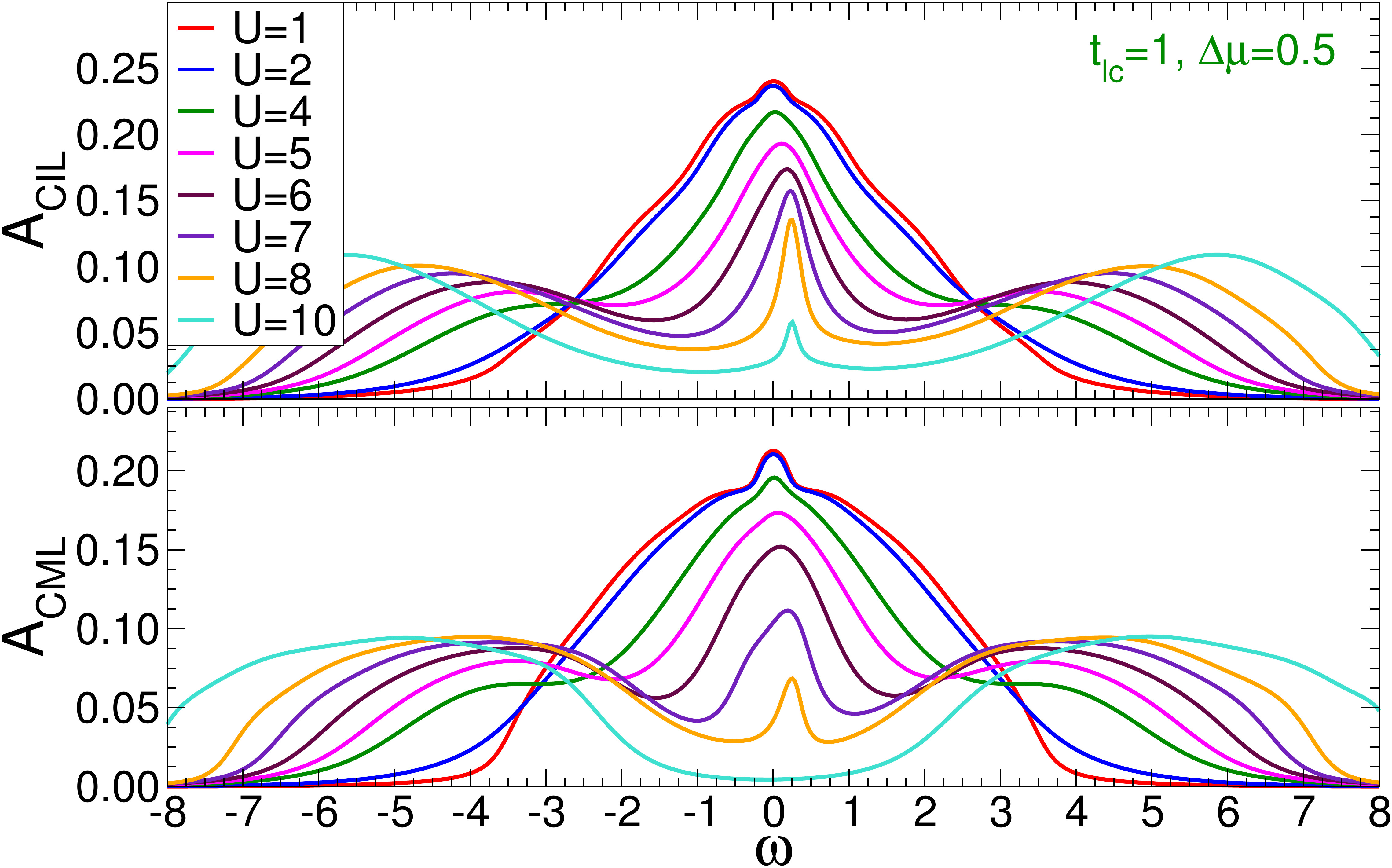}
\end{minipage}
}
\\
\caption{(Color online) 
Steady state spectral function for different Hubbard interactions for the CIL, upper panel, and central middle layer (CML), lower panel. Same set up as in Fig.~\ref{Fig:N_Phi2Phi0p5}. (a) $\Delta \mu=2$ and \LCJ coupling $t_{lc}=t_{rc}=0.2$, (b) $\Delta \mu =0.5$ and \LCJ coupling $t_{lc}=t_{rc}=1$.}
\label{Fig:Green}
\end{figure*}
\end{center}

\begin{center}
\begin{figure}[t!]
\subfigure[]{
\label{Fig:N_vs_z_Lc40Phi2}
\begin{minipage}[b]{0.45\textwidth}
\centering \includegraphics[width=1\textwidth]{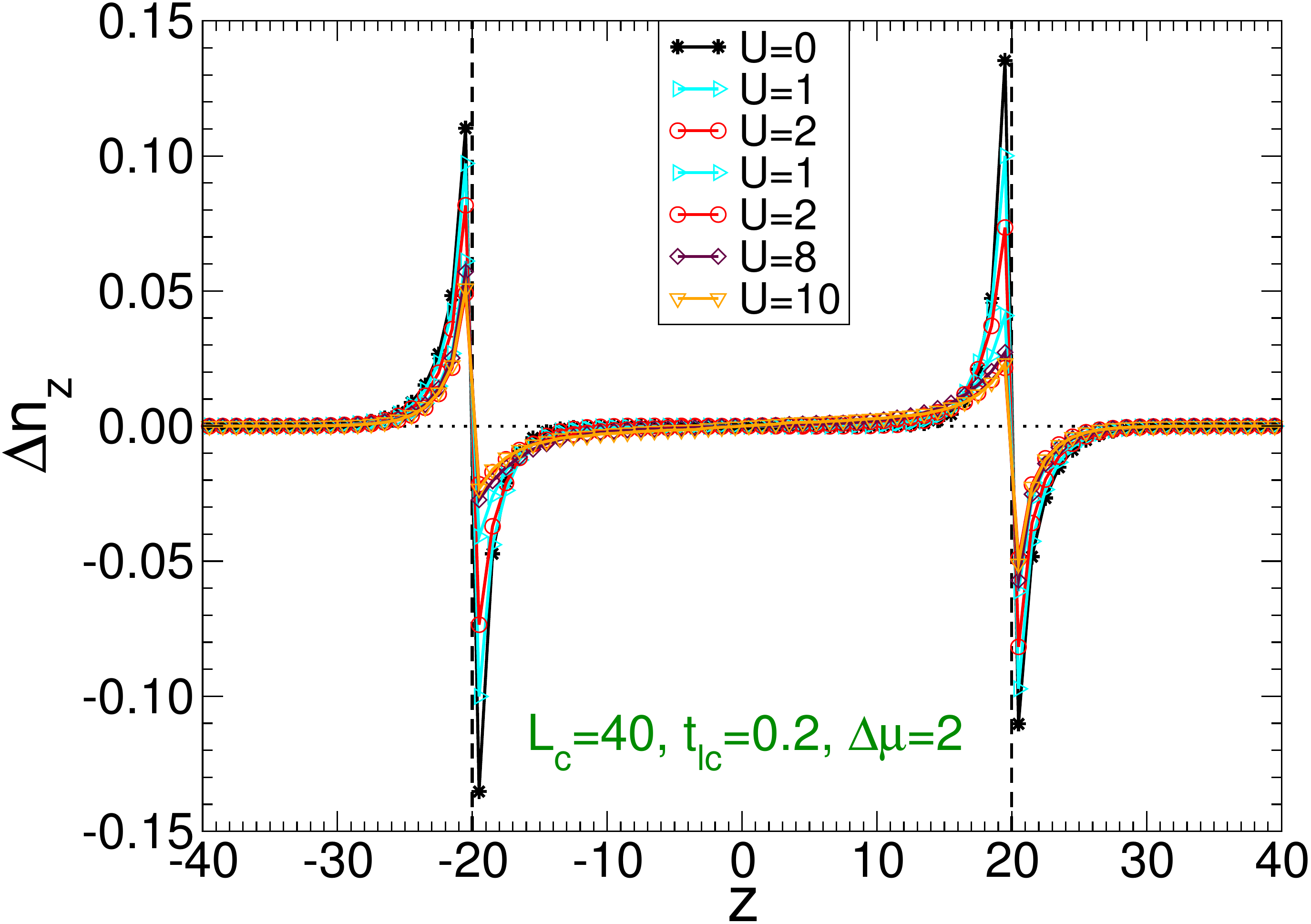}
\end{minipage}}
\hspace{0.025\textwidth}
\subfigure[]{
\label{Fig:V_vs_z_Lc40Phi2}
\begin{minipage}[b]{0.45\textwidth}
\centering \includegraphics[width=1\textwidth]{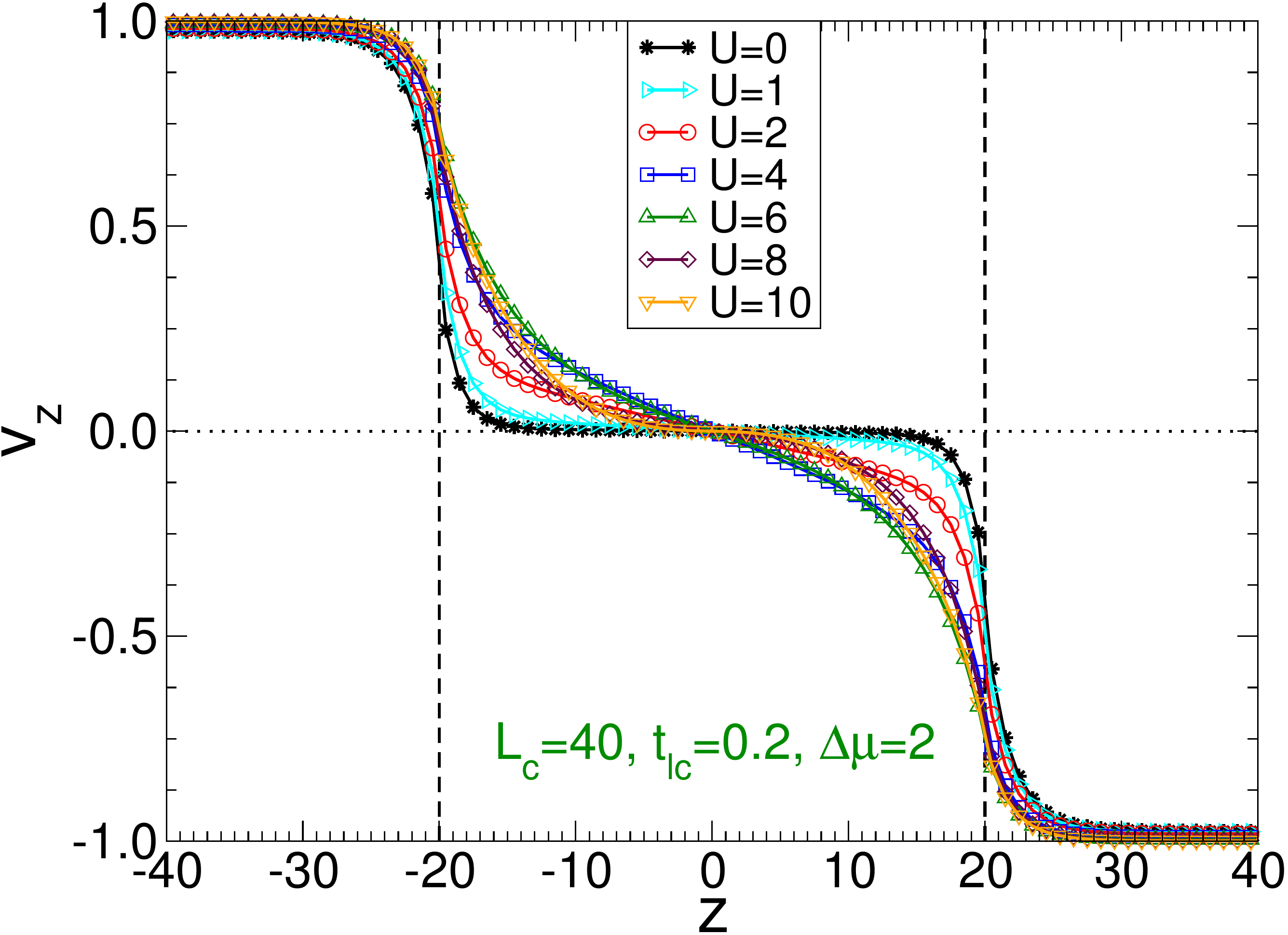}
\end{minipage}}
\hspace{0.025\textwidth}
\subfigure[]{
\label{Fig:N_vs_U_Lc40Phi2}
\begin{minipage}[b]{0.45\textwidth}
\centering \includegraphics[width=1\textwidth]{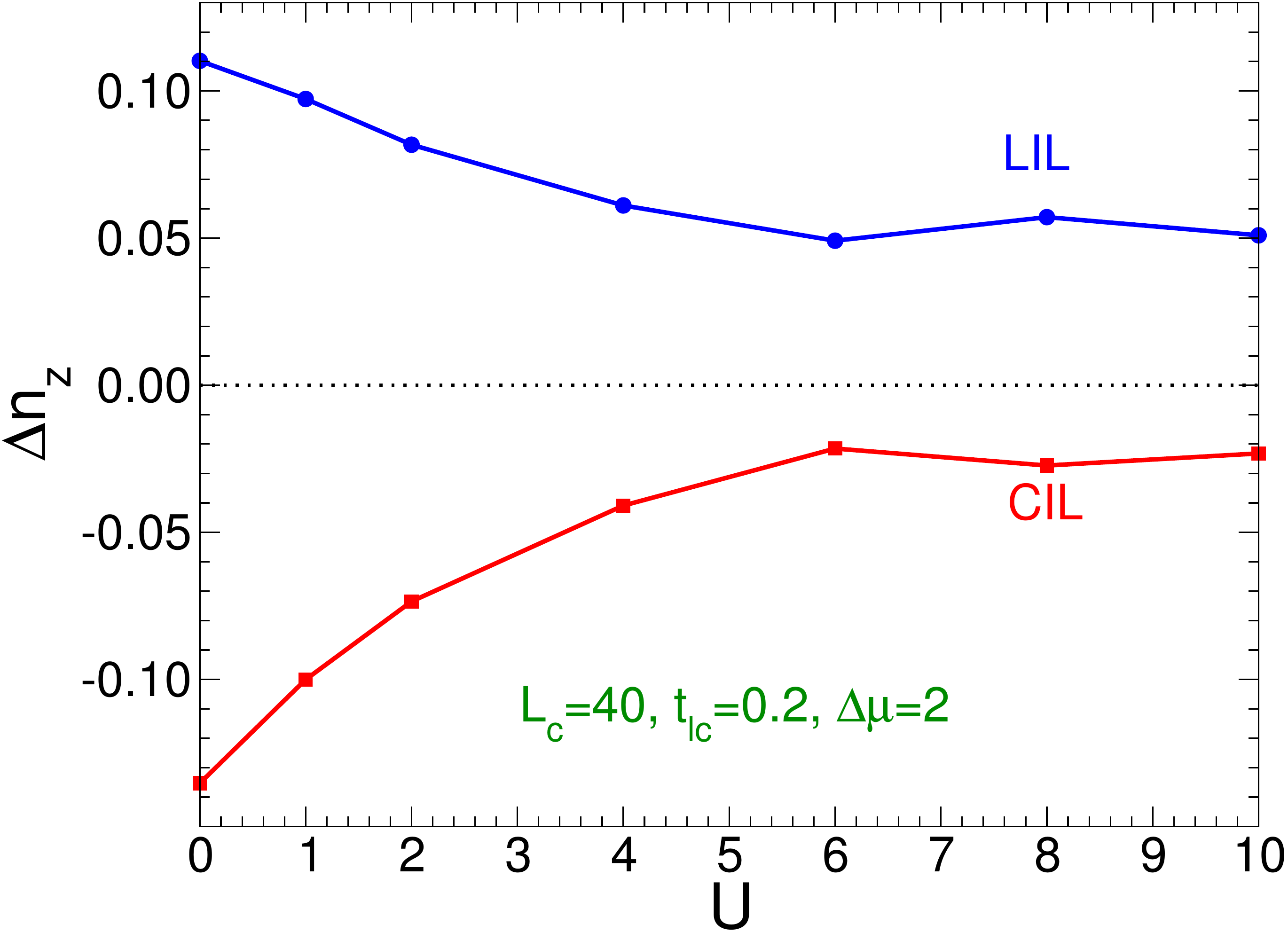}
\end{minipage}}
\\
\caption{(Color online) $\Delta n_z$ (a) and onsite energies $v_z$ (b) as a function of layer index $z$ for $L_c=40$ correlated layers, $t_{lc}=t_{rc}=0.2$, $\Delta \mu =2$ and different values of Hubbard interaction $U$. Total number of layers $L=100$.
Calculations are performed with $N_b=4$. Other parameters are the same as in Fig.~\ref{Fig:N_U4U8tlc0p2tlc1}. (c) $\Delta n_z$ for the LIL (blue curve) and CIL (red curve) as a function of the Hubbard interaction $U$.}
\label{Fig:N_Lc40Phi2}
\end{figure}
\end{center}

\section{Results }\label{Results}

As mentioned in the introduction, the emphasis of the present work lies on the influence of electronic correlations on the charge redistribution in a nonequilibrium situation. To this end, we consider the heterostructure sketched in Fig.~\ref{schematicp} which is driven out of equilibrium by an applied bias voltage.

To understand the behavior of the charge distribution, for finite \LCJ coupling ($t_{lc}>0$) it is  instructive to begin with a qualitative discussion of the expected behavior in the limit in which the \cc region is isolated from the leads ($t_{lc}=0$),
but still capacitively coupled to them via the long range Coulomb interaction. In that case, when the \cc region is metallic, i.e. for weak to intermediate Hubbard interactions, the system consists of two capacitors (one at each \LCJ) connected in series. On the other hand, when the \cc region is insulating, i.e. for large values of the Hubbard interaction,  it can be viewed as one capacitor with a dielectric material placed between two conducting materials. Applying a bias voltage will cause in both cases opposite charging of the facing surface layers of the lead and the \cc region, which can be viewed as dipole-like layers.
For definiteness, we will refer to them as lead interface layer (LIL) and the \cc interface layer (CIL), respectively (see Fig.~\ref{schematicp}).

We perform calculations for $L_{c}=4$  and $L_c=40$ correlated layers, with a homogeneous local Hubbard interaction $U_{z}=U$.  For $L_{c}=4$ ($L_c=40$), we explicitly consider $L_{\rm lead}=23$ ($L_{\rm lead}=30$) non-interacting, $U_z=0$, layers for each lead, to allow for proper charge redistribution in the leads as well. Therefore, in total, the \ec region, where the long range Coulomb interaction is accounted for, contains $L=50$ ($L=100)$ layers. The infinite region  outside of this range is treated exactly, whereby we take the charge and the Coulomb potential to be equal to its asymptotic bulk values. This is justified, as can be seen from Figs.~\ref{Fig:N_vs_z_U4U8tlc0p2tlc1}, \ref{Fig:N_vs_z_U4U8Phi2}, \ref{Fig:N_vs_z_Phi2}, \ref{Fig:N_vs_z_Phi0.5},  \ref{Fig:N_vs_z_Lc40Phi2}, and \ref{Fig:V_vs_z_Lc40Phi2}. To work at particle-hole symmetry, we set the bare onsite energies $v_{z}^{(0)}=-U_z/2$ and the asymptotic lead charge densities $n_{z=\pm \infty}=1$. The hopping between nearest-neighbor \cc region sites is taken as unit of energy, $t_c=1$, and the hopping between nearest-neighbor sites of the leads is $t_l=t_r=2$. Further, to investigate the effect of the coupling strength of \LCJ on the behavior of the system, we perform calculations for different values of $t_{lc}=0.2, 0.4, \ldots, 1$. All calculations are performed at ambient temperature $T_l=T_r=0.025$ and we consider an isotropic Coulomb parameter with the moderate value $c_z=c=1.5$.

Due to particle-hole symmetry, properties of the $z$-th and $(-z)$-th layer are connected by a particle-hole transformation. For the self-energies, the relation reads 
\begin{align}\label{eq: Sigma ph relations}
\Sigma_{z}^R(\omega)&=-[\Sigma_{-z}^R(-\omega)]^*+U_z\\
\Sigma_{z}^K(\omega)&=[\Sigma_{-z}^K(-\omega)]^*. 
\end{align}
Consequently, we need to calculate the self-energies only for half the system, i.e. $z<0$.
Finally, all results for $L_c=4$  are obtained with $N_b=6$ bath sites in the AMEA, while for $L_c=40$ we considered $N_B=4$ due to the increased numerical effort.~\cite{about_AMEA}

\subsection{Effect of the bias voltage}\label{Results V}

First, we investigate the effect of an applied bias voltage for intermediate, $U=4$, and strong, $U=8$,
Hubbard interaction, as well as small ($t_{lc}=0.2$) and large ($t_{lc}=1$) 
coupling strengths between the leads and the \cc region.

 Our calculations show that at the \LCJ the system still hosts dipole-like layers for small but non-zero \LCJ coupling strengths. Fig.~\ref{Fig:N_vs_z_U4U8tlc0p2tlc1} indeed shows for 
 $t_{lc}=0.2$, that the charge density deviations from half-filling, $\Delta n_z=n_z-1$, for the CIL and the LIL have opposite signs and their absolute values increase with  bias voltage $V$ (see Fig.~\ref{Fig:N_vs_Phi_U4U8tlc0p2tlc1}) for both considered Hubbard interactions. So, similar to $t_{lc}=0$, also for $t_{lc}=0.2$ LIL and CIL can be viewed as dipole-like layers.
 
 On the other hand the behavior is qualitatively different for large values of the \LCJ coupling ($t_{lc}=1$) and in particular sensitive to the value of the Hubbard interaction. For strong interaction (U=8), we obtain that $\Delta n_z$ of the LIL and CIL have the same sign  and their absolute values increase with the bias voltage. When considering a weaker interaction (U=4) this stays true for $\Delta n_{\rm LIL}$ (charge density deviation from half-filling for the  LIL), while $\Delta n_{\rm CIL}$ (charge density deviation from half-filling for the CIL) shows non-monotonic behavior and a sign change as a function of the bias voltage. So, in contrast to small values of the \LCJ coupling strength, for large ones dipole-like layers are only present at the \LCJ for weak to intermediate $U$ and low bias voltages. 
 
Remember that the bias voltage $V=v_l-v_r$ and the difference between the chemical potentials $\Delta \mu=\mu_l-\mu_r$ differ from each other. As we have already discussed in Sec. \ref{Self-consistency_loop}, it is numerically more convenient to perform calculations for fixed $\Delta \mu$ and evaluate $V$ {\it a posteriori}. For weak values of the \LCJ coupling or for
large value of $U$, the difference between $V$ and $\Delta \mu$ is negligible ($1\%$ or smaller). However, there is a significant deviation for the case of $t_{lc}=1$ and $U=4$, see Fig.~\ref{Fig:tildePhi_vs_Bias_U4U8tlc0p2tlc1}.
This is due the fact that when increasing $t_{lc}$ the flow of particles from the left lead  to the right one increases. As a result, there is a depletion of particles on the left lead which, if one wants to keep both leads at half filling, has to be compensated by increasing $\mu_l$. The opposite situation obviously occurs on the right lead.

\subsection{Effect of the \LCJ coupling strength} \label{Results tlc}

We further investigate the effect of the \LCJ coupling strength $t_{lc}$ between the leads and the \cc region. We perform calculations for several values of $t_{lc}$, fixing $\Delta \mu=2$ and again considering $U=4,\,8$.

When the \LCJ coupling strength is increased, the current through the heterostructure rises. Thus, we expect that more charge is transferred from the left lead to the \cc region. Indeed our results, Fig.~\ref{Fig:N_vs_z_U4U8Phi2} and \ref{Fig:N_vs_tlc_Phi2}, show that the charging of the LIL and the CIL are first decreasing as $t_{lc}$ is increased. With further increase of $t_{lc}$ this trend holds true for the LIL, while interestingly, for the CIL, $\Delta n_{\rm CIL}$  changes sign at some $U$-dependent value $t_{lc}^*$. 
Furthermore, we find that $t_{lc}^*$ decreases with increasing $U$ and for non-interacting \cc region ($U=0$) $\Delta n_{\rm CIL}$ is negative for all values of $t_{lc}$ we have considered. From here, it follows then that correlations lead to an earlier disappearance of dipole-like layers with respect to the \LCJ coupling strength. This can be understood by the following:

For $U=0$, the behavior of the system can be intuitively understood by the hydraulic analogy, where a fluid takes over the role of the electric charge and pipes represent wires. In this picture larger $t_{lc}$ translates into a bigger diameter of the ``\LCJ-pipe". For the behavior of the LIL, this means that less fluid gets jammed at the interface.
When thinking about the behavior of the left-CIL in the hydraulic picture it is easiest to consider the jam created at the right-CIL, since the two are connected by particle-hole symmetry, which will also get decreased with increasing $t_{lc}$.
This means, that the trends observed in Fig.~\ref{Fig:N_vs_tlc_Phi2} are consistent with the Hydraulic analogy.

Coming back to the reason why for stronger Hubbard interaction $t_{lc}^*$ is lowered, we can thus interpret the slope of $\Delta n_{\rm CIL}(t_{lc})$, for low $t_{lc}$, to originate from the $U=0$ behavior and thus the value of $t_{lc}^*$ is mainly influenced by the starting value $\Delta n_{\rm CIL}(t_{lc}=0)$ which is suppressed by the Hubbard interaction leading to the decrease of $t_{lc}^*$ as a function of $U$.

\subsection{Effect of the local interaction} \label{Results U}

Finally we investigate the effect of the interaction $U$ for small ($t_{lc}=0.2$) and larger ($t_{lc}=1$) values of the \LCJ coupling strength. We consider differences between the chemical potentials, $\Delta \mu =2$ and $\Delta \mu =0.5$, respectively. These values are chosen such that for small interactions the opposite charging of the LIL and CIL is most pronounced, see Fig.~\ref{Fig:N_vs_Phi_U4U8tlc0p2tlc1}. Furthermore, to better resolve the charge distribution, we also present results for a system with a larger \cc region ($L_c=40$), in addition to the case with $L_c=4$.
When studying the charging dependence as a function of $U$, we should expect that in the limit of large $U$, $\Delta n_z$ vanishes for the \cc region, since in this limit any double occupation is extinguished.

\subsubsection{Small correlated region ($L_c=4$)} \label{Small_system}

First, we discus the effect of the interaction for weak \LCJ coupling  ($t_{lc}=0.2)$ and $\Delta \mu=2$.
Fig.~\ref{Fig:N_vs_z_Phi2} and \ref{Fig:N_vs_U_Phi2} show that the opposite charging of the interface layers is suppressed by the Hubbard interaction. Further, $\Delta n_z$ for LIL converges monotonically to some finite value for $U\to \infty$, while for the CIL it converges to  $0$ as expected.
In order to investigate the behavior of the boundary charge, we fit them (for $U \geq 2$) with  exponential functions (see Fig.~\ref{Fig:N_vs_U_Phi2}), namely  $\Delta n_{\rm CIL}=A_1 \exp(-A_0 U)$ and $\Delta n_{\rm LIL}=B_2+B_1 \exp(-B_0 U)$. The resulting fit parameters are given in the figure caption. Notice that both fits give approximately the same exponent, that is $A_0\approx B_0$. 

For small \LCJ coupling ($t_{lc} = 0.2$) and increasing  interaction strength $U$, as we already mentioned above, the charging of the LIL and CIL is exponentially suppressed, but these layers still have opposite sign and for any finite $U$, while being reduced, the dipole-like layers are still there.

On the other hand, this is no longer the case for  stronger \LCJ coupling strength ($t_{lc}=1$) and $\Delta \mu=0.5$, as can be anticipated based on the results presented in previous subsections. Indeed, from Fig.~\ref{Fig:N_vs_z_Phi0.5} and Fig.~\ref{Fig:N_vs_U_Phi0.5} we can see that $\Delta n_z$ is non-monotonic for both surface layers and in addition the CIL displays a sign change at $U\approx 5$ which approaches zero only for higher values of the interaction.

To understand this behavior, it is important to recall that the results presented in Figs.~\ref{Fig:N_vs_z_Phi0.5} and \ref{Fig:N_vs_U_Phi0.5} are performed for fixed $\Delta \mu=0.5$, which corresponds to different bias voltages $V$ (see inset of Fig.~\ref{Fig:N_vs_U_Phi0.5}).
When examining Fig.~\ref{Fig:N_vs_U_Phi0.5} more closely, one can see that the shape of $\Delta n_{\rm LIL}(U)$ for $U>4$  resembles that of $V(U)$ from the inset. Moreover, from Fig.~\ref{Fig:N_vs_Phi_U4U8tlc0p2tlc1} we know that $\Delta n_{\rm LIL}(V)$ is just proportional to $V$ and almost insensitive to $U$. 
Based on that, to exclude the dependence on the bias voltage we plot $n(U)=0.04V(U)$, where the coefficient of proportionality is extracted from Fig.~\ref{Fig:N_vs_Phi_U4U8tlc0p2tlc1}, see green line in Fig.~\ref{Fig:N_vs_U_Phi0.5}. One indeed finds that the behavior of $\Delta n_{\rm LIL}$ for $U>4$ is controlled by the  $V(U)$ dependency. We thus expect that  the curve of  $\Delta n_{\rm LIL}$ vs. $U$ for fixed $V$  would continue its downward trend also for $U>4$ and converge to some value as in the case of the smaller \LCJ coupling strength $t_{lc}=0.2$.
In contrast to the behavior of $\Delta n_{\rm LIL}$, fixing $V$  would not affect qualitatively the behavior of $\Delta n_{\rm CIL}$ versus $U$. As a matter of fact, taking the dependence on $V$ into account, one would expect an even more pronounced maximum in the behavior of $\Delta n_{\rm CIL}$ (see red curve in Fig.~\ref{Fig:N_vs_U_Phi0.5}). 

We also investigate the double occupancy $d_z= \langle n_{z,{\bf r},\uparrow} n_{z, {\bf r}\downarrow}\rangle$. Our calculations show that both for small as well as for large \LCJ coupling strength, the double occupancies $d_z$ for the correlated sites are monotonically decreasing as expected (see Figs. \ref{Fig:d_vs_U_Phi2} and \ref{Fig:d_vs_U_Phi0p5}). For  weak \LCJ coupling strength, the double occupancy $d_z$ of the CIL is always larger compared to the one of the correlated middle layer (CML), while for  large \LCJ coupling strength this is only true for $U \lesssim 5$. This can be explained by the fact that for $U \lesssim 5$ the filling in the CML is larger than the filling in the CIL.


A different behavior of the system between the regimes of weak and strong \LCJ coupling strengths can be also seen by considering the steady state spectral functions $A_z(\omega)=-\frac{1}{\pi}\Im m G^R_z(\omega)$ (see Fig.~\ref{Fig:Green}). For $t_{lc}=0.2$ and $\Delta \mu=2$ the spectral function does not show a Kondo-like peak at $\omega=\mu_l=1$. We attribute this fact to a combined effect of the width of the Kondo-like peak being so small that we are not able to resolve it as well as the substantial bias voltage present in the system leading to decoherence which suppresses the resonance. In contrast, for large values of the \LCJ coupling strength there is a clear Kondo-like peak for the CIL (at $\omega=\mu_l=0.25$) up to interactions as strong as $U=10$. This is not surprising, because the width of the Kondo-like peak is proportional to $t_{lc}^2$ and correspondingly the difference between these two cases is $O(100)$ and in addition the considered $\Delta \mu$ is a factor of four smaller. 
Fig.~\ref{Fig:Green} also shows the spectral function for the 
CML featuring, as expected\cite{ho.bu.00} due to the increased distance to the leads, a less pronounced Kondo-like peak compared to the CIL which is already destroyed for $U=10$. 

It appears that the Kondo-like peak in the spectral density occurs whenever $\Delta n_\text{LIL}$
and $\Delta n_\text{CIL}$ have the same sign, which indicates that the mobility within the \cc region is small as compared to $t_{lc}$.

\begin{figure}[t]

\includegraphics[width=0.99\columnwidth]{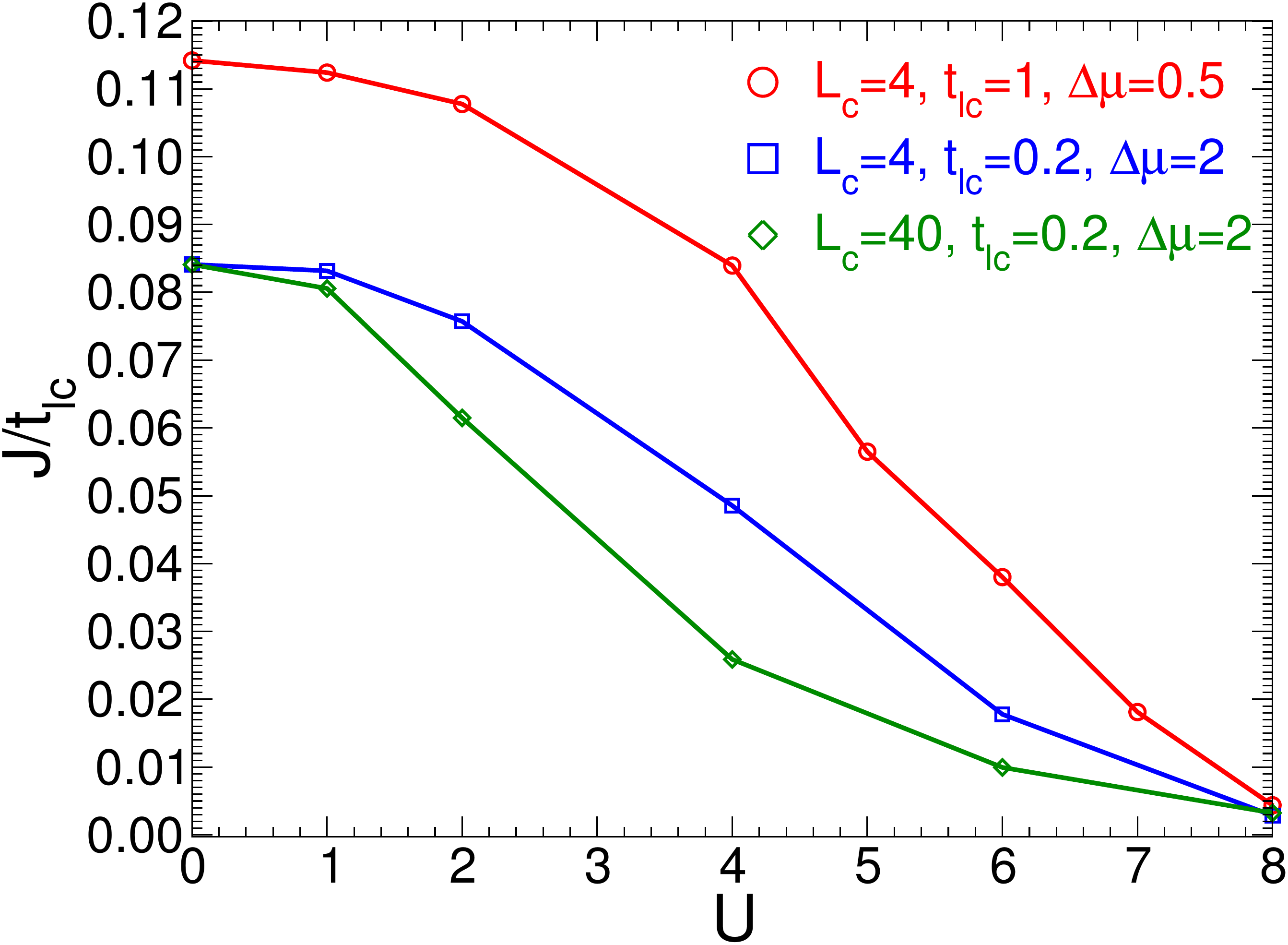}
\caption{(Color online) Current density $J$ as a function of the interaction for different parameter sets.  Blue line with squares (red line with circles) correspond to a system with $L_c=4$ correlated layers, with \LCJ coupling strength $t_{lc}=0.2$ ($t_{lc}=1$) and $\Delta \mu=2$ ($\Delta \mu=0.5$). Green diamonds correspond to a system with $L_c=40$ correlated sites with $t_{lc}=0.2$ and $\Delta\mu=2$.  Other parameters are the same as in Fig.~\ref{Fig:N_U4U8tlc0p2tlc1}.
}
\label{Fig:Current}
\end{figure} 

\subsubsection{Large correlated region ($L_c=40$)}

We now  want to investigate how far the charging of the interface region extends into a bulk system. To this end, we enlarge the \cc region to  $L_c=40$. Results are obtained with $N_b=4$ auxiliary bath sites in the AMEA impurity solver.\cite{about_AMEA}  
Due to the heavy numerical calculations the  convergence of the DMFT self-consistency is quite slow, especially for the strong interactions.

At this point it is worth noting that for a metallic material, one would expect that only the surface is charged with an exponential tail into the bulk since the induced charge on the surface will compensate the electric field in the bulk. Indeed, our results for $\Delta n_z$ and $v_z$, presented in Figs.~\ref{Fig:N_vs_z_Lc40Phi2} and \ref{Fig:V_vs_z_Lc40Phi2} respectively, show that the charging and onsite energies behave as expected and fall off exponentially into the bulk. Further, we find that the corresponding penetration depth for charging, although increasing with $U$, depends only weakly on $U$ and that this dependence is more pronounced for the onsite energies. Note that the system is still metallic for all values of the interaction $U\leq 10$ and the exponential suppression can therefore be attributed to screening. The trend that the penetration depth increases with $U$ can thus be interpreted as less effective screening due to the lower density of states around $\omega\approx 0$.

As in the previous results for $L_c=4$ the main effect of the interaction  is to reduce the absolute value of the charging at the interface between the correlated and uncorrelated region. As can be seen from Fig.~\ref{Fig:N_vs_U_Lc40Phi2} the behavior agrees qualitatively with the ones observed for $L_c=4$, see also Fig.~\ref{Fig:N_vs_U_Phi2}. The fact that the exponential dependence on $U$  is not so obvious in Fig.~\ref{Fig:N_vs_U_Lc40Phi2} can be attributed to the lower accuracy due to the increased numerical challenge to converge the self-consistent equations.

\subsubsection{Current}

We also investigate the effect of the interaction on the steady-state current density through the correlated interface. The latter can be calculated using  off-diagonal elements of the Keldysh Green's function\cite{okam.07,about_Current}
\begin{equation}
\label{Current}
J=J_{z,z+1}=t_{z,z+1}\int_{-\infty}^{\infty}\frac{d\omega}{2\pi}\int\limits_{\rm BZ} \frac{d^2{\kxy}}{(2\pi)^2}\left({\bf G}_{z+1,z}^K - {\bf G}_{z,z+1}^K \right)  \, , 
\end{equation}
where summation over spin is implicitly assumed.

Results are shown in Fig. \ref{Fig:Current} where we plot a rescaled current density $J/t_{lc}$
in order to present the curves on the same plot. 
As expected our calculations show that for all considered system parameters the current density is strongly suppressed when increasing the interaction strength $U$.~\footnote{Results for $U=10$ are not shown because the current density is so low in this case that it lies below our numerical uncertainty.}
For the system with a smaller correlated region ($L_c=4$) the  qualitative form of the suppression 
as a function of $U$ seems rather independent of $t_{lc}$ and $\Delta \mu$.
Nevertheless, from the figure it appears that the scaling behavior of the current density is stronger than $\propto t_{lc}$.
 This is because the stronger hybridization leads to a more pronounced resonance peak making the central region more metallic especially around $\omega=0$ resulting in more spectral weight within the Fermi-window of the leads already for small voltages. See also Fig. \ref{Fig:Aomega_tlc0p2Phi2} and \ref{Fig:Aomega_tlc1Phi0p5}.

Furthermore, we compare the steady-state current density for the small ($L_c=4$) and the large ($L_c=40$) correlated regions (see Fig. \ref{Fig:Current}). We observe that the difference between them is marginal for weak interactions while for intermediate to strong interactions we have a substantial suppression for $L_c=40$.
This is due to a reduced electron mobility induced by the loss of metallicity of the correlated region. However, this cannot be simply generically described by a decreased conductivity but rather by the fact that for $U\gtrsim 2$ the penetration depth of the electric field exceeds the size of the small correlated region $L_c=4$, 
see also Fig.~\ref{Fig:V_vs_z_Lc40Phi2}

\begin{figure}[t]
\includegraphics[width=0.99\columnwidth]{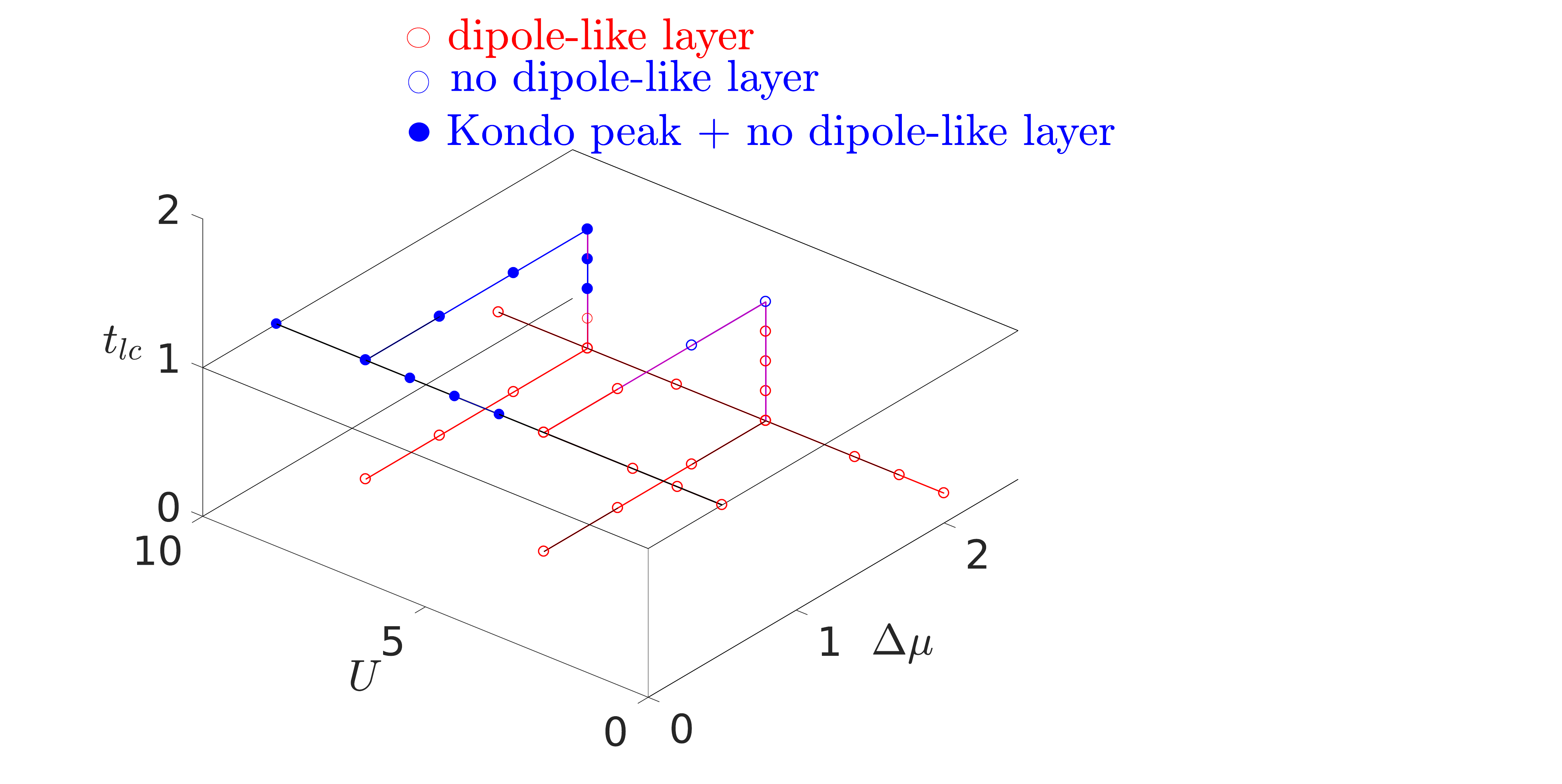}
\caption{(Color online) Three dimensional representation of the regions in $t_{lc}-\Delta \mu-U$-space
in which the system exhibits(lacks) dipole-like layers at the \LCJ red (blue) open circles. In addition, parameter combinations where a Kondo-like peak can be clearly identified, are marked by full circles.
}
\label{Fig:3Dsummary}
\end{figure} 

\section{Conclusions}\label{Conclusions}

We addressed the steady-state properties of a system consisting of a multilayer \cc region attached to two metallic leads. The model was solved by  nonequilibrium R-DMFT whereby AMEA~\cite{ar.kn.13, do.nu.14, ti.do.15} was used as impurity solver. We studied the charge redistribution in the system induced by  the local Hubbard and the long-range Coulomb interactions in the presence of a bias voltage. We find that its behavior  is very different for weak and strong \LCJ coupling strengths, especially for strong local interactions. The influence of $U$ on the lead layers is due to the proximity effect and therefore less pronounced in the lead compared to the \cc region.

Our results indicate that the charges (considered with respect to the bulk value) on opposite sides of the \LCJ, can have equal or opposite signs depending on the system parameters. The case of opposite signs can be interpreted as the formation of a dipole-like layer. In particular, these dipole-like layers are present for small but finite \LCJ coupling strengths. In contrast, for stronger values of the \LCJ coupling strength this is only true for intermediate to weak interactions at low bias voltages. For strong interactions, as well as for intermediate to weak interactions at moderate to high values of the bias voltage, the dipole-like layers are destroyed and the charging of the LIL and CIL have the same sign. The dependence of $\Delta n_{\rm CIL}$ on the local Hubbard interaction $U$ is quite peculiar, being  exponentially decreasing for small $t_{lc}$ while for large $t_{lc}$ it displays a non-monotonic behavior and even changes sign as a function of $U$.

This behavior can be understood from the fact that the dipole-like layers are formed if the charges flow faster out of the transition region than they flow in, i.e.
\begin{equation}
\label{eq:dipole_uneq}
\frac{t_{c,{\rm eff}}}{t_{lc}}>1
\end{equation}
where $t_{c,{\rm eff}}$ is the effective hopping for the \cc region. Indeed, we observe that for sufficiently large $t_{lc}$ the dipole-like layers get destroyed in accordance with Eq.~\ref{eq:dipole_uneq}. Obviously, increasing the Hubbard interaction effectively decreases the mobility in the \cc region. We also observe that a Kondo-like peak is present in the spectral function of the CIL for large values of the \LCJ coupling strength $t_{lc}$ and the Hubbard interaction $U$. This suggests that the dipole-like layers has the tendency to suppress the Kondo-like peak. As summary of our results is reported in the  three-dimensional plot Fig.~\ref{Fig:3Dsummary}.

Finally, we want to emphasize that the results presented in this work obtained for the Hubbard interaction differ from the ones for the Falicov-Kimball model for large values of the \LCJ coupling ($t_{lc}=1$) in Ref. \onlinecite{ha.fr.12}. In the latter, the LIL and CIL are always oppositely charged. This indicates, that the sign change of $\Delta n_{\rm CIL}$ is not a generic feature of strong local correlations paired with long-range coulomb forces.  Rather it is a combined effect of strong local Hubbard interactions together with long-range coulomb forces.

\begin{acknowledgments}

We thank Walter Hofstetter and Martin Eckstein for valuable discussions. This work was supported by the Austrian Science Fund (FWF):  P26508, as well as SfB-ViCoM project F04103, and NaWi Graz. The calculations were partly performed on the D-Cluster Graz 
and on the VSC-3 cluster Vienna. 
\end{acknowledgments}

\appendix
\section{Poisson equation}\label{appendix}

Here, we present the details of the self-consistent solution of Eq.~\eqref{Poisson}-\eqref{n_z}. As mentioned in the main text, we employ the Newton-Raphson method to find the root of
\begin{equation}
\label{app:Phij}
\Phi_z(\vec{v})=c_z\left[\frac{\partial}{\partial z}\left(\frac{1}{c_z}\frac{\partial v_z}{\partial z}\right) +\left(n_z - n^{\rm bulk}\right)\right]\,.
\end{equation}
First, we discretize the derivative. Setting the lattice constant  $a=1$, we get
\begin{eqnarray}
\label{app:Phij_discret}
&&\hspace{-0.6cm}\Phi_z({\vec{v}})=v_{z+1}-2v_{z}+v_{z-1}
+\frac{\varepsilon_{r,z+1}-\varepsilon_{r,z-1}}{2\varepsilon_{r,z}}\frac{v_{z+1}-v_{z-1}}{2}
\nonumber\\
&&\hspace{0.35cm}+\frac{1}{\varepsilon_{r,z}\varepsilon_{0}}\left(n_z - n^{\rm bulk}\right)
\,.
\end{eqnarray}

Following Newton-Raphson, we expand
\begin{equation}
\label{app:Phi_expansion}
\Phi_{j}(\vec{v} + \Delta \vec{v}) = \Phi_{j}(\vec{v}) + \sum_{i} \frac{\partial \Phi_{j}(\vec{v})}{\partial v_{i}}  \Delta v_{i} \, .
\end{equation}
Here $\Delta \vec{v}={\vec{v}}^{(n+1)}-{\vec{v}}^{(n)}$ is the difference between two consecutive iterations in the Poisson loop. Assuming $\Phi_{z}(\vec{v} + \Delta \vec{v})\overset{!}{=} 0$, we obtain
\begin{equation}
\label{app:Deltav}
\Phi_{j}(\vec{v}) =-M_{ji} (v^{(n+1)}_{i}- v^{(n)}_{i}) \, , 
\end{equation}
with
\begin{equation}
\label{app:Mji}
M_{ji}=\frac{\partial}{\partial v_{i}} \Phi_{j}(\vec{v}) \, ,
\end{equation}
which leads to the final iteration scheme
\begin{equation}
\label{app:iteration_scheme}
\vec{v}^{(n+1)}=\vec{v}^{(n)}-M^{-1}\vv \Phi(\vec{v}) \, .
\end{equation}

\subsection{Expressions for  the matrix Elements $M_{ji}$}
Plugging  Eq.~\eqref{app:Phij} into Eq.~\eqref{app:Mji}, we obtain
\begin{equation}
\label{app:Mji0}
M_{ji} = 
\underbrace{
\frac{\partial }{\partial v_{i}}\left[\frac{1}{\varepsilon_{r,j}}\frac{\partial}{\partial z}\left(\varepsilon_{r,j} \frac{\partial v_{j}}{\partial z}\right)\right] 
}_{\equiv M^{(1)}_{ji}} 
+ \underbrace{
\frac{\partial }{\partial v_{i}} \left[\frac{1}{\varepsilon_{r,j}\varepsilon_{0}} n_{j}(\vec{v})\right]}_{M^{(2)}_{ji}} \, .
\end{equation}
Here, we used $c_z\equiv \frac{1}{\varepsilon_0 \varepsilon_{r,z}}$ and the fact that $n^{\rm bulk}$ does not depend on $\vec{v}$ and therefore $\partial n^{\rm bulk}/\partial v_{i}=0$.

After some simple manipulations, we arrive at
\begin{align}
\label{app:Mji_1} 
M^{(1)}_{ji} &\equiv \frac{\partial }{\partial v_{i}} \left[\frac{1}{\varepsilon_{r,j}}\frac{\partial}{\partial z}\left(\varepsilon_{r,j} \frac{\partial v_{j}}{\partial z}\right) \right] \\
&=\left(1-\frac{\delta\varepsilon_{r,j}}{4\varepsilon_{r,j}}\right) \delta_{i,j-1}- 2\delta_{i,j}
	     +\left(1+\frac{\delta\varepsilon_{r,j}}{4\varepsilon_{r,j}}\right) \delta_{i,j+1}
              \, \nonumber
\end{align}
with $\delta\varepsilon_{r,j}=\varepsilon_{r,j+1}-\varepsilon_{r,j-1}$.

This leaves us with the evaluation of the matrix elements for $M^{(2)}$ which involves the dependence of the charge density on the onsite energies. Using, the defining equations Eqs.~\eqref{G_loc},~\eqref{n_z} in Eq.~\eqref{app:Mji0}, we obtain
\begin{align}
\label{app:Mji_2_a} 
M^{(2)}_{ji}&=\frac{e^2}{\varepsilon_{r,j} \varepsilon_{0}} \frac{\partial }{\partial v_{i}} n_{j}\\
	    &= \frac{e^2}{\varepsilon_{r,j} \varepsilon_{0}}\int\limits_{\rm BZ}\frac{d^2 \vv k}{(2\pi)^2} \int \frac{d\omega}{2\pi} \Im m\bigg(\frac{\partial }{\partial v_{i}} G^{K}_{jj}(\omega, \vv k)
\bigg)\;. \nonumber
\end{align}

Next, using the Keldysh inversion formula, $G^K = -G^R [G^{-1}]^K G^A$, we can expand the derivative 
\begin{align}
\label{app:dGKdv_a}
\frac{\partial }{\partial v_{i}} G^{K}_{jj}&=
-\frac{\partial G^{R}_{jl}}{\partial v_i}\left[G^{-1}\right]^{K}_{ll'}G^{A}_{l'j}
-G^{R}_{jl}\left[G^{-1}\right]^{K}_{ll'}\frac{\partial G^{A}_{l'j}}{\partial v_i} \nonumber\\
&-G^{R}_{jl}\frac{\partial}{\partial v_i}\left[G^{-1}\right]^{K}_{ll'}G^{A}_{l'j} \, .
\end{align}
Here and below, all indices appearing twice are summed over. 
Relating the derivative $\frac{\partial G^{\gamma=R,A}_{jl}}{\partial v_i}$ to the derivative of its inverse, given by Eq.~\eqref{g0R},\footnote{In principle this expression is missing the self-energy due to the interaction, however recall that the Poissonian loop is performed for fixed self-energy and thus this term does not contribute to the sort for derivative} leads to
\begin{align}
\label{app:dGgammadv}
\frac{\partial G^{\gamma=R,A}_{jl}}{\partial v_i}&=-G^{\gamma}_{jl'}\frac{\partial}{\partial v_{i}}[G^{\gamma}]^{-1}_{l'l''}G^{\gamma}_{l''l} \nonumber\\
&=G^{\gamma}_{ji}G^{\gamma}_{il}+G^{\gamma}_{ji}G^{\gamma}_{il}\frac{\partial}{\partial v_{i}}\Sigma^\gamma_{{\rm hyb},i}\, .
\end{align}
and recalling Eq.~\eqref{g0K}, we also have
\begin{align}
\frac{\partial}{\partial v_i}\left[G^{-1}\right]^{K}_{ll'} =  -\delta_{li}\delta_{l'i}
\frac{\partial }{\partial v_{i}} \Sigma^\gamma_{{\rm hyb},i} \, .
\end{align}
Thus, Eq.~\eqref{app:dGKdv_a} now reads
\begin{align}
\label{app:dGKdv_b}
\frac{\partial }{\partial v_{i}} G^{K}_{jj}&=
G^{R}_{ji} G^{K}_{ij} +G^{K}_{ji} G^{A}_{ij}\nonumber\\
&+ G^{R}_{ji} G^{K}_{ij}\frac{\partial}{\partial v_{i}}\Sigma^R_{{\rm hyb},i}+G^{K}_{ji} G^{A}_{ij}\frac{\partial}{\partial v_{i}}\Sigma^A_{{\rm hyb},i}
\nonumber\\
&+G^{R}_{ji}G^{A}_{ij}\frac{\partial}{\partial v_{i}}\Sigma^K_{{\rm hyb},i} \, .
\end{align}
which based on the symmetries of the Green's function and the fluctuation dissipation theorem for $\Sigma^K_{{\rm hyb},i}$ allows the simplification to the final form
\begin{align}
\label{app:dGKdv_fin}
\frac{\partial }{\partial v_{i}} G^{K}_{jj}&=
2i\Im m\left[G^{R}_{ji} G^{K}_{ij}\right]
+N_{j \kappa}^{b}\delta_{\kappa 1}+N_{j \kappa}^{b}\delta_{\kappa L} \; .
\end{align}
Here
\begin{align}
\label{app:Njkappa}
N_{j\kappa}^{b}&=2i\Im m \bigg[
\eta_{\kappa}\left(G^{R}_{j \kappa} G^{K}_{\kappa j}
+ \big|G^{R}_{j \kappa}\big|^{2}\big(1-2 f_{\kappa} \big)\right)\bigg] 
\end{align}
and
\begin{align}
\label{app:eta}
\eta_{\kappa}&=- \frac{1}{2}
\bigg(1 + i\frac{\omega-v_{\kappa}-v_{\kappa}^{(0)}-E({\vv k})}{\sqrt{4t_\kappa^{2}-\left(\omega-v_{\kappa}-v_{\kappa}^{(0)}-E({\vv k})\right)^{2}}}\bigg) \, .
\end{align}
Moreover $f_{\kappa=1,L}$ stands for the Fermi function in left and right leads respectively. 

To speed up the convergence, we can use the fact that the electron density in the first and last site will converge to their bulk values and therefore we consider them fixed, which also means $\partial n_{j=1,l}/\partial v_i=0$ and correspondingly $M^{(2)}_{ji}=0$ for $j\in \{1,L\}$.


\begin{thebibliography}{87}%
\makeatletter
\providecommand \@ifxundefined [1]{%
 \@ifx{#1\undefined}
}%
\providecommand \@ifnum [1]{%
 \ifnum #1\expandafter \@firstoftwo
 \else \expandafter \@secondoftwo
 \fi
}%
\providecommand \@ifx [1]{%
 \ifx #1\expandafter \@firstoftwo
 \else \expandafter \@secondoftwo
 \fi
}%
\providecommand \natexlab [1]{#1}%
\providecommand \enquote  [1]{``#1''}%
\providecommand \bibnamefont  [1]{#1}%
\providecommand \bibfnamefont [1]{#1}%
\providecommand \citenamefont [1]{#1}%
\providecommand \href@noop [0]{\@secondoftwo}%
\providecommand \href [0]{\begingroup \@sanitize@url \@href}%
\providecommand \@href[1]{\@@startlink{#1}\@@href}%
\providecommand \@@href[1]{\endgroup#1\@@endlink}%
\providecommand \@sanitize@url [0]{\catcode `\\12\catcode `\$12\catcode
  `\&12\catcode `\#12\catcode `\^12\catcode `\_12\catcode `\%12\relax}%
\providecommand \@@startlink[1]{}%
\providecommand \@@endlink[0]{}%
\providecommand \url  [0]{\begingroup\@sanitize@url \@url }%
\providecommand \@url [1]{\endgroup\@href {#1}{\urlprefix }}%
\providecommand \urlprefix  [0]{URL }%
\providecommand \Eprint [0]{\href }%
\providecommand \doibase [0]{http://dx.doi.org/}%
\providecommand \selectlanguage [0]{\@gobble}%
\providecommand \bibinfo  [0]{\@secondoftwo}%
\providecommand \bibfield  [0]{\@secondoftwo}%
\providecommand \translation [1]{[#1]}%
\providecommand \BibitemOpen [0]{}%
\providecommand \bibitemStop [0]{}%
\providecommand \bibitemNoStop [0]{.\EOS\space}%
\providecommand \EOS [0]{\spacefactor3000\relax}%
\providecommand \BibitemShut  [1]{\csname bibitem#1\endcsname}%
\let\auto@bib@innerbib\@empty
\bibitem [{\citenamefont {Izumi}\ \emph {et~al.}(2001)\citenamefont {Izumi},
  \citenamefont {Ogimoto}, \citenamefont {Konishi}, \citenamefont {Manako},
  \citenamefont {Kawasaki},\ and\ \citenamefont {Tokura}}]{is.og.01}%
  \BibitemOpen
  \bibfield  {author} {\bibinfo {author} {\bibfnamefont {M.}~\bibnamefont
  {Izumi}}, \bibinfo {author} {\bibfnamefont {Y.}~\bibnamefont {Ogimoto}},
  \bibinfo {author} {\bibfnamefont {Y.}~\bibnamefont {Konishi}}, \bibinfo
  {author} {\bibfnamefont {T.}~\bibnamefont {Manako}}, \bibinfo {author}
  {\bibfnamefont {M.}~\bibnamefont {Kawasaki}}, \ and\ \bibinfo {author}
  {\bibfnamefont {Y.}~\bibnamefont {Tokura}},\ }\href {\doibase
  10.1016/S0921-5107(01)00569-4} {\bibfield  {journal} {\bibinfo  {journal}
  {Materials Science and Engineering: B}\ }\textbf {\bibinfo {volume} {84}},\
  \bibinfo {pages} {53 } (\bibinfo {year} {2001})}\BibitemShut {NoStop}%
\bibitem [{\citenamefont {Ohtomo}\ \emph {et~al.}(2002)\citenamefont {Ohtomo},
  \citenamefont {Muller}, \citenamefont {Grazul},\ and\ \citenamefont
  {Hwang}}]{oh.mu.02}%
  \BibitemOpen
  \bibfield  {author} {\bibinfo {author} {\bibfnamefont {A.}~\bibnamefont
  {Ohtomo}}, \bibinfo {author} {\bibfnamefont {D.~A.}\ \bibnamefont {Muller}},
  \bibinfo {author} {\bibfnamefont {J.~L.}\ \bibnamefont {Grazul}}, \ and\
  \bibinfo {author} {\bibfnamefont {H.~Y.}\ \bibnamefont {Hwang}},\ }\href
  {\doibase 10.1038/nature00977} {\bibfield  {journal} {\bibinfo  {journal}
  {Nature}\ }\textbf {\bibinfo {volume} {419}},\ \bibinfo {pages} {378}
  (\bibinfo {year} {2002})}\BibitemShut {NoStop}%
\bibitem [{\citenamefont {Gariglio}\ \emph {et~al.}(2002)\citenamefont
  {Gariglio}, \citenamefont {Ahn}, \citenamefont {Matthey},\ and\ \citenamefont
  {Triscone}}]{ga.ah.02}%
  \BibitemOpen
  \bibfield  {author} {\bibinfo {author} {\bibfnamefont {S.}~\bibnamefont
  {Gariglio}}, \bibinfo {author} {\bibfnamefont {C.~H.}\ \bibnamefont {Ahn}},
  \bibinfo {author} {\bibfnamefont {D.}~\bibnamefont {Matthey}}, \ and\
  \bibinfo {author} {\bibfnamefont {J.-M.}\ \bibnamefont {Triscone}},\ }\href
  {\doibase 10.1103/PhysRevLett.88.067002} {\bibfield  {journal} {\bibinfo
  {journal} {Phys. Rev. Lett.}\ }\textbf {\bibinfo {volume} {88}},\ \bibinfo
  {pages} {067002} (\bibinfo {year} {2002})}\BibitemShut {NoStop}%
\bibitem [{\citenamefont {Ahn}\ \emph {et~al.}(1999)\citenamefont {Ahn},
  \citenamefont {Gariglio}, \citenamefont {Paruch}, \citenamefont {Tybell},
  \citenamefont {Antognazza},\ and\ \citenamefont {Triscone}}]{an.ga.99}%
  \BibitemOpen
  \bibfield  {author} {\bibinfo {author} {\bibfnamefont {C.~H.}\ \bibnamefont
  {Ahn}}, \bibinfo {author} {\bibfnamefont {S.}~\bibnamefont {Gariglio}},
  \bibinfo {author} {\bibfnamefont {P.}~\bibnamefont {Paruch}}, \bibinfo
  {author} {\bibfnamefont {T.}~\bibnamefont {Tybell}}, \bibinfo {author}
  {\bibfnamefont {L.}~\bibnamefont {Antognazza}}, \ and\ \bibinfo {author}
  {\bibfnamefont {J.-M.}\ \bibnamefont {Triscone}},\ }\href {\doibase
  10.1126/science.284.5417.1152} {\bibfield  {journal} {\bibinfo  {journal}
  {Science}\ }\textbf {\bibinfo {volume} {284}},\ \bibinfo {pages} {1152}
  (\bibinfo {year} {1999})}\BibitemShut {NoStop}%
\bibitem [{\citenamefont {Ohtomo}\ and\ \citenamefont
  {Hwang}(2004)}]{oh.hw.04}%
  \BibitemOpen
  \bibfield  {author} {\bibinfo {author} {\bibfnamefont {A.}~\bibnamefont
  {Ohtomo}}\ and\ \bibinfo {author} {\bibfnamefont {H.~Y.}\ \bibnamefont
  {Hwang}},\ }\href {\doibase 10.1038/nature02308} {\bibfield  {journal}
  {\bibinfo  {journal} {Nature}\ }\textbf {\bibinfo {volume} {427}},\ \bibinfo
  {pages} {423} (\bibinfo {year} {2004})}\BibitemShut {NoStop}%
\bibitem [{\citenamefont {Zhu}\ \emph {et~al.}(2012)\citenamefont {Zhu},
  \citenamefont {Wang}, \citenamefont {Zhao}, \citenamefont {Li}, \citenamefont
  {Wang}, \citenamefont {Luo}, \citenamefont {Chan},\ and\ \citenamefont
  {Zheng}}]{zh.wa.12}%
  \BibitemOpen
  \bibfield  {author} {\bibinfo {author} {\bibfnamefont {Q.~X.}\ \bibnamefont
  {Zhu}}, \bibinfo {author} {\bibfnamefont {W.}~\bibnamefont {Wang}}, \bibinfo
  {author} {\bibfnamefont {X.~Q.}\ \bibnamefont {Zhao}}, \bibinfo {author}
  {\bibfnamefont {X.~M.}\ \bibnamefont {Li}}, \bibinfo {author} {\bibfnamefont
  {Y.}~\bibnamefont {Wang}}, \bibinfo {author} {\bibfnamefont {H.~S.}\
  \bibnamefont {Luo}}, \bibinfo {author} {\bibfnamefont {H.~L.~W.}\
  \bibnamefont {Chan}}, \ and\ \bibinfo {author} {\bibfnamefont {R.~K.}\
  \bibnamefont {Zheng}},\ }\href {\doibase 10.1063/1.4716188} {\bibfield
  {journal} {\bibinfo  {journal} {Journal of Applied Physics}\ }\textbf
  {\bibinfo {volume} {111}},\ \bibinfo {eid} {103702} (\bibinfo {year}
  {2012})}\BibitemShut {NoStop}%
\bibitem [{\citenamefont {Zenia}\ \emph {et~al.}(2009)\citenamefont {Zenia},
  \citenamefont {Freericks}, \citenamefont {Krishnamurthy},\ and\ \citenamefont
  {Pruschke}}]{ze.fr.09}%
  \BibitemOpen
  \bibfield  {author} {\bibinfo {author} {\bibfnamefont {H.}~\bibnamefont
  {Zenia}}, \bibinfo {author} {\bibfnamefont {J.~K.}\ \bibnamefont
  {Freericks}}, \bibinfo {author} {\bibfnamefont {H.~R.}\ \bibnamefont
  {Krishnamurthy}}, \ and\ \bibinfo {author} {\bibfnamefont {T.}~\bibnamefont
  {Pruschke}},\ }\href {\doibase 10.1103/PhysRevLett.103.116402} {\bibfield
  {journal} {\bibinfo  {journal} {Phys. Rev. Lett.}\ }\textbf {\bibinfo
  {volume} {103}},\ \bibinfo {pages} {116402} (\bibinfo {year}
  {2009})}\BibitemShut {NoStop}%
\bibitem [{\citenamefont {Hale}\ and\ \citenamefont
  {Freericks}(2011)}]{ha.fr.11}%
  \BibitemOpen
  \bibfield  {author} {\bibinfo {author} {\bibfnamefont {S.~T.~F.}\
  \bibnamefont {Hale}}\ and\ \bibinfo {author} {\bibfnamefont {J.~K.}\
  \bibnamefont {Freericks}},\ }\href {\doibase 10.1103/PhysRevB.83.035102}
  {\bibfield  {journal} {\bibinfo  {journal} {Phys. Rev. B}\ }\textbf {\bibinfo
  {volume} {83}},\ \bibinfo {pages} {035102} (\bibinfo {year}
  {2011})}\BibitemShut {NoStop}%
\bibitem [{\citenamefont {Knap}, \citenamefont {{von der Linden}},\ and\
  \citenamefont {Arrigoni}(2011)}]{kn.li.11}%
  \BibitemOpen
  \bibfield  {author} {\bibinfo {author} {\bibfnamefont {M.}~\bibnamefont
  {Knap}}, \bibinfo {author} {\bibfnamefont {W.}~\bibnamefont {{von der
  Linden}}}, \ and\ \bibinfo {author} {\bibfnamefont {E.}~\bibnamefont
  {Arrigoni}},\ }\href {\doibase 10.1103/PhysRevB.84.115145} {\bibfield
  {journal} {\bibinfo  {journal} {Phys. Rev. B}\ }\textbf {\bibinfo {volume}
  {84}},\ \bibinfo {pages} {115145} (\bibinfo {year} {2011})}\BibitemShut
  {NoStop}%
\bibitem [{\citenamefont {Mazza}\ \emph {et~al.}(2015)\citenamefont {Mazza},
  \citenamefont {Amaricci}, \citenamefont {Capone},\ and\ \citenamefont
  {Fabrizio}}]{ma.am.15}%
  \BibitemOpen
  \bibfield  {author} {\bibinfo {author} {\bibfnamefont {G.}~\bibnamefont
  {Mazza}}, \bibinfo {author} {\bibfnamefont {A.}~\bibnamefont {Amaricci}},
  \bibinfo {author} {\bibfnamefont {M.}~\bibnamefont {Capone}}, \ and\ \bibinfo
  {author} {\bibfnamefont {M.}~\bibnamefont {Fabrizio}},\ }\href {\doibase
  10.1103/PhysRevB.91.195124} {\bibfield  {journal} {\bibinfo  {journal} {Phys.
  Rev. B}\ }\textbf {\bibinfo {volume} {91}},\ \bibinfo {pages} {195124}
  (\bibinfo {year} {2015})}\BibitemShut {NoStop}%
\bibitem [{\citenamefont {Ribeiro}, \citenamefont {Antipov},\ and\
  \citenamefont {Rubtsov}(2016)}]{ri.an.16}%
  \BibitemOpen
  \bibfield  {author} {\bibinfo {author} {\bibfnamefont {P.}~\bibnamefont
  {Ribeiro}}, \bibinfo {author} {\bibfnamefont {A.~E.}\ \bibnamefont
  {Antipov}}, \ and\ \bibinfo {author} {\bibfnamefont {A.~N.}\ \bibnamefont
  {Rubtsov}},\ }\href {\doibase 10.1103/PhysRevB.93.144305} {\bibfield
  {journal} {\bibinfo  {journal} {Phys. Rev. B}\ }\textbf {\bibinfo {volume}
  {93}},\ \bibinfo {pages} {144305} (\bibinfo {year} {2016})}\BibitemShut
  {NoStop}%
\bibitem [{\citenamefont {Okamoto}(2007)}]{okam.07}%
  \BibitemOpen
  \bibfield  {author} {\bibinfo {author} {\bibfnamefont {S.}~\bibnamefont
  {Okamoto}},\ }\href {\doibase 10.1103/PhysRevB.76.035105} {\bibfield
  {journal} {\bibinfo  {journal} {Phys. Rev. B}\ }\textbf {\bibinfo {volume}
  {76}},\ \bibinfo {pages} {035105} (\bibinfo {year} {2007})}\BibitemShut
  {NoStop}%
\bibitem [{\citenamefont {Okamoto}(2008)}]{okam.08}%
  \BibitemOpen
  \bibfield  {author} {\bibinfo {author} {\bibfnamefont {S.}~\bibnamefont
  {Okamoto}},\ }\href {\doibase 10.1103/PhysRevLett.101.116807} {\bibfield
  {journal} {\bibinfo  {journal} {Phys. Rev. Lett.}\ }\textbf {\bibinfo
  {volume} {101}},\ \bibinfo {pages} {116807} (\bibinfo {year}
  {2008})}\BibitemShut {NoStop}%
\bibitem [{\citenamefont {Eckstein}\ and\ \citenamefont
  {Werner}(2014)}]{ec.we.14}%
  \BibitemOpen
  \bibfield  {author} {\bibinfo {author} {\bibfnamefont {M.}~\bibnamefont
  {Eckstein}}\ and\ \bibinfo {author} {\bibfnamefont {P.}~\bibnamefont
  {Werner}},\ }\href {\doibase 10.1103/PhysRevLett.113.076405} {\bibfield
  {journal} {\bibinfo  {journal} {Phys. Rev. Lett.}\ }\textbf {\bibinfo
  {volume} {113}},\ \bibinfo {pages} {076405} (\bibinfo {year}
  {2014})}\BibitemShut {NoStop}%
\bibitem [{\citenamefont {Sorantin}\ \emph {et~al.}(2018)\citenamefont
  {Sorantin}, \citenamefont {Dorda}, \citenamefont {Held},\ and\ \citenamefont
  {Arrigoni}}]{so.do.18}%
  \BibitemOpen
  \bibfield  {author} {\bibinfo {author} {\bibfnamefont {M.~E.}\ \bibnamefont
  {Sorantin}}, \bibinfo {author} {\bibfnamefont {A.}~\bibnamefont {Dorda}},
  \bibinfo {author} {\bibfnamefont {K.}~\bibnamefont {Held}}, \ and\ \bibinfo
  {author} {\bibfnamefont {E.}~\bibnamefont {Arrigoni}},\ }\href {\doibase
  10.1103/PhysRevB.97.115113} {\bibfield  {journal} {\bibinfo  {journal} {Phys.
  Rev. B}\ }\textbf {\bibinfo {volume} {97}},\ \bibinfo {pages} {115113}
  (\bibinfo {year} {2018})}\BibitemShut {NoStop}%
\bibitem [{\citenamefont {Eckstein}\ and\ \citenamefont
  {Werner}(2013)}]{ec.we.13}%
  \BibitemOpen
  \bibfield  {author} {\bibinfo {author} {\bibfnamefont {M.}~\bibnamefont
  {Eckstein}}\ and\ \bibinfo {author} {\bibfnamefont {P.}~\bibnamefont
  {Werner}},\ }\href {\doibase 10.1103/PhysRevB.88.075135} {\bibfield
  {journal} {\bibinfo  {journal} {Phys. Rev. B}\ }\textbf {\bibinfo {volume}
  {88}},\ \bibinfo {pages} {075135} (\bibinfo {year} {2013})}\BibitemShut
  {NoStop}%
\bibitem [{\citenamefont {Titvinidze}\ \emph {et~al.}(2016)\citenamefont
  {Titvinidze}, \citenamefont {Dorda}, \citenamefont {von~der Linden},\ and\
  \citenamefont {Arrigoni}}]{ti.do.16}%
  \BibitemOpen
  \bibfield  {author} {\bibinfo {author} {\bibfnamefont {I.}~\bibnamefont
  {Titvinidze}}, \bibinfo {author} {\bibfnamefont {A.}~\bibnamefont {Dorda}},
  \bibinfo {author} {\bibfnamefont {W.}~\bibnamefont {von~der Linden}}, \ and\
  \bibinfo {author} {\bibfnamefont {E.}~\bibnamefont {Arrigoni}},\ }\href
  {\doibase 10.1103/PhysRevB.94.245142} {\bibfield  {journal} {\bibinfo
  {journal} {Phys. Rev. B}\ }\textbf {\bibinfo {volume} {94}},\ \bibinfo
  {pages} {245142} (\bibinfo {year} {2016})}\BibitemShut {NoStop}%
\bibitem [{\citenamefont {Steffen}, \citenamefont {Fr\'esard},\ and\
  \citenamefont {Kopp}(2017)}]{st.fr.17}%
  \BibitemOpen
  \bibfield  {author} {\bibinfo {author} {\bibfnamefont {K.}~\bibnamefont
  {Steffen}}, \bibinfo {author} {\bibfnamefont {R.}~\bibnamefont {Fr\'esard}},
  \ and\ \bibinfo {author} {\bibfnamefont {T.}~\bibnamefont {Kopp}},\ }\href
  {\doibase 10.1103/PhysRevB.95.035143} {\bibfield  {journal} {\bibinfo
  {journal} {Phys. Rev. B}\ }\textbf {\bibinfo {volume} {95}},\ \bibinfo
  {pages} {035143} (\bibinfo {year} {2017})}\BibitemShut {NoStop}%
\bibitem [{\citenamefont {Bakalov}, \citenamefont {Ydens},\ and\ \citenamefont
  {Locquet}(2014)}]{ba.yd.14}%
  \BibitemOpen
  \bibfield  {author} {\bibinfo {author} {\bibfnamefont {P.}~\bibnamefont
  {Bakalov}}, \bibinfo {author} {\bibfnamefont {B.}~\bibnamefont {Ydens}}, \
  and\ \bibinfo {author} {\bibfnamefont {J.}~\bibnamefont {Locquet}},\ }\href
  {\doibase 10.1002/pssa.201300418} {\bibfield  {journal} {\bibinfo  {journal}
  {physica status solidi (a)}\ }\textbf {\bibinfo {volume} {211}},\ \bibinfo
  {pages} {440} (\bibinfo {year} {2014})},\ \Eprint
  {http://arxiv.org/abs/https://onlinelibrary.wiley.com/doi/pdf/10.1002/pssa.201300418}
  {https://onlinelibrary.wiley.com/doi/pdf/10.1002/pssa.201300418} \BibitemShut
  {NoStop}%
\bibitem [{\citenamefont {Hale}\ and\ \citenamefont
  {Freericks}(2012)}]{ha.fr.12}%
  \BibitemOpen
  \bibfield  {author} {\bibinfo {author} {\bibfnamefont {S.~T.~F.}\
  \bibnamefont {Hale}}\ and\ \bibinfo {author} {\bibfnamefont {J.~K.}\
  \bibnamefont {Freericks}},\ }\href {\doibase 10.1103/PhysRevB.85.205444}
  {\bibfield  {journal} {\bibinfo  {journal} {Phys. Rev. B}\ }\textbf {\bibinfo
  {volume} {85}},\ \bibinfo {pages} {205444} (\bibinfo {year}
  {2012})}\BibitemShut {NoStop}%
\bibitem [{\citenamefont {Freericks}\ and\ \citenamefont
  {Zlati\'{c}}(2007)}]{fr.zl.07}%
  \BibitemOpen
  \bibfield  {author} {\bibinfo {author} {\bibfnamefont {J.~K.}\ \bibnamefont
  {Freericks}}\ and\ \bibinfo {author} {\bibfnamefont {V.}~\bibnamefont
  {Zlati\'{c}}},\ }\href {\doibase 10.1002/pssb.200674611} {\bibfield
  {journal} {\bibinfo  {journal} {physica status solidi (b)}\ }\textbf
  {\bibinfo {volume} {244}},\ \bibinfo {pages} {2351} (\bibinfo {year}
  {2007})}\BibitemShut {NoStop}%
\bibitem [{\citenamefont {Bakalov}\ \emph {et~al.}(2016)\citenamefont
  {Bakalov}, \citenamefont {Nasr~Esfahani}, \citenamefont {Covaci},
  \citenamefont {Peeters}, \citenamefont {Tempere},\ and\ \citenamefont
  {Locquet}}]{ba.es.16}%
  \BibitemOpen
  \bibfield  {author} {\bibinfo {author} {\bibfnamefont {P.}~\bibnamefont
  {Bakalov}}, \bibinfo {author} {\bibfnamefont {D.}~\bibnamefont
  {Nasr~Esfahani}}, \bibinfo {author} {\bibfnamefont {L.}~\bibnamefont
  {Covaci}}, \bibinfo {author} {\bibfnamefont {F.~M.}\ \bibnamefont {Peeters}},
  \bibinfo {author} {\bibfnamefont {J.}~\bibnamefont {Tempere}}, \ and\
  \bibinfo {author} {\bibfnamefont {J.-P.}\ \bibnamefont {Locquet}},\ }\href
  {\doibase 10.1103/PhysRevB.93.165112} {\bibfield  {journal} {\bibinfo
  {journal} {Phys. Rev. B}\ }\textbf {\bibinfo {volume} {93}},\ \bibinfo
  {pages} {165112} (\bibinfo {year} {2016})}\BibitemShut {NoStop}%
\bibitem [{\citenamefont {Georges}\ \emph {et~al.}(1996)\citenamefont
  {Georges}, \citenamefont {Kotliar}, \citenamefont {Krauth},\ and\
  \citenamefont {Rozenberg}}]{ge.ko.96}%
  \BibitemOpen
  \bibfield  {author} {\bibinfo {author} {\bibfnamefont {A.}~\bibnamefont
  {Georges}}, \bibinfo {author} {\bibfnamefont {G.}~\bibnamefont {Kotliar}},
  \bibinfo {author} {\bibfnamefont {W.}~\bibnamefont {Krauth}}, \ and\ \bibinfo
  {author} {\bibfnamefont {M.~J.}\ \bibnamefont {Rozenberg}},\ }\href {\doibase
  10.1103/RevModPhys.68.13} {\bibfield  {journal} {\bibinfo  {journal} {Rev.
  Mod. Phys.}\ }\textbf {\bibinfo {volume} {68}},\ \bibinfo {pages} {13}
  (\bibinfo {year} {1996})}\BibitemShut {NoStop}%
\bibitem [{\citenamefont {Vollhardt}(2010)}]{voll.10}%
  \BibitemOpen
  \bibfield  {author} {\bibinfo {author} {\bibfnamefont {D.}~\bibnamefont
  {Vollhardt}},\ }in\ \href {\doibase 10.1063/1.3518901} {\emph {\bibinfo
  {booktitle} {Lecture Notes on the Physics of Strongly Correlated Systems}}},\
  \bibinfo {series} {AIP Conf. Proc.}, Vol.\ \bibinfo {volume} {1297},\
  \bibinfo {editor} {edited by\ \bibinfo {editor} {\bibfnamefont
  {A.}~\bibnamefont {Avella}}\ and\ \bibinfo {editor} {\bibfnamefont
  {F.}~\bibnamefont {Mancini}}}\ (\bibinfo {address} {AIP, New York},\ \bibinfo
  {year} {2010})\ pp.\ \bibinfo {pages} {339--403}\BibitemShut {NoStop}%
\bibitem [{\citenamefont {Metzner}\ and\ \citenamefont
  {Vollhardt}(1989)}]{me.vo.89}%
  \BibitemOpen
  \bibfield  {author} {\bibinfo {author} {\bibfnamefont {W.}~\bibnamefont
  {Metzner}}\ and\ \bibinfo {author} {\bibfnamefont {D.}~\bibnamefont
  {Vollhardt}},\ }\href {\doibase 10.1103/PhysRevLett.62.324} {\bibfield
  {journal} {\bibinfo  {journal} {Phys. Rev. Lett.}\ }\textbf {\bibinfo
  {volume} {62}},\ \bibinfo {pages} {324} (\bibinfo {year} {1989})}\BibitemShut
  {NoStop}%
\bibitem [{\citenamefont {Potthoff}\ and\ \citenamefont
  {Nolting}(1999{\natexlab{a}})}]{po.no.99.sm}%
  \BibitemOpen
  \bibfield  {author} {\bibinfo {author} {\bibfnamefont {M.}~\bibnamefont
  {Potthoff}}\ and\ \bibinfo {author} {\bibfnamefont {W.}~\bibnamefont
  {Nolting}},\ }\href {\doibase 10.1103/PhysRevB.59.2549} {\bibfield  {journal}
  {\bibinfo  {journal} {Phys. Rev. B}\ }\textbf {\bibinfo {volume} {59}},\
  \bibinfo {pages} {2549} (\bibinfo {year} {1999}{\natexlab{a}})}\BibitemShut
  {NoStop}%
\bibitem [{\citenamefont {Potthoff}\ and\ \citenamefont
  {Nolting}(1999{\natexlab{b}})}]{po.no.99.ms}%
  \BibitemOpen
  \bibfield  {author} {\bibinfo {author} {\bibfnamefont {M.}~\bibnamefont
  {Potthoff}}\ and\ \bibinfo {author} {\bibfnamefont {W.}~\bibnamefont
  {Nolting}},\ }\href {\doibase 10.1103/PhysRevB.60.7834} {\bibfield  {journal}
  {\bibinfo  {journal} {Phys. Rev. B}\ }\textbf {\bibinfo {volume} {60}},\
  \bibinfo {pages} {7834} (\bibinfo {year} {1999}{\natexlab{b}})}\BibitemShut
  {NoStop}%
\bibitem [{\citenamefont {Potthoff}\ and\ \citenamefont
  {Nolting}(1999{\natexlab{c}})}]{po.no.99.ldmft}%
  \BibitemOpen
  \bibfield  {author} {\bibinfo {author} {\bibfnamefont {M.}~\bibnamefont
  {Potthoff}}\ and\ \bibinfo {author} {\bibfnamefont {W.}~\bibnamefont
  {Nolting}},\ }\href {\doibase 10.1007/s100510050722} {\bibfield  {journal}
  {\bibinfo  {journal} {The European Physical Journal B - Condensed Matter and
  Complex Systems}\ }\textbf {\bibinfo {volume} {8}},\ \bibinfo {pages} {555}
  (\bibinfo {year} {1999}{\natexlab{c}})}\BibitemShut {NoStop}%
\bibitem [{\citenamefont {Potthoff}\ and\ \citenamefont
  {Nolting}(1999{\natexlab{d}})}]{po.no.99.emfl}%
  \BibitemOpen
  \bibfield  {author} {\bibinfo {author} {\bibfnamefont {M.}~\bibnamefont
  {Potthoff}}\ and\ \bibinfo {author} {\bibfnamefont {W.}~\bibnamefont
  {Nolting}},\ }\href {\doibase
  http://dx.doi.org/10.1016/S0921-4526(98)01126-0} {\bibfield  {journal}
  {\bibinfo  {journal} {Physica B: Condensed Matter}\ }\textbf {\bibinfo
  {volume} {259-261}},\ \bibinfo {pages} {760 } (\bibinfo {year}
  {1999}{\natexlab{d}})}\BibitemShut {NoStop}%
\bibitem [{\citenamefont {Freericks}(2006)}]{free.06}%
  \BibitemOpen
  \bibfield  {author} {\bibinfo {author} {\bibfnamefont {J.~K.}\ \bibnamefont
  {Freericks}},\ }\href@noop {} {\emph {\bibinfo {title} {Transport in
  multilayered nanostructurs}}}\ (\bibinfo  {publisher} {Imperial College
  Press},\ \bibinfo {address} {London},\ \bibinfo {year} {2006})\BibitemShut
  {NoStop}%
\bibitem [{\citenamefont {Nourafkan}, \citenamefont {Marsiglio},\ and\
  \citenamefont {Capone}(2010)}]{no.ma.10}%
  \BibitemOpen
  \bibfield  {author} {\bibinfo {author} {\bibfnamefont {R.}~\bibnamefont
  {Nourafkan}}, \bibinfo {author} {\bibfnamefont {F.}~\bibnamefont
  {Marsiglio}}, \ and\ \bibinfo {author} {\bibfnamefont {M.}~\bibnamefont
  {Capone}},\ }\href {\doibase 10.1103/PhysRevB.82.115127} {\bibfield
  {journal} {\bibinfo  {journal} {Phys. Rev. B}\ }\textbf {\bibinfo {volume}
  {82}},\ \bibinfo {pages} {115127} (\bibinfo {year} {2010})}\BibitemShut
  {NoStop}%
\bibitem [{\citenamefont {Ishida}\ and\ \citenamefont
  {Liebsch}(2009)}]{is.li.09}%
  \BibitemOpen
  \bibfield  {author} {\bibinfo {author} {\bibfnamefont {H.}~\bibnamefont
  {Ishida}}\ and\ \bibinfo {author} {\bibfnamefont {A.}~\bibnamefont
  {Liebsch}},\ }\href {\doibase 10.1103/PhysRevB.79.045130} {\bibfield
  {journal} {\bibinfo  {journal} {Phys. Rev. B}\ }\textbf {\bibinfo {volume}
  {79}},\ \bibinfo {pages} {045130} (\bibinfo {year} {2009})}\BibitemShut
  {NoStop}%
\bibitem [{\citenamefont {Nourafkan}\ and\ \citenamefont
  {Marsiglio}(2011)}]{no.ma.11}%
  \BibitemOpen
  \bibfield  {author} {\bibinfo {author} {\bibfnamefont {R.}~\bibnamefont
  {Nourafkan}}\ and\ \bibinfo {author} {\bibfnamefont {F.}~\bibnamefont
  {Marsiglio}},\ }\href {\doibase 10.1103/PhysRevB.83.155116} {\bibfield
  {journal} {\bibinfo  {journal} {Phys. Rev. B}\ }\textbf {\bibinfo {volume}
  {83}},\ \bibinfo {pages} {155116} (\bibinfo {year} {2011})}\BibitemShut
  {NoStop}%
\bibitem [{\citenamefont {Okamoto}(2011)}]{okam11}%
  \BibitemOpen
  \bibfield  {author} {\bibinfo {author} {\bibfnamefont {S.}~\bibnamefont
  {Okamoto}},\ }\href {\doibase 10.1103/PhysRevB.84.201305} {\bibfield
  {journal} {\bibinfo  {journal} {Phys. Rev. B}\ }\textbf {\bibinfo {volume}
  {84}},\ \bibinfo {pages} {201305} (\bibinfo {year} {2011})}\BibitemShut
  {NoStop}%
\bibitem [{\citenamefont {Miller}\ and\ \citenamefont
  {Freericks}(2001)}]{mi.fr.01}%
  \BibitemOpen
  \bibfield  {author} {\bibinfo {author} {\bibfnamefont {P.}~\bibnamefont
  {Miller}}\ and\ \bibinfo {author} {\bibfnamefont {J.~K.}\ \bibnamefont
  {Freericks}},\ }\href@noop {} {\bibfield  {journal} {\bibinfo  {journal} {J.
  Phys.: Condens. Matter}\ }\textbf {\bibinfo {volume} {13}},\ \bibinfo {pages}
  {3187} (\bibinfo {year} {2001})}\BibitemShut {NoStop}%
\bibitem [{\citenamefont {Freericks}(2004)}]{free.04}%
  \BibitemOpen
  \bibfield  {author} {\bibinfo {author} {\bibfnamefont {J.~K.}\ \bibnamefont
  {Freericks}},\ }\href {\doibase 10.1103/PhysRevB.70.195342} {\bibfield
  {journal} {\bibinfo  {journal} {Phys. Rev. B}\ }\textbf {\bibinfo {volume}
  {70}},\ \bibinfo {pages} {195342} (\bibinfo {year} {2004})}\BibitemShut
  {NoStop}%
\bibitem [{\citenamefont {Okamoto}\ and\ \citenamefont
  {Millis}(2004)}]{ok.mi.04.ldmft}%
  \BibitemOpen
  \bibfield  {author} {\bibinfo {author} {\bibfnamefont {S.}~\bibnamefont
  {Okamoto}}\ and\ \bibinfo {author} {\bibfnamefont {A.~J.}\ \bibnamefont
  {Millis}},\ }\href {\doibase 10.1103/PhysRevB.70.241104} {\bibfield
  {journal} {\bibinfo  {journal} {Phys. Rev. B}\ }\textbf {\bibinfo {volume}
  {70}},\ \bibinfo {pages} {241104} (\bibinfo {year} {2004})}\BibitemShut
  {NoStop}%
\bibitem [{\citenamefont {Dobrosavljevi\ifmmode~\acute{c}\else \'{c}\fi{}}\
  and\ \citenamefont {Kotliar}(1997)}]{do.ko.97}%
  \BibitemOpen
  \bibfield  {author} {\bibinfo {author} {\bibfnamefont {V.}~\bibnamefont
  {Dobrosavljevi\ifmmode~\acute{c}\else \'{c}\fi{}}}\ and\ \bibinfo {author}
  {\bibfnamefont {G.}~\bibnamefont {Kotliar}},\ }\href {\doibase
  10.1103/PhysRevLett.78.3943} {\bibfield  {journal} {\bibinfo  {journal}
  {Phys. Rev. Lett.}\ }\textbf {\bibinfo {volume} {78}},\ \bibinfo {pages}
  {3943} (\bibinfo {year} {1997})}\BibitemShut {NoStop}%
\bibitem [{\citenamefont {Dobrosavljevi{\'c}}\ and\ \citenamefont
  {Kotliar}(1998)}]{do.ko.98}%
  \BibitemOpen
  \bibfield  {author} {\bibinfo {author} {\bibfnamefont {V.}~\bibnamefont
  {Dobrosavljevi{\'c}}}\ and\ \bibinfo {author} {\bibfnamefont
  {G.}~\bibnamefont {Kotliar}},\ }\href@noop {} {\bibfield  {journal} {\bibinfo
   {journal} {Philosophical Transactions of the Royal Society of London A:
  Mathematical, Physical and Engineering Sciences}\ }\textbf {\bibinfo {volume}
  {356}},\ \bibinfo {pages} {57} (\bibinfo {year} {1998})}\BibitemShut
  {NoStop}%
\bibitem [{\citenamefont {Song}, \citenamefont {Wortis},\ and\ \citenamefont
  {Atkinson}(2008)}]{so.yu.08}%
  \BibitemOpen
  \bibfield  {author} {\bibinfo {author} {\bibfnamefont {Y.}~\bibnamefont
  {Song}}, \bibinfo {author} {\bibfnamefont {R.}~\bibnamefont {Wortis}}, \ and\
  \bibinfo {author} {\bibfnamefont {W.~A.}\ \bibnamefont {Atkinson}},\ }\href
  {\doibase 10.1103/PhysRevB.77.054202} {\bibfield  {journal} {\bibinfo
  {journal} {Phys. Rev. B}\ }\textbf {\bibinfo {volume} {77}},\ \bibinfo
  {pages} {054202} (\bibinfo {year} {2008})}\BibitemShut {NoStop}%
\bibitem [{\citenamefont {Wernsdorfer}\ \emph {et~al.}(2011)\citenamefont
  {Wernsdorfer}, \citenamefont {Harder}, \citenamefont {Schollwoeck},\ and\
  \citenamefont {Hofstetter}}]{we.ha.11u}%
  \BibitemOpen
  \bibfield  {author} {\bibinfo {author} {\bibfnamefont {J.}~\bibnamefont
  {Wernsdorfer}}, \bibinfo {author} {\bibfnamefont {G.}~\bibnamefont {Harder}},
  \bibinfo {author} {\bibfnamefont {U.}~\bibnamefont {Schollwoeck}}, \ and\
  \bibinfo {author} {\bibfnamefont {W.}~\bibnamefont {Hofstetter}},\ }\href
  {http://arxiv.org/abs/1108.6057} {\enquote {\bibinfo {title} {Signatures of
  delocalization in the fermionic 1d hubbard model with box disorder:
  Comparative study with dmrg and r-dmft},}\ } (\bibinfo {year} {2011}),\
  \bibinfo {note} {arXiv:1108.6057}\BibitemShut {NoStop}%
\bibitem [{\citenamefont {Helmes}, \citenamefont {Costi},\ and\ \citenamefont
  {Rosch}(2008)}]{he.co.08}%
  \BibitemOpen
  \bibfield  {author} {\bibinfo {author} {\bibfnamefont {R.~W.}\ \bibnamefont
  {Helmes}}, \bibinfo {author} {\bibfnamefont {T.~A.}\ \bibnamefont {Costi}}, \
  and\ \bibinfo {author} {\bibfnamefont {A.}~\bibnamefont {Rosch}},\ }\href
  {\doibase 10.1103/PhysRevLett.100.056403} {\bibfield  {journal} {\bibinfo
  {journal} {Phys. Rev. Lett.}\ }\textbf {\bibinfo {volume} {100}},\ \bibinfo
  {pages} {056403} (\bibinfo {year} {2008})}\BibitemShut {NoStop}%
\bibitem [{\citenamefont {Koga}\ \emph {et~al.}(2008)\citenamefont {Koga},
  \citenamefont {Higashiyama}, \citenamefont {Inaba}, \citenamefont {Suga},\
  and\ \citenamefont {Kawakami}}]{ko.hi.08}%
  \BibitemOpen
  \bibfield  {author} {\bibinfo {author} {\bibfnamefont {A.}~\bibnamefont
  {Koga}}, \bibinfo {author} {\bibfnamefont {T.}~\bibnamefont {Higashiyama}},
  \bibinfo {author} {\bibfnamefont {K.}~\bibnamefont {Inaba}}, \bibinfo
  {author} {\bibfnamefont {S.}~\bibnamefont {Suga}}, \ and\ \bibinfo {author}
  {\bibfnamefont {N.}~\bibnamefont {Kawakami}},\ }\href {\doibase
  10.1143/JPSJ.77.073602} {\bibfield  {journal} {\bibinfo  {journal} {Journal
  of the Physical Society of Japan}\ }\textbf {\bibinfo {volume} {77}},\
  \bibinfo {pages} {073602} (\bibinfo {year} {2008})}\BibitemShut {NoStop}%
\bibitem [{\citenamefont {Koga}\ \emph {et~al.}(2009)\citenamefont {Koga},
  \citenamefont {Higashiyama}, \citenamefont {Inaba}, \citenamefont {Suga},\
  and\ \citenamefont {Kawakami}}]{ko.hi.09}%
  \BibitemOpen
  \bibfield  {author} {\bibinfo {author} {\bibfnamefont {A.}~\bibnamefont
  {Koga}}, \bibinfo {author} {\bibfnamefont {T.}~\bibnamefont {Higashiyama}},
  \bibinfo {author} {\bibfnamefont {K.}~\bibnamefont {Inaba}}, \bibinfo
  {author} {\bibfnamefont {S.}~\bibnamefont {Suga}}, \ and\ \bibinfo {author}
  {\bibfnamefont {N.}~\bibnamefont {Kawakami}},\ }\href {\doibase
  10.1103/PhysRevA.79.013607} {\bibfield  {journal} {\bibinfo  {journal} {Phys.
  Rev. A}\ }\textbf {\bibinfo {volume} {79}},\ \bibinfo {pages} {013607}
  (\bibinfo {year} {2009})}\BibitemShut {NoStop}%
\bibitem [{\citenamefont {Noda}\ \emph {et~al.}(2009)\citenamefont {Noda},
  \citenamefont {Koga}, \citenamefont {Kawakami},\ and\ \citenamefont
  {Pruschke}}]{no.ka.09}%
  \BibitemOpen
  \bibfield  {author} {\bibinfo {author} {\bibfnamefont {K.}~\bibnamefont
  {Noda}}, \bibinfo {author} {\bibfnamefont {A.}~\bibnamefont {Koga}}, \bibinfo
  {author} {\bibfnamefont {N.}~\bibnamefont {Kawakami}}, \ and\ \bibinfo
  {author} {\bibfnamefont {T.}~\bibnamefont {Pruschke}},\ }\href {\doibase
  10.1103/PhysRevA.80.063622} {\bibfield  {journal} {\bibinfo  {journal} {Phys.
  Rev. A}\ }\textbf {\bibinfo {volume} {80}},\ \bibinfo {pages} {063622}
  (\bibinfo {year} {2009})}\BibitemShut {NoStop}%
\bibitem [{\citenamefont {Koga}\ \emph {et~al.}(2011)\citenamefont {Koga},
  \citenamefont {Bauer}, \citenamefont {Werner},\ and\ \citenamefont
  {Pruschke}}]{ko.ba.11}%
  \BibitemOpen
  \bibfield  {author} {\bibinfo {author} {\bibfnamefont {A.}~\bibnamefont
  {Koga}}, \bibinfo {author} {\bibfnamefont {J.}~\bibnamefont {Bauer}},
  \bibinfo {author} {\bibfnamefont {P.}~\bibnamefont {Werner}}, \ and\ \bibinfo
  {author} {\bibfnamefont {T.}~\bibnamefont {Pruschke}},\ }\href {\doibase
  http://dx.doi.org/10.1016/j.physe.2010.07.032} {\bibfield  {journal}
  {\bibinfo  {journal} {Physica E: Low-dimensional Systems and Nanostructures}\
  }\textbf {\bibinfo {volume} {43}},\ \bibinfo {pages} {697 } (\bibinfo {year}
  {2011})},\ \bibinfo {note} {nanoPHYS 09Proceedings of the International
  Symposium on Nanoscience and Quantum Physics}\BibitemShut {NoStop}%
\bibitem [{\citenamefont {Bl\"umer}\ and\ \citenamefont
  {Gorelik}(2011)}]{bl.go.11}%
  \BibitemOpen
  \bibfield  {author} {\bibinfo {author} {\bibfnamefont {N.}~\bibnamefont
  {Bl\"umer}}\ and\ \bibinfo {author} {\bibfnamefont {E.}~\bibnamefont
  {Gorelik}},\ }\href {\doibase http://dx.doi.org/10.1016/j.cpc.2010.07.011}
  {\bibfield  {journal} {\bibinfo  {journal} {Computer Physics Communications}\
  }\textbf {\bibinfo {volume} {182}},\ \bibinfo {pages} {115 } (\bibinfo {year}
  {2011})},\ \bibinfo {note} {computer Physics Communications Special Edition
  for Conference on Computational Physics Kaohsiung, Taiwan, Dec 15-19,
  2009}\BibitemShut {NoStop}%
\bibitem [{\citenamefont {Kim}\ \emph {et~al.}(2011)\citenamefont {Kim},
  \citenamefont {Kinnunen}, \citenamefont {Martikainen},\ and\ \citenamefont
  {T\"orm\"a}}]{ki.ki.11}%
  \BibitemOpen
  \bibfield  {author} {\bibinfo {author} {\bibfnamefont {D.-H.}\ \bibnamefont
  {Kim}}, \bibinfo {author} {\bibfnamefont {J.~J.}\ \bibnamefont {Kinnunen}},
  \bibinfo {author} {\bibfnamefont {J.-P.}\ \bibnamefont {Martikainen}}, \ and\
  \bibinfo {author} {\bibfnamefont {P.}~\bibnamefont {T\"orm\"a}},\ }\href
  {\doibase 10.1103/PhysRevLett.106.095301} {\bibfield  {journal} {\bibinfo
  {journal} {Phys. Rev. Lett.}\ }\textbf {\bibinfo {volume} {106}},\ \bibinfo
  {pages} {095301} (\bibinfo {year} {2011})}\BibitemShut {NoStop}%
\bibitem [{\citenamefont {Aulbach}, \citenamefont {Assaad},\ and\ \citenamefont
  {Potthoff}(2015)}]{au.as.15}%
  \BibitemOpen
  \bibfield  {author} {\bibinfo {author} {\bibfnamefont {M.~W.}\ \bibnamefont
  {Aulbach}}, \bibinfo {author} {\bibfnamefont {F.~F.}\ \bibnamefont {Assaad}},
  \ and\ \bibinfo {author} {\bibfnamefont {M.}~\bibnamefont {Potthoff}},\
  }\href {\doibase 10.1103/PhysRevB.92.235131} {\bibfield  {journal} {\bibinfo
  {journal} {Phys. Rev. B}\ }\textbf {\bibinfo {volume} {92}},\ \bibinfo
  {pages} {235131} (\bibinfo {year} {2015})}\BibitemShut {NoStop}%
\bibitem [{\citenamefont {Snoek}\ \emph {et~al.}(2008)\citenamefont {Snoek},
  \citenamefont {Titvinidze}, \citenamefont {Take}, \citenamefont {Byczuk},\
  and\ \citenamefont {Hofstetter}}]{sn.ti.08}%
  \BibitemOpen
  \bibfield  {author} {\bibinfo {author} {\bibfnamefont {M.}~\bibnamefont
  {Snoek}}, \bibinfo {author} {\bibfnamefont {I.}~\bibnamefont {Titvinidze}},
  \bibinfo {author} {\bibfnamefont {C.}~\bibnamefont {Take}}, \bibinfo {author}
  {\bibfnamefont {K.}~\bibnamefont {Byczuk}}, \ and\ \bibinfo {author}
  {\bibfnamefont {W.}~\bibnamefont {Hofstetter}},\ }\href
  {http://stacks.iop.org/1367-2630/10/i=9/a=093008} {\bibfield  {journal}
  {\bibinfo  {journal} {New Journal of Physics}\ }\textbf {\bibinfo {volume}
  {10}},\ \bibinfo {pages} {093008} (\bibinfo {year} {2008})}\BibitemShut
  {NoStop}%
\bibitem [{\citenamefont {Snoek}, \citenamefont {Titvinidze},\ and\
  \citenamefont {Hofstetter}(2011)}]{sn.ti.11}%
  \BibitemOpen
  \bibfield  {author} {\bibinfo {author} {\bibfnamefont {M.}~\bibnamefont
  {Snoek}}, \bibinfo {author} {\bibfnamefont {I.}~\bibnamefont {Titvinidze}}, \
  and\ \bibinfo {author} {\bibfnamefont {W.}~\bibnamefont {Hofstetter}},\
  }\href {\doibase 10.1103/PhysRevB.83.054419} {\bibfield  {journal} {\bibinfo
  {journal} {Phys. Rev. B}\ }\textbf {\bibinfo {volume} {83}},\ \bibinfo
  {pages} {054419} (\bibinfo {year} {2011})}\BibitemShut {NoStop}%
\bibitem [{\citenamefont {Titvinidze}\ \emph {et~al.}(2012)\citenamefont
  {Titvinidze}, \citenamefont {Schwabe}, \citenamefont {Rother},\ and\
  \citenamefont {Potthoff}}]{ti.sc.12}%
  \BibitemOpen
  \bibfield  {author} {\bibinfo {author} {\bibfnamefont {I.}~\bibnamefont
  {Titvinidze}}, \bibinfo {author} {\bibfnamefont {A.}~\bibnamefont {Schwabe}},
  \bibinfo {author} {\bibfnamefont {N.}~\bibnamefont {Rother}}, \ and\ \bibinfo
  {author} {\bibfnamefont {M.}~\bibnamefont {Potthoff}},\ }\href {\doibase
  10.1103/PhysRevB.86.075141} {\bibfield  {journal} {\bibinfo  {journal} {Phys.
  Rev. B}\ }\textbf {\bibinfo {volume} {86}},\ \bibinfo {pages} {075141}
  (\bibinfo {year} {2012})}\BibitemShut {NoStop}%
\bibitem [{\citenamefont {Gorelik}\ \emph {et~al.}(2010)\citenamefont
  {Gorelik}, \citenamefont {Titvinidze}, \citenamefont {Hofstetter},
  \citenamefont {Snoek},\ and\ \citenamefont {Bl\"umer}}]{go.ti.10}%
  \BibitemOpen
  \bibfield  {author} {\bibinfo {author} {\bibfnamefont {E.~V.}\ \bibnamefont
  {Gorelik}}, \bibinfo {author} {\bibfnamefont {I.}~\bibnamefont {Titvinidze}},
  \bibinfo {author} {\bibfnamefont {W.}~\bibnamefont {Hofstetter}}, \bibinfo
  {author} {\bibfnamefont {M.}~\bibnamefont {Snoek}}, \ and\ \bibinfo {author}
  {\bibfnamefont {N.}~\bibnamefont {Bl\"umer}},\ }\href {\doibase
  10.1103/PhysRevLett.105.065301} {\bibfield  {journal} {\bibinfo  {journal}
  {Phys. Rev. Lett.}\ }\textbf {\bibinfo {volume} {105}},\ \bibinfo {pages}
  {065301} (\bibinfo {year} {2010})}\BibitemShut {NoStop}%
\bibitem [{\citenamefont {Schwabe}, \citenamefont {Titvinidze},\ and\
  \citenamefont {Potthoff}(2013)}]{sc.ti.13}%
  \BibitemOpen
  \bibfield  {author} {\bibinfo {author} {\bibfnamefont {A.}~\bibnamefont
  {Schwabe}}, \bibinfo {author} {\bibfnamefont {I.}~\bibnamefont {Titvinidze}},
  \ and\ \bibinfo {author} {\bibfnamefont {M.}~\bibnamefont {Potthoff}},\
  }\href {\doibase 10.1103/PhysRevB.88.121107} {\bibfield  {journal} {\bibinfo
  {journal} {Phys. Rev. B}\ }\textbf {\bibinfo {volume} {88}},\ \bibinfo
  {pages} {121107} (\bibinfo {year} {2013})}\BibitemShut {NoStop}%
\bibitem [{\citenamefont {Aulbach}, \citenamefont {Titvinidze},\ and\
  \citenamefont {Potthoff}(2015)}]{au.ti.15}%
  \BibitemOpen
  \bibfield  {author} {\bibinfo {author} {\bibfnamefont {M.~W.}\ \bibnamefont
  {Aulbach}}, \bibinfo {author} {\bibfnamefont {I.}~\bibnamefont {Titvinidze}},
  \ and\ \bibinfo {author} {\bibfnamefont {M.}~\bibnamefont {Potthoff}},\
  }\href {\doibase 10.1103/PhysRevB.91.174420} {\bibfield  {journal} {\bibinfo
  {journal} {Phys. Rev. B}\ }\textbf {\bibinfo {volume} {91}},\ \bibinfo
  {pages} {174420} (\bibinfo {year} {2015})}\BibitemShut {NoStop}%
\bibitem [{\citenamefont {Aoki}\ \emph {et~al.}(2014)\citenamefont {Aoki},
  \citenamefont {Tsuji}, \citenamefont {Eckstein}, \citenamefont {Kollar},
  \citenamefont {Oka},\ and\ \citenamefont {Werner}}]{ao.ts.14}%
  \BibitemOpen
  \bibfield  {author} {\bibinfo {author} {\bibfnamefont {H.}~\bibnamefont
  {Aoki}}, \bibinfo {author} {\bibfnamefont {N.}~\bibnamefont {Tsuji}},
  \bibinfo {author} {\bibfnamefont {M.}~\bibnamefont {Eckstein}}, \bibinfo
  {author} {\bibfnamefont {M.}~\bibnamefont {Kollar}}, \bibinfo {author}
  {\bibfnamefont {T.}~\bibnamefont {Oka}}, \ and\ \bibinfo {author}
  {\bibfnamefont {P.}~\bibnamefont {Werner}},\ }\href {\doibase
  10.1103/RevModPhys.86.779} {\bibfield  {journal} {\bibinfo  {journal} {Rev.
  Mod. Phys.}\ }\textbf {\bibinfo {volume} {86}},\ \bibinfo {pages} {779}
  (\bibinfo {year} {2014})}\BibitemShut {NoStop}%
\bibitem [{\citenamefont {Schmidt}\ and\ \citenamefont {Monien}()}]{sc.mo.02u}%
  \BibitemOpen
  \bibfield  {author} {\bibinfo {author} {\bibfnamefont {P.}~\bibnamefont
  {Schmidt}}\ and\ \bibinfo {author} {\bibfnamefont {H.}~\bibnamefont
  {Monien}},\ }\href {http://arxiv.org/abs/cond-mat/0202046} {\enquote
  {\bibinfo {title} {Nonequilibrium dynamical mean -- field theory of a
  strongly correlated system},}\ }\bibinfo {note}
  {ArXiv:cond-mat/0202046}\BibitemShut {NoStop}%
\bibitem [{\citenamefont {Freericks}, \citenamefont {Turkowski},\ and\
  \citenamefont {Zlati{\'{c}}}(2006)}]{fr.tu.06}%
  \BibitemOpen
  \bibfield  {author} {\bibinfo {author} {\bibfnamefont {J.~K.}\ \bibnamefont
  {Freericks}}, \bibinfo {author} {\bibfnamefont {V.~M.}\ \bibnamefont
  {Turkowski}}, \ and\ \bibinfo {author} {\bibfnamefont {V.}~\bibnamefont
  {Zlati{\'{c}}}},\ }\href {\doibase 10.1103/PhysRevLett.97.266408} {\bibfield
  {journal} {\bibinfo  {journal} {Phys. Rev. Lett.}\ }\textbf {\bibinfo
  {volume} {97}},\ \bibinfo {pages} {266408} (\bibinfo {year}
  {2006})}\BibitemShut {NoStop}%
\bibitem [{\citenamefont {Freericks}(2008)}]{free.08}%
  \BibitemOpen
  \bibfield  {author} {\bibinfo {author} {\bibfnamefont {J.~K.}\ \bibnamefont
  {Freericks}},\ }\href {\doibase 10.1103/PhysRevB.77.075109} {\bibfield
  {journal} {\bibinfo  {journal} {Phys. Rev. B}\ }\textbf {\bibinfo {volume}
  {77}},\ \bibinfo {pages} {075109} (\bibinfo {year} {2008})}\BibitemShut
  {NoStop}%
\bibitem [{\citenamefont {Joura}, \citenamefont {Freericks},\ and\
  \citenamefont {Pruschke}(2008)}]{jo.fr.08}%
  \BibitemOpen
  \bibfield  {author} {\bibinfo {author} {\bibfnamefont {A.~V.}\ \bibnamefont
  {Joura}}, \bibinfo {author} {\bibfnamefont {J.~K.}\ \bibnamefont
  {Freericks}}, \ and\ \bibinfo {author} {\bibfnamefont {T.}~\bibnamefont
  {Pruschke}},\ }\href {\doibase 10.1103/PhysRevLett.101.196401} {\bibfield
  {journal} {\bibinfo  {journal} {Phys. Rev. Lett.}\ }\textbf {\bibinfo
  {volume} {101}},\ \bibinfo {pages} {196401} (\bibinfo {year}
  {2008})}\BibitemShut {NoStop}%
\bibitem [{\citenamefont {Eckstein}, \citenamefont {Kollar},\ and\
  \citenamefont {Werner}(2009)}]{ec.ko.09}%
  \BibitemOpen
  \bibfield  {author} {\bibinfo {author} {\bibfnamefont {M.}~\bibnamefont
  {Eckstein}}, \bibinfo {author} {\bibfnamefont {M.}~\bibnamefont {Kollar}}, \
  and\ \bibinfo {author} {\bibfnamefont {P.}~\bibnamefont {Werner}},\ }\href
  {\doibase 10.1103/PhysRevLett.103.056403} {\bibfield  {journal} {\bibinfo
  {journal} {Phys. Rev. Lett.}\ }\textbf {\bibinfo {volume} {103}},\ \bibinfo
  {pages} {056403} (\bibinfo {year} {2009})}\BibitemShut {NoStop}%
\bibitem [{\citenamefont {Arrigoni}, \citenamefont {Knap},\ and\ \citenamefont
  {von~der Linden}(2013)}]{ar.kn.13}%
  \BibitemOpen
  \bibfield  {author} {\bibinfo {author} {\bibfnamefont {E.}~\bibnamefont
  {Arrigoni}}, \bibinfo {author} {\bibfnamefont {M.}~\bibnamefont {Knap}}, \
  and\ \bibinfo {author} {\bibfnamefont {W.}~\bibnamefont {von~der Linden}},\
  }\href {\doibase 10.1103/PhysRevLett.110.086403} {\bibfield  {journal}
  {\bibinfo  {journal} {Phys. Rev. Lett.}\ }\textbf {\bibinfo {volume} {110}},\
  \bibinfo {pages} {086403} (\bibinfo {year} {2013})}\BibitemShut {NoStop}%
\bibitem [{\citenamefont {Titvinidze}\ \emph {et~al.}(2015)\citenamefont
  {Titvinidze}, \citenamefont {Dorda}, \citenamefont {von~der Linden},\ and\
  \citenamefont {Arrigoni}}]{ti.do.15}%
  \BibitemOpen
  \bibfield  {author} {\bibinfo {author} {\bibfnamefont {I.}~\bibnamefont
  {Titvinidze}}, \bibinfo {author} {\bibfnamefont {A.}~\bibnamefont {Dorda}},
  \bibinfo {author} {\bibfnamefont {W.}~\bibnamefont {von~der Linden}}, \ and\
  \bibinfo {author} {\bibfnamefont {E.}~\bibnamefont {Arrigoni}},\ }\href
  {\doibase 10.1103/PhysRevB.92.245125} {\bibfield  {journal} {\bibinfo
  {journal} {Phys. Rev. B}\ }\textbf {\bibinfo {volume} {92}},\ \bibinfo
  {pages} {245125} (\bibinfo {year} {2015})}\BibitemShut {NoStop}%
\bibitem [{\citenamefont {Dorda}, \citenamefont {Titvinidze},\ and\
  \citenamefont {Arrigoni}(2016)}]{do.ti.16}%
  \BibitemOpen
  \bibfield  {author} {\bibinfo {author} {\bibfnamefont {A.}~\bibnamefont
  {Dorda}}, \bibinfo {author} {\bibfnamefont {I.}~\bibnamefont {Titvinidze}}, \
  and\ \bibinfo {author} {\bibfnamefont {E.}~\bibnamefont {Arrigoni}},\ }\href
  {http://stacks.iop.org/1742-6596/696/i=1/a=012003} {\bibfield  {journal}
  {\bibinfo  {journal} {Journal of Physics: Conference Series}\ }\textbf
  {\bibinfo {volume} {696}},\ \bibinfo {pages} {012003} (\bibinfo {year}
  {2016})}\BibitemShut {NoStop}%
\bibitem [{\citenamefont {Kubo}(1957)}]{kubo.57}%
  \BibitemOpen
  \bibfield  {author} {\bibinfo {author} {\bibfnamefont {R.}~\bibnamefont
  {Kubo}},\ }\href {\doibase 10.1143/JPSJ.12.570} {\bibfield  {journal}
  {\bibinfo  {journal} {Journal of the Physical Society of Japan}\ }\textbf
  {\bibinfo {volume} {12}},\ \bibinfo {pages} {570} (\bibinfo {year}
  {1957})}\BibitemShut {NoStop}%
\bibitem [{\citenamefont {Schwinger}(1961)}]{schw.61}%
  \BibitemOpen
  \bibfield  {author} {\bibinfo {author} {\bibfnamefont {J.}~\bibnamefont
  {Schwinger}},\ }\href {http://dx.doi.org/10.1063/1.1703727} {\bibfield
  {journal} {\bibinfo  {journal} {J. Math. Phys.}\ }\textbf {\bibinfo {volume}
  {2}},\ \bibinfo {pages} {407} (\bibinfo {year} {1961})}\BibitemShut {NoStop}%
\bibitem [{\citenamefont {Baym}\ and\ \citenamefont
  {Kadanoff}(1961)}]{ba.ka.61}%
  \BibitemOpen
  \bibfield  {author} {\bibinfo {author} {\bibfnamefont {G.}~\bibnamefont
  {Baym}}\ and\ \bibinfo {author} {\bibfnamefont {L.~P.}\ \bibnamefont
  {Kadanoff}},\ }\href {\doibase 10.1103/PhysRev.124.287} {\bibfield  {journal}
  {\bibinfo  {journal} {Phys. Rev.}\ }\textbf {\bibinfo {volume} {124}},\
  \bibinfo {pages} {287} (\bibinfo {year} {1961})}\BibitemShut {NoStop}%
\bibitem [{\citenamefont {Kadanoff}\ and\ \citenamefont
  {Baym}(1962)}]{kad.baym}%
  \BibitemOpen
  \bibfield  {author} {\bibinfo {author} {\bibfnamefont {L.~P.}\ \bibnamefont
  {Kadanoff}}\ and\ \bibinfo {author} {\bibfnamefont {G.}~\bibnamefont
  {Baym}},\ }\href@noop {} {\emph {\bibinfo {title} {Quantum Statistical
  Mechanics: Green's Function Methods in Equilibrium and Nonequilibrium
  Problems}}}\ (\bibinfo  {publisher} {Addison-Wesley},\ \bibinfo {address}
  {Redwood City, CA},\ \bibinfo {year} {1962})\BibitemShut {NoStop}%
\bibitem [{\citenamefont {Keldysh}(1965)}]{keld.65}%
  \BibitemOpen
  \bibfield  {author} {\bibinfo {author} {\bibfnamefont {L.~V.}\ \bibnamefont
  {Keldysh}},\ }\href
  {http://www.jetp.ac.ru/cgi-bin/e/index/e/20/4/p1018?a=list} {\bibfield
  {journal} {\bibinfo  {journal} {Sov. Phys. JETP}\ }\textbf {\bibinfo {volume}
  {20}},\ \bibinfo {pages} {1018} (\bibinfo {year} {1965})}\BibitemShut
  {NoStop}%
\bibitem [{\citenamefont {Anderson}(1961)}]{ande.61}%
  \BibitemOpen
  \bibfield  {author} {\bibinfo {author} {\bibfnamefont {P.~W.}\ \bibnamefont
  {Anderson}},\ }\href {\doibase 10.1103/PhysRev.124.41} {\bibfield  {journal}
  {\bibinfo  {journal} {Phys. Rev.}\ }\textbf {\bibinfo {volume} {124}},\
  \bibinfo {pages} {41} (\bibinfo {year} {1961})}\BibitemShut {NoStop}%
\bibitem [{\citenamefont {Dorda}\ \emph {et~al.}(2014)\citenamefont {Dorda},
  \citenamefont {Nuss}, \citenamefont {von~der Linden},\ and\ \citenamefont
  {Arrigoni}}]{do.nu.14}%
  \BibitemOpen
  \bibfield  {author} {\bibinfo {author} {\bibfnamefont {A.}~\bibnamefont
  {Dorda}}, \bibinfo {author} {\bibfnamefont {M.}~\bibnamefont {Nuss}},
  \bibinfo {author} {\bibfnamefont {W.}~\bibnamefont {von~der Linden}}, \ and\
  \bibinfo {author} {\bibfnamefont {E.}~\bibnamefont {Arrigoni}},\ }\href
  {\doibase 10.1103/PhysRevB.89.165105} {\bibfield  {journal} {\bibinfo
  {journal} {Phys. Rev. B}\ }\textbf {\bibinfo {volume} {89}},\ \bibinfo
  {pages} {165105} (\bibinfo {year} {2014})}\BibitemShut {NoStop}%
\bibitem [{\citenamefont {POTZ}(1989)}]{potz.89}%
  \BibitemOpen
  \bibfield  {author} {\bibinfo {author} {\bibfnamefont {W.}~\bibnamefont
  {POTZ}},\ }\href {http://link.aip.org/link/JAPIAU/v66/i6/p2458/s1} {\bibfield
   {journal} {\bibinfo  {journal} {JOURNAL OF APPLIED PHYSICS}\ }\textbf
  {\bibinfo {volume} {66}},\ \bibinfo {pages} {2458} (\bibinfo {year}
  {1989})}\BibitemShut {NoStop}%
\bibitem [{\citenamefont {Haug}\ and\ \citenamefont {Jauho}(1998)}]{ha.ja}%
  \BibitemOpen
  \bibfield  {author} {\bibinfo {author} {\bibfnamefont {H.}~\bibnamefont
  {Haug}}\ and\ \bibinfo {author} {\bibfnamefont {A.-P.}\ \bibnamefont
  {Jauho}},\ }\href@noop {} {\emph {\bibinfo {title} {Quantum Kinetics in
  Transport and Optics of Semiconductors}}}\ (\bibinfo  {publisher}
  {Springer},\ \bibinfo {address} {Heidelberg},\ \bibinfo {year}
  {1998})\BibitemShut {NoStop}%
\bibitem [{\citenamefont {Rammer}\ and\ \citenamefont
  {Smith}(1986)}]{ra.sm.86}%
  \BibitemOpen
  \bibfield  {author} {\bibinfo {author} {\bibfnamefont {J.}~\bibnamefont
  {Rammer}}\ and\ \bibinfo {author} {\bibfnamefont {H.}~\bibnamefont {Smith}},\
  }\href {\doibase 10.1103/RevModPhys.58.323} {\bibfield  {journal} {\bibinfo
  {journal} {Rev. Mod. Phys.}\ }\textbf {\bibinfo {volume} {58}},\ \bibinfo
  {pages} {323} (\bibinfo {year} {1986})}\BibitemShut {NoStop}%
\bibitem [{\citenamefont {Dorda}\ \emph {et~al.}(2017)\citenamefont {Dorda},
  \citenamefont {Sorantin}, \citenamefont {von~der Linden},\ and\ \citenamefont
  {Arrigoni}}]{do.so.17}%
  \BibitemOpen
  \bibfield  {author} {\bibinfo {author} {\bibfnamefont {A.}~\bibnamefont
  {Dorda}}, \bibinfo {author} {\bibfnamefont {M.}~\bibnamefont {Sorantin}},
  \bibinfo {author} {\bibfnamefont {W.}~\bibnamefont {von~der Linden}}, \ and\
  \bibinfo {author} {\bibfnamefont {E.}~\bibnamefont {Arrigoni}},\ }\href
  {http://stacks.iop.org/1367-2630/19/i=6/a=063005} {\bibfield  {journal}
  {\bibinfo  {journal} {New Journal of Physics}\ }\textbf {\bibinfo {volume}
  {19}},\ \bibinfo {pages} {063005} (\bibinfo {year} {2017})}\BibitemShut
  {NoStop}%
\bibitem [{\citenamefont {Haydock}(1980)}]{hayd.80}%
  \BibitemOpen
  \bibfield  {author} {\bibinfo {author} {\bibfnamefont {R.}~\bibnamefont
  {Haydock}},\ }\href@noop {} {\emph {\bibinfo {title} {Solid State Physics,
  Advances in Research and Applications, edited by H. Ehrenreich, F. Seitz, and
  D. Turnbull}}},\ Vol.~\bibinfo {volume} {35}\ (\bibinfo  {publisher}
  {Academic, London},\ \bibinfo {address} {Academic},\ \bibinfo {year}
  {1980})\BibitemShut {NoStop}%
\bibitem [{\citenamefont {Thouless}\ and\ \citenamefont
  {Kirkpatrick}(1981)}]{th.ki.81}%
  \BibitemOpen
  \bibfield  {author} {\bibinfo {author} {\bibfnamefont {D.~J.}\ \bibnamefont
  {Thouless}}\ and\ \bibinfo {author} {\bibfnamefont {S.}~\bibnamefont
  {Kirkpatrick}},\ }\href {http://stacks.iop.org/0022-3719/14/i=3/a=007}
  {\bibfield  {journal} {\bibinfo  {journal} {Journal of Physics C: Solid State
  Physics}\ }\textbf {\bibinfo {volume} {14}},\ \bibinfo {pages} {235}
  (\bibinfo {year} {1981})}\BibitemShut {NoStop}%
\bibitem [{\citenamefont {Lewenkopf}\ and\ \citenamefont
  {Mucciolo}(2013)}]{le.ca.13}%
  \BibitemOpen
  \bibfield  {author} {\bibinfo {author} {\bibfnamefont {C.~H.}\ \bibnamefont
  {Lewenkopf}}\ and\ \bibinfo {author} {\bibfnamefont {E.~R.}\ \bibnamefont
  {Mucciolo}},\ }\href {\doibase 10.1007/s10825-013-0458-7} {\bibfield
  {journal} {\bibinfo  {journal} {Journal of Computational Electronics}\
  }\textbf {\bibinfo {volume} {12}},\ \bibinfo {pages} {203} (\bibinfo {year}
  {2013})}\BibitemShut {NoStop}%
\bibitem [{\citenamefont {Dorda}\ \emph {et~al.}(2015)\citenamefont {Dorda},
  \citenamefont {Ganahl}, \citenamefont {Evertz}, \citenamefont {von~der
  Linden},\ and\ \citenamefont {Arrigoni}}]{do.ga.15}%
  \BibitemOpen
  \bibfield  {author} {\bibinfo {author} {\bibfnamefont {A.}~\bibnamefont
  {Dorda}}, \bibinfo {author} {\bibfnamefont {M.}~\bibnamefont {Ganahl}},
  \bibinfo {author} {\bibfnamefont {H.~G.}\ \bibnamefont {Evertz}}, \bibinfo
  {author} {\bibfnamefont {W.}~\bibnamefont {von~der Linden}}, \ and\ \bibinfo
  {author} {\bibfnamefont {E.}~\bibnamefont {Arrigoni}},\ }\href {\doibase
  10.1103/PhysRevB.92.125145} {\bibfield  {journal} {\bibinfo  {journal} {Phys.
  Rev. B}\ }\textbf {\bibinfo {volume} {92}},\ \bibinfo {pages} {125145}
  (\bibinfo {year} {2015})}\BibitemShut {NoStop}%
\bibitem [{\citenamefont {Breuer}, \citenamefont {Kappler},\ and\ \citenamefont
  {Petruccione}(1997)}]{br.ka.97}%
  \BibitemOpen
  \bibfield  {author} {\bibinfo {author} {\bibfnamefont {H.-P.}\ \bibnamefont
  {Breuer}}, \bibinfo {author} {\bibfnamefont {B.}~\bibnamefont {Kappler}}, \
  and\ \bibinfo {author} {\bibfnamefont {F.}~\bibnamefont {Petruccione}},\
  }\href {\doibase 10.1103/PhysRevA.56.2334} {\bibfield  {journal} {\bibinfo
  {journal} {Phys. Rev. A}\ }\textbf {\bibinfo {volume} {56}},\ \bibinfo
  {pages} {2334} (\bibinfo {year} {1997})}\BibitemShut {NoStop}%
\bibitem [{Qua()}]{Quantum_Jumps}%
  \BibitemOpen
  \href@noop {} {}\bibinfo {note} {M. Sorantin, D. Fugger, A. Dorda, W. von der
  Linden, and E. Arrigoni in preperation}\BibitemShut {NoStop}%
\bibitem [{z_d()}]{z_dependent_layers}%
  \BibitemOpen
  \href@noop {} {}\bibinfo {note} {Our formalism can be easily generalized in
  the case when filling of the layers are different from each other. In this
  case we should just change $n^{\rm bulk}$ by $n_z^{\rm bulk}$.}\BibitemShut
  {Stop}%
\bibitem [{rem()}]{reminder_nz_depends_on_v}%
  \BibitemOpen
  \href@noop {} {}\bibinfo {note} {Recall that $\vec{n}$ depend on $\vec{v}$
  and therefore needs to be recalculated every time $\vec{v}$ is
  updated.}\BibitemShut {Stop}%
\bibitem [{end()}]{endwith_Sigma}%
  \BibitemOpen
  \href@noop {} {}\bibinfo {note} {Alternatively, one can also consider
  differences in the self-energies.}\BibitemShut {Stop}%
\bibitem [{abo({\natexlab{a}})}]{about_AMEA}%
  \BibitemOpen
  \href@noop {} {}  \bibinfo {note} {for details of AMEA and
  the accuracy expected by taking a certain number $N_b$ of bath sites see,
  e.g. ~\cite{do.nu.14,ti.do.15,do.so.17}. Notice that the accuracy obtained
  with a certain $N_b$ in AMEA is expected to be the same or better than with
  twice the same number of bath sites ($2 N_b$) in exact diagonalisation.
  Furthermore, the AMEA spectrum is continuous without the need of an
  artificial broadening.}\BibitemShut {Stop}%
\bibitem [{\citenamefont {Hofstetter}, \citenamefont {Bulla},\ and\
  \citenamefont {Vollhardt}(2000)}]{ho.bu.00}%
  \BibitemOpen
  \bibfield  {author} {\bibinfo {author} {\bibfnamefont {W.}~\bibnamefont
  {Hofstetter}}, \bibinfo {author} {\bibfnamefont {R.}~\bibnamefont {Bulla}}, \
  and\ \bibinfo {author} {\bibfnamefont {D.}~\bibnamefont {Vollhardt}},\ }\href
  {\doibase 10.1103/PhysRevLett.84.4417} {\bibfield  {journal} {\bibinfo
  {journal} {Phys. Rev. Lett.}\ }\textbf {\bibinfo {volume} {84}},\ \bibinfo
  {pages} {4417} (\bibinfo {year} {2000})}\BibitemShut {NoStop}%
\bibitem [{abo({\natexlab{b}})}]{about_Current}%
  \BibitemOpen
  \href@noop {} {} \bibinfo {note} {due to our approximation
  in the impurity solver there is a small deviation from current conservation
  on the correlated bonds. Therefore, our results show the average of the
  current over the correlated bonds and one of the uncorrelated ones. However,
  we do not show explicit error-bars since they are within the symbol size of
  the presented results , Fig. \ref{Fig:Current}.}\BibitemShut {Stop}%
\end{thebibliography}
%

\end{document}